\documentclass[11pt]{article}
\usepackage[utf8]{inputenc}
\usepackage[T1]{fontenc}
\pdfoutput = 1
\usepackage{dsdshorthand}
\usepackage{myoptionsgravityfactorized}
\usepackage{upgreek}

\usepackage{amsmath}
\usepackage{amssymb}
\usepackage{bbm}
\usepackage{bbold}
\usepackage{esint}
\usepackage{braket}
\allowdisplaybreaks
\usepackage{physics}
\usepackage{caption}
\usepackage{cite}
\usepackage{color}
    \definecolor{darkgreen}{rgb}{0,0.5,0}
    \definecolor{darkred}{rgb}{0.5,0,0}
    \definecolor{darkblue}{rgb}{0,0,0.6}
    \definecolor{purple}{rgb}{0.4,.2,0.7}
\usepackage[mathcal]{eucal}
\usepackage{float}
\usepackage{graphicx}

\usepackage{tikz}
\usetikzlibrary{arrows,positioning} 
\tikzset{
    >=stealth',
    punkt/.style={
           rectangle,
           rounded corners,
           draw=black, very thick,
           text width=15em,
           minimum height=2em,
           text centered},
    pil/.style={
           ->,
           thick,
           shorten <=2pt,
           shorten >=2pt,}
}

\usepackage[margin = 2.2cm]{geometry}
\renewcommand{\d}{\mathrm{d}}
\renewcommand{\i}{\mathrm{i}}
\newcommand{\Ss}{\textsf{S}_0}
\newcommand{\Hh}{{\textsf{H}_0}}
\newcommand{\HJT}{{\textsf{H}_\text{JT}}}

\newcommand{\average}[1]{\left\langle #1 \right\rangle}

\def\nn{\nonumber}

\def\ba{\begin{array}}
\def\ea{\end{array}}

\def\td{\tilde}

\def\dalemb#1#2{{\vbox{\hrule height .#2pt
      \hbox{\vrule width.#2pt height#1pt \kern#1pt
              \vrule width.#2pt}
      \hrule height.#2pt}}}

\def\cont{{{\mathcal{C}}}}
\def\mo{{{\mathcal{O}}}}

\def\bn{\boldsymbol{n}}

\let\tilde=\widetilde

\def\fHH{{\text{HH}_{\rm fact}}}

\def\brane{\text{brane}}

\def\S{\textsf{S}_0}

\renewcommand{\d}{\mathrm{d}}
\renewcommand{\i}{\mathrm{i}}
\renewcommand{\tilde}{\widetilde}

\numberwithin{equation}{section}

\begin{document}

\thispagestyle{empty}
\begin{center}
    ~\vspace{5mm}
    
         {\LARGE \bf Alpha states demystified\\\vspace{0.15in}\Large\bf Towards microscopic models of AdS$_2$ holography}

    \vspace{0.4in}
    
    {\bf Andreas Blommaert$^1$, Luca V. Iliesiu$^2$ and Jorrit Kruthoff$\,^2$}

    \vspace{0.4in}
    {$^1$SISSA, Via Bonomea 265, 34127 Trieste, Italy\\$^2$Department of Physics, Stanford University, Stanford, CA 94305, USA}
    \vspace{0.1in}
    
    {\tt ablommae@sissa.it, liliesiu@stanford.edu, kruthoff@stanford.edu}
\end{center}

\vspace{0.4in}

\begin{abstract}
\noindent We continue our study of factorizing theories of dilaton gravity, characterized by a universal bilocal interaction. All such factorizing theories can be shown to have discrete spectra, distinguished only by their local dilaton potentials. We show how such theories can be used to construct all alpha-states in the Hilbert space of baby universes of ordinary JT gravity. Large classes of these theories with different local potentials are found to be non-perturbatively equivalent and have identical discrete spectra. This is a concrete example of how different bulk descriptions can give rise to the same boundary theory. Such equivalences manifest themselves as null states, which have to be quotiented out in order to construct a proper baby universe Hilbert space. Our results also allow us to revisit the mechanism discussed by Coleman, Giddings and Strominger and concretely link ensemble averaging to the appearance or disappearance of spacetime wormholes.   

We then investigate JT gravity deformed only by the universal bilocal interaction. In this theory, the only terms that do not cancel in a topological expansion are disks, which capture perturbative fluctuations around a two-dimensional black hole saddle.
We find that this theory of black holes has an evenly spaced spectrum, instead of a quantum chaotic one. We present a dual quantum mechanical system with exactly the same discrete spectrum, and propose that this is an example of a new holographic duality between a two-dimensional theory of quantum gravity and a conventional  quantum mechanics. 
\end{abstract}

\pagebreak
\setcounter{page}{1}
\tableofcontents

\newpage

\section{Introduction and summary}\label{sec:intro}

There is a widely held belief that UV complete theories of quantum gravity in AdS are dual to individual non-gravitational quantum mechanical systems on the boundary of AdS \cite{Maldacena:1997re}. On the other hand, if we consider two copies of such a boundary quantum system, in gravity, we are instructed to sum over all geometries that end on two copies of such asymptotically AdS boundary. This includes summing over wormhole geometries connecting the two boundaries. Those have desirable properties, explaining many non-trivial predictions of the boundary quantum mechanics from the bulk point of view, like the Page curve \cite{Almheiri:2019qdq,Penington:2019kki, Almheiri:2020cfm}, the non-decaying behavior of correlation functions \cite{Saad:2018bqo,Saad:2019pqd,Blommaert:2019hjr}, complexity \cite{Iliesiu:2021ari} and the lack of global symmetries \cite{Hsin:2020mfa,Chen:2020ojn}. To explore such features of gravity we seemingly need wormholes. However wormholes also introduce a tension with the dual quantum mechanics, known as the factorization puzzle - naively, wormholes imply a nontrivial variance for say the two-boundary partition function $Z(\beta_1,\beta_2)\neq Z(\beta_1)Z(\beta_2)$, in contradiction with the dual quantum mechanics where partition functions are just ordinary numbers. This means that  gravity systems are naively dual to ensembles \cite{Saad:2019lba,Saad:2019pqd,Almheiri:2019qdq,Penington:2019kki,Marolf:2020xie,Stanford:2020wkf,Blommaert:2019wfy,Blommaert:2020seb,Pollack:2020gfa,Afkhami-Jeddi:2020ezh,Maloney:2020nni,Belin:2020hea,Cotler:2020ugk,Anous:2020lka,Chen:2020tes,Liu:2020jsv,Marolf:2021kjc,Meruliya:2021utr,Giddings:2020yes,Stanford:2019vob,Okuyama:2019xbv,Belin:2020jxr,Verlinde:2021jwu,Cotler:2020lxj, Collier:2021rsn,Betzios:2021fnm,Belin:2021ryy,Saad:2021uzi}, unless something cancels the variance in this ensemble. Since we believe that UV complete theories are  dual to individual quantum mechanical systems and therefore do factorize, we should identify the effect that cancels the variance due to wormholes.

In \cite{Blommaert:2021fob} we investigated, in a bottom-up approach, how factorization of UV complete models trickles down to the low energy EFT. In models that reduce to JT gravity in the IR, we found that factorization demands the existence of a tiny but universal bilocal  correction (in spacetime), which can counter the variance due to wormholes. Schematically
\begin{equation}
    \label{eq:factorization-leading-order}
         \begin{tikzpicture}[baseline={([yshift=-.5ex]current bounding box.center)}, scale=0.65]
 \pgftext{\includegraphics[scale=0.55]{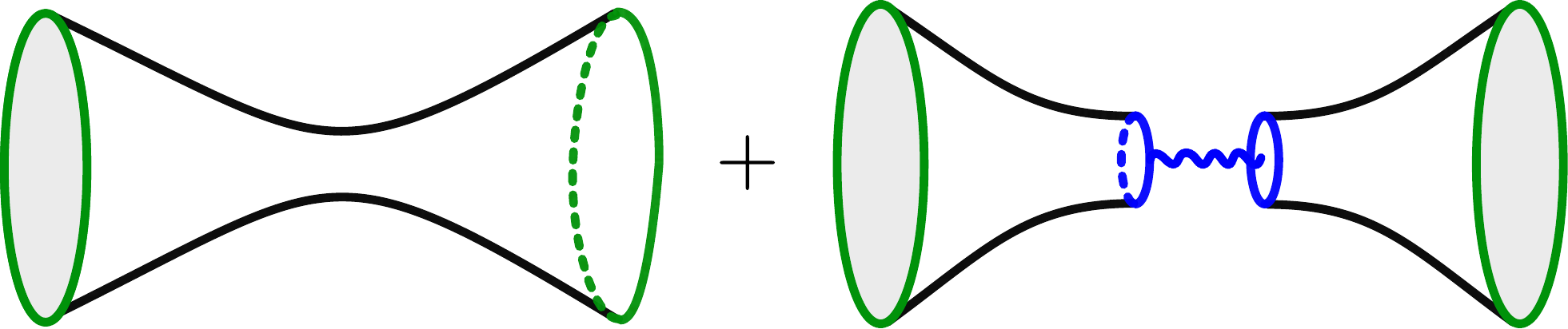}} at (0,0);
  \end{tikzpicture} \,\,=\,\, 0\,,
\end{equation}
where the wiggly blue line represent the effect of the bilocal correction which makes seemingly disconnected parts of the spacetime interact. 
The bilocal is universal because the same interaction results in factorization for all the dilaton gravities obtained by adding any possible local dilaton potential to JT \cite{Witten:2020ert,Maxfield:2020ale}
\be
I[U_\text{local}]= -\frac{1}{2}\int_\Sigma \d^2 x \sqrt{g} \,\Phi(R + 2) -\frac{1}{2}\int_\Sigma \d^2 x \sqrt{g}\, U_\text{local}(\Phi)+\text{universal bilocal.}\label{theories}
\ee
We find that this bilocal interaction always results in a discrete spectrum in the gravitational path integral of \eqref{theories}: in other words, at least in our models, factorization always implies spectral discreteness. This discrete spectrum of black hole states is determined entirely by the local dilaton potential $U_\text{local}(\Phi)$. 

The construction of such theories gives us the opportunity to make a few widely discussed concepts in quantum gravity rather concrete, which will be the purpose of this paper. Our main results are the following:
\begin{enumerate}
    \item \textbf{Explicit construction of alpha-states in JT gravity.} The baby universe Hilbert space of any theory can be constructed by acting on a state with no-boundaries, typically called the Hartle-Hawking state, with boundary creation operators. Alpha-states are the special states in this space that are eigenfunctions of all such operators, leading to inner products in the baby universe Hilbert space that factorize.  Marolf and Maxfield \cite{Marolf:2020xie} constructed such states in a 2d topological model that does not exhibit any of the rich physics of black holes. One of the achievement in this paper is to give a concrete geometric construction for all the alpha-states of JT gravity, a model which has served as a great exploration ground for black hole physics. 
 We find that each factorizing model of dilation gravity \eqref{theories} can be mapped to an \emph{explicit} operator $e^{-\hat I_{\rm deform}[U_\text{local}]}$ \eqref{eq:boundary-creation-operator}, which is formed from boundary creation operators and can thus act on the Hartle-Hawking state. When acting on this state, such operators create the alpha-states of JT gravity
\be 
e^{-I[U_\text{local}]} = e^{-I_{\rm JT} - I_{\rm local} - I_{\text{nonlocal}}}  \quad\Leftrightarrow \quad \ket{\alpha }= \cN_{\alpha}\, e^{-\hat I_{\rm deform}[U_\text{local}]} \ket{\text{HH}}\,.
\label{eq:O-alpha-to-alpha}
\ee
Each alpha-state $\ket{\alpha}$ is associated to an energy spectrum that is the same as that of the factorizing dilaton gravity with potential $U_\text{local}$. All spectra can be obtained by choosing suitable potentials $U_\text{local}$ and, therefore, we obtain a basis of alpha-states that spans the entire baby universe Hilbert space. Formula \eqref{eq:O-alpha-to-alpha} shows two perspectives on alpha-states in JT. Either an alpha-state provides a complicated prescription for the path integral where we sum over all geometries that end on the many boundaries, present in $e^{-\hat I_{\rm deform}[U_\text{local}]}$. Alternatively, a more physical picture is that an alpha-state can be viewed as not including additional boundaries, but as specifying the spacetime action to $I[U_\text{local}]$ in \eqref{theories} of a factorizing dilaton gravity theory. The fact that these two perspectives are the same is a type of open-closed duality that we shall explain in \textbf{section \ref{sect:alpha}}. Having the exact form of all alpha-states \eqref{eq:O-alpha-to-alpha} allows to explicitly revisit and shed light on the original proposals of Coleman, Giddings and Strominger \cite{coleman1988black, giddings1988loss}: wormholes can be ``integrated-out'' by taking an ensemble average over couplings in bulk theories that preserve locality,\footnote{Not to be confused with the ensemble average typically discussed in the boundary theory.} or they can be ''integrated-in'' by considering an average in bulk theories where the universal non-local term is present.\footnote{This latter ensemble average turns out to be equivalent to the ensemble typically discussed in the boundary theory.}

\item \textbf{Coexistence of different bulk descriptions yields null states.} As described above, theories with different $U_\text{local}$ can have identical spectra and can therefore be thought of as quantum gravities with different actions that have identical non-perturbative UV completions.  In \textbf{section~\ref{sect:null}} we present and investigate large classes of such examples. Denoting the local potentials of two theories with an identical energy spectrum by $U_\text{local}^{(1)}$ and  $U_\text{local}^{(2)}$, the associated alpha-states will be indistinguishable by any measurement, which means they are physically equivalent. This causes the difference of these states to be null (meaning it has zero overlap with all states, including itself)
\begin{equation}
  \left(\cN_{\alpha^{(1)}} e^{-\hat I_{\rm deform}[U_\text{local}^{(1)}]} - \cN_{\alpha^{(2)}} e^{-\hat I_{\rm deform}[U_\text{local}^{(2)}]} \right)\ket{\text{HH}}\sim 0\,.
\end{equation}
The physical Hilbert space is obtained after quotienting out such null states \cite{Marolf:2020xie}. This highlights a physical interpretation of null states, which was not obvious in the multi-boundary description of Marolf and Maxfield. They describe redundancies in the spacetime action of the gravitational path integral: theories with different actions can be non-perturbatively equivalent, even if they have different semiclassical descriptions. For instance, they can even have different black hole solutions:
\begin{equation} \begin{tikzpicture}[baseline={([yshift=0cm]current bounding box.center)}, scale=0.6 ]
 \pgftext{\includegraphics[scale=0.45]{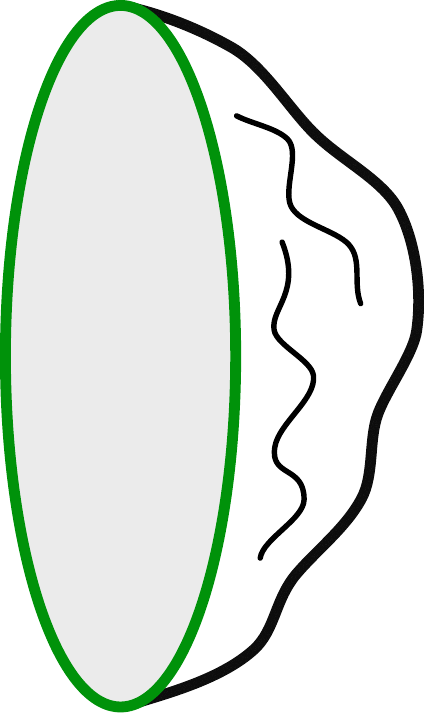}} at (0,0);
  \end{tikzpicture}\hspace{0.3cm}+\hspace{0.3cm}\text{subleading corrections} \quad
 =
  \quad \begin{tikzpicture}[baseline={([yshift=0cm]current bounding box.center)}, scale=0.6]
 \pgftext{\includegraphics[scale=0.4]{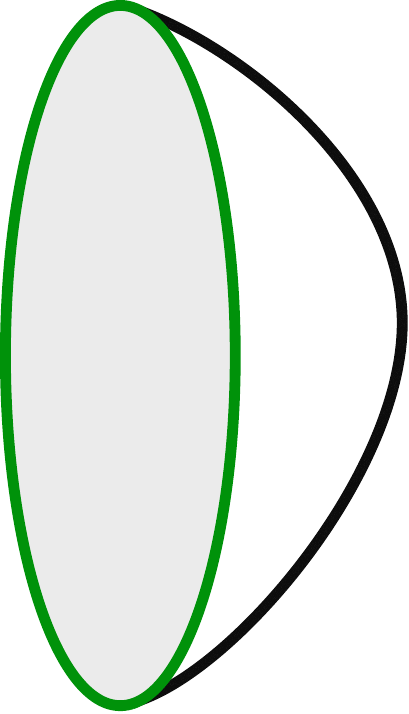}} at (0,0);
  \end{tikzpicture}\hspace{0.3cm}+\hspace{0.3cm}\text{ subleading corrections, } 
\end{equation}
where on the left, we have schematically represented in the semi-classical contribution to the partition function in theory 1, while on the right we represented it in theory 2. The meaning of the subleading corrections will be concretely addressed. 
This is an explicit example showing that \emph{multiple bulk descriptions can coexist}. This is a concrete realization in gravity of this idea put forward in \cite{Saad:2021rcu} in the context of the SYK model. 
    
\item \textbf{A quantum mechanical dual for black holes without a chaotic spectrum.} In \textbf{section \ref{sec:gravity_without_chaos}} we study \emph{canonical} JT gravity, the factorized version of old-school JT gravity: \eqref{theories} with $U_\text{local}=0$. In the geometric expansion, it only receives contributions from the disk geometry
\begin{equation}
\label{eq:disk+nothing}
    Z(\beta)= \quad \begin{tikzpicture}[baseline={([yshift=0cm]current bounding box.center)}, scale=0.6]
 \pgftext{\includegraphics[scale=0.4]{disk.pdf}} at (0,0);
  \end{tikzpicture}\hspace{0.3cm}+\hspace{0.3cm}\text{ non-perturbative effects}\,,
\end{equation}
where this geometry captures fluctuations around JT gravity's black hole saddle.\footnote{This makes it similar to what \cite{Harlow:2018tqv, harlowUCSBTalk} call canonical JT gravity. There the disk was defined to be the sole contribution. The important distinction between our approach and that of \cite{Harlow:2018tqv, harlowUCSBTalk} is that we have a matrix integral description which allows us to explore the discreteness of the spectrum and the precise quantum mechanical dual of the bulk theory. } Nevertheless, we show via the deformed matrix integral that our version of canonical JT gravity has a \emph{discrete} spectrum, instead of a continuous one \cite{Harlow:2018tqv}. This spectrum is obtained from the following QM problem\footnote{This Hamiltonian was also recently discussed in \cite{Johnson:2022wsr}, however that paper concerns only the ensemble averaged version of JT gravity, which does not factorize, nor have a discrete spectrum. The point of \cite{Johnson:2022wsr} is that in some sense, even within the ensemble, there is a ``preferred'' spectrum, whereas we are studying the microscopic theory with that spectrum which can be explicitly determined from the path integral in the gravitational theory.}
    \begin{equation}
       \HJT\,\psi(x,\l)= \l\, \psi(x,\l)\,,\quad \HJT=-e^{-2\Ss}\frac{\d^2}{\d x^2}+u(x)\,,\quad \psi(0,\l)=0\,,
    \end{equation}
    where $u(x)$ is obtained by solving the JT string equation, which to leading order is $\sqrt{u} I_1(2\pi \sqrt{u}) = 2\pi x$. Apart from the crucial Dirichlet boundary condition at $x=0$ this is the same QM problem that one considers in the orthogonal polynomials approach to matrix models \cite{Gross1989nonperturbative, Banks:1989df} (see \cite{Johnson:2019eik, Johnson:2020exp, Johnson:2021zuo, Johnson:2022wsr} for more recent discussions of this approach in the context of JT gravity). Surprisingly, since this should be a theory of black holes \cite{Cotler:2016fpe}, this spectrum does not have random matrix level statistics, but rather, adjacent eigenvalues behave as,\footnote{Here, $\rho_{0, {\rm JT}}(\l)$ is the leading density of stats in JT gravity.} 
    \begin{equation}
        \l_{i+1}-\l_{i}=\frac{1}{\rho_{0,{\rm JT}}(\l)}\,,
    \end{equation}
    and are thus (locally) evenly spaced. We also study the null deformations of this theory in detail in the semiclassical limit and see that they give rise to different semiclassical bulk descriptions.
\end{enumerate}


\section{Review of gravity factorized}\label{sect:review} 

The idea of \cite{Blommaert:2021fob} was to imagine starting from a factorizing UV complete bulk theory. After integrating out all degrees of freedom in the decoupling limit of a large near extremal black hole in such a theory, except for the 2d metric $g$ and the dilaton $\Phi$, one should also obtain a factorizing  theory in 2d. Such a theory is given by JT gravity at leading order in $e^{\S}$, with $\S$ the entropy of the black hole
\begin{align}
-I_{\rm JT}=\frac{1}{2} \int_{\Sigma} \d^2 x \sqrt{g}\, \Phi\,(R+2) + \int_{\partial \Sigma} \d u \sqrt{h}\, \Phi\,(K-1) +\S\chi(\Sigma)\,,
\end{align}
however it should also have both local and non-local corrections in its effective action. 
The goal of \cite{Blommaert:2021fob} was to find what kind of non-local terms are required in order for this 2d theory to factorize.

Surprisingly, the only non-local terms that result in a theory that factorizes to all orders in a genus expansion was \textit{universal} (it was independent of the detailed spectrum of black hole microstates that the resulting factorizing theory could have) and  \textit{unique} (other models do not factorize). Explicitly, the theory factorizes exactly if the non-local interaction is purely bilocal
\begin{align}
-I_{\text{nonlocal}}&= -\frac{1}{2}\,e^{-2\Ss}\,\int_0^\infty \d b\, b \int_{\Sigma_1} \d^2 x_1\sqrt{g_1}\int_{\Sigma_2} \d^2 x_2\sqrt{g_2}\,e^{-2\pi \left( \Phi_1 + \Phi_2\right)}\,\cos(b\Phi_1)\,\cos(b\Phi_2)\nn\\&=-\frac{1}{2}\int_0^\infty \d b\, b\,\mo_\text{G}(b,\Phi)\,\mo_\text{G}(b,\Phi)\,,
\label{eq:non-local-deformation}
\end{align}
where we introduced a basis of functions $\mo_\text{G}(b,\Phi)$, insertions of which correspond with inserting geodesic boundaries in the gravitational path integral \cite{Blommaert:2021fob} 
\be
\mo_\text{G}(b,\Phi) =  e^{-\S}\int \d^2 x \sqrt{g}\, e^{-2\pi \Phi(x)} \cos\left(b\Phi(x)\right)\quad \Leftrightarrow\quad \raisebox{-5.5mm}{\includegraphics[scale=0.125]{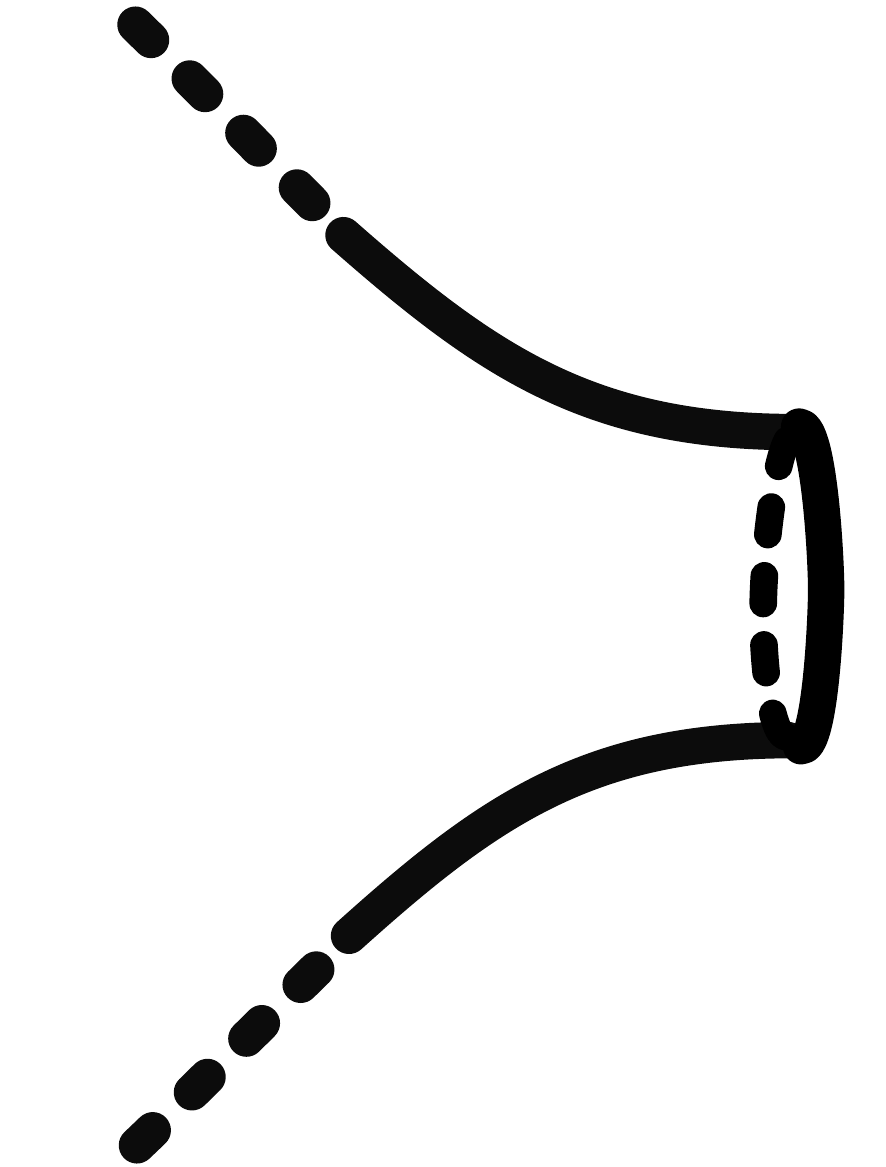}}\,\,\,b
\label{eq:OGb}
\ee
The most general factorizing dilaton gravity has besides this bilocal also a local dilaton potential, which can be expanded in the same basis
\be 
\label{eq:local-deformation}
-I_{\rm local}-I_\text{nonlocal} =\int_0^\infty \d b\, b\, Z_{\rm brane}(b) \, \mo_\text{G}(b,\Phi)-\frac{1}{2}\int_0^\infty \d b\, b\,\mo_\text{G}(b,\Phi)\,\mo_\text{G}(b,\Phi) \,.
\ee
As we shall see, the expansion coefficient $Z_{\rm brane}(b)$ does not influence factorization, or the fact that the theory becomes discrete. Its role will be to determine the specific discrete spectrum of the theory and thus encode its ``microstructure''.

\subsection{Geometric argument}
Suppose we compute the partition function in this deformed theory
\be
Z(\beta) =  \frac{1}{\mathcal Z^{\rm factorized}}\sum_{\substack{\text{geometries with}\\\text{one boundary}}}
    \int\mathcal{D}g\,\mathcal{D}\Phi\,e^{-I_{JT} - I_{\rm local} - I_{\rm nonlocal}}\,,
\label{eq:grav-path-integral-local-and-nonlocal}
\ee
where $\mathcal Z^{\rm factorized}$ is a normalization factor given by the gravitational path integral performed on geometries with no asymptotic boundary, 
\be
\mathcal Z^{\rm factorized} = \sum_{\substack{\text{geometries with}\\\text{no boundaries}}}
    \int\mathcal{D}g\,\mathcal{D}\Phi\,e^{-I_{JT} - I_{\rm local} - I_{\rm nonlocal}}\,.
\ee
We now expand out $ I_{\rm local} + I_{\rm nonlocal}$, just like one does with interaction vertices when computing Feynman diagrams in an interacting QFT. The Feynman rules are that we insert geodesic boundaries with \emph{smeared} boundary conditions, from expanding out $I_\text{local}$
\begin{equation}
   \int_0^\infty \d b\, b\, Z_{\rm brane}(b) \, \mo_\text{G}(b,\Phi)\quad  \Leftrightarrow\quad\raisebox{-6mm}{\includegraphics[scale=0.12]{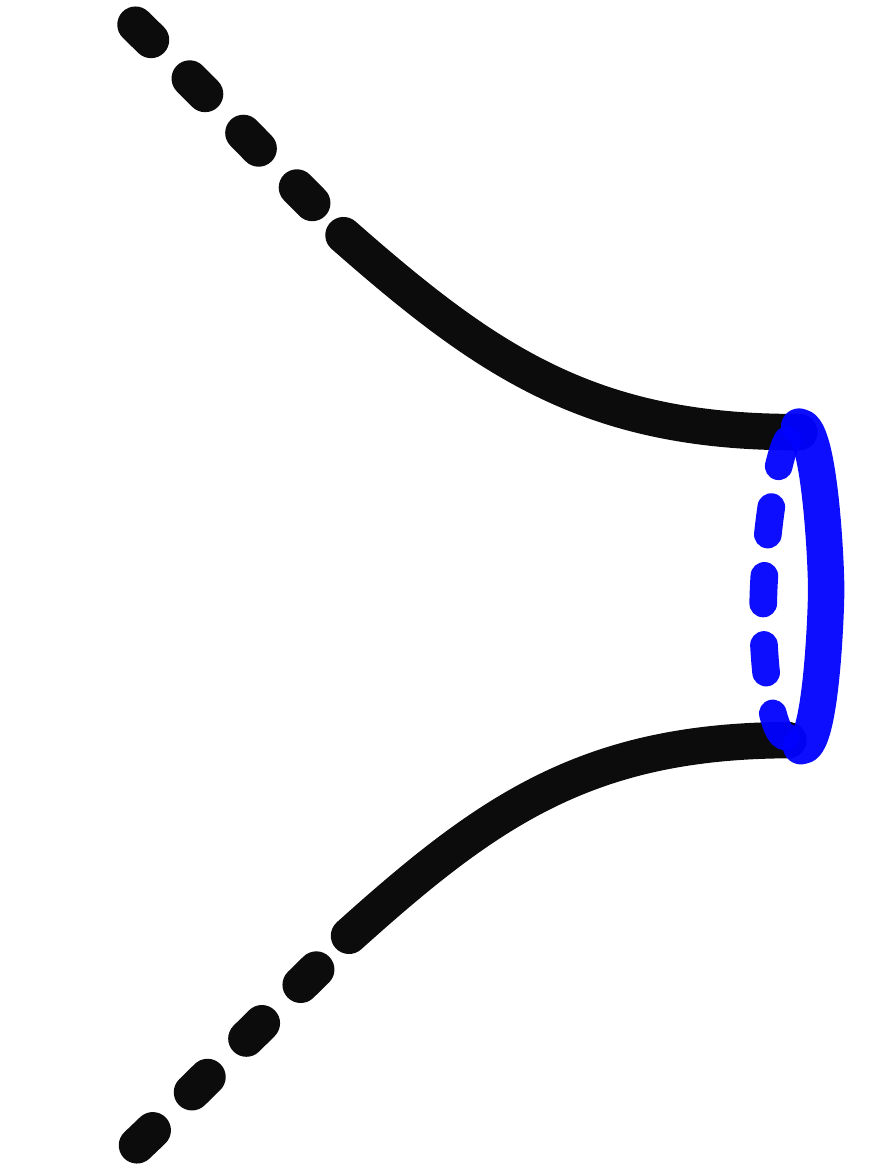}}\label{brane}
\end{equation}
Expanding out $I_{\rm nonlocal}$ gives \emph{correlated} geodesic boundaries, or correlated branes
\begin{align}
    -\frac{1}{2}\int_0^\infty \d b\, b\,\mo_\text{G}(b,\Phi)\,\mo_\text{G}(b,\Phi)\quad \Leftrightarrow \quad \raisebox{-7mm}{\includegraphics[scale=0.35]{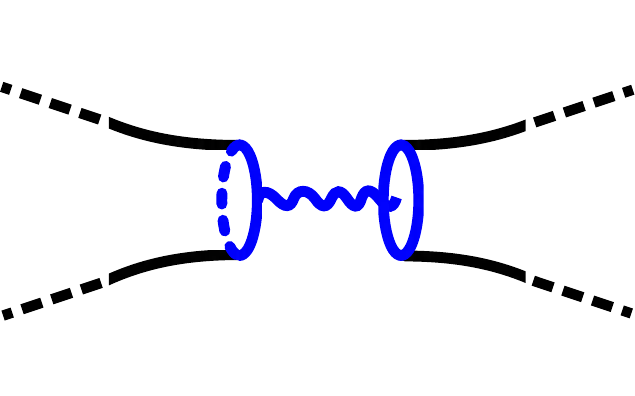}}\label{bilocal}
\end{align}
Expanding local interactions, one sums over all possible brane insertions \eqref{brane}, schematically 
\be 
Z(\beta) = \raisebox{-10mm}{\includegraphics[width=0.72\textwidth]{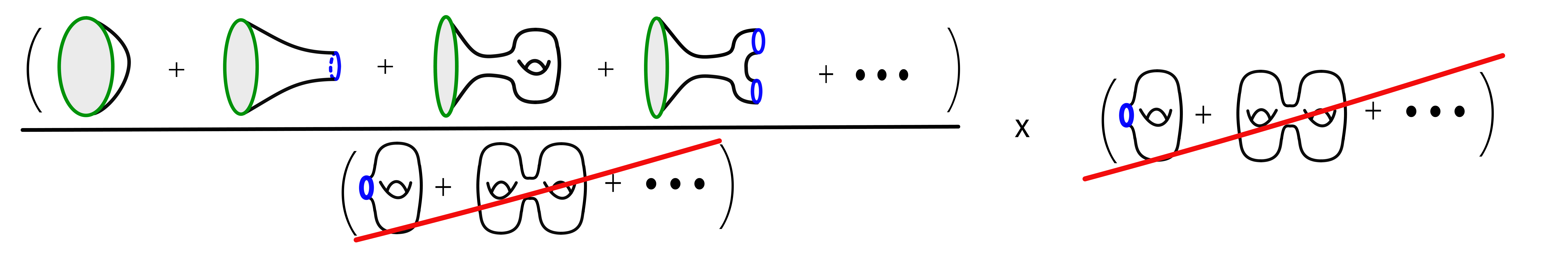}}\hspace{-0.3cm}+\text{bilocal contrib.}\,
\ee
It is important to emphasize that one must in principle include disconnected closed spacetimes, with no asymptotic boundaries. When only local interactions are present, the partition function is normalized by dividing by such closed spacetimes. However, the bilocal interactions \eqref{bilocal} can connect closed universes to universes that have an asymptotic boundary, and we end up with an expansion of the type
\be 
\label{eq:blue-connections}
Z(\beta) = \raisebox{-9mm}{\includegraphics[width=0.75\textwidth]{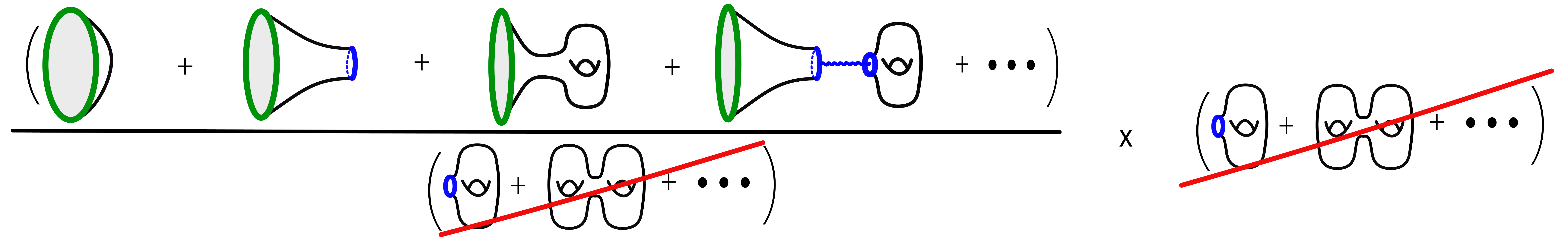}} \hspace{-0.3cm}\,.
\ee
Crucially, the bilocal can also connect disconnected components of spacetimes with different asymptotic boundaries, as in \eqref{eq:factorization-leading-order}, this is the key mechanism for factorization, as we now review.

Consider the sum of all connected contributions to the $n$-boundary gravitational partition function. We can organize this sum by grouping together geometries that share the part $\Sigma$ below
\be
   \sum_{k=0}^n  \sum_{\s_k^n}
    \;\begin{tikzpicture}[baseline={([yshift=-.5ex]current bounding box.center)}, scale=0.90]
 \pgftext{\includegraphics[scale=0.38]{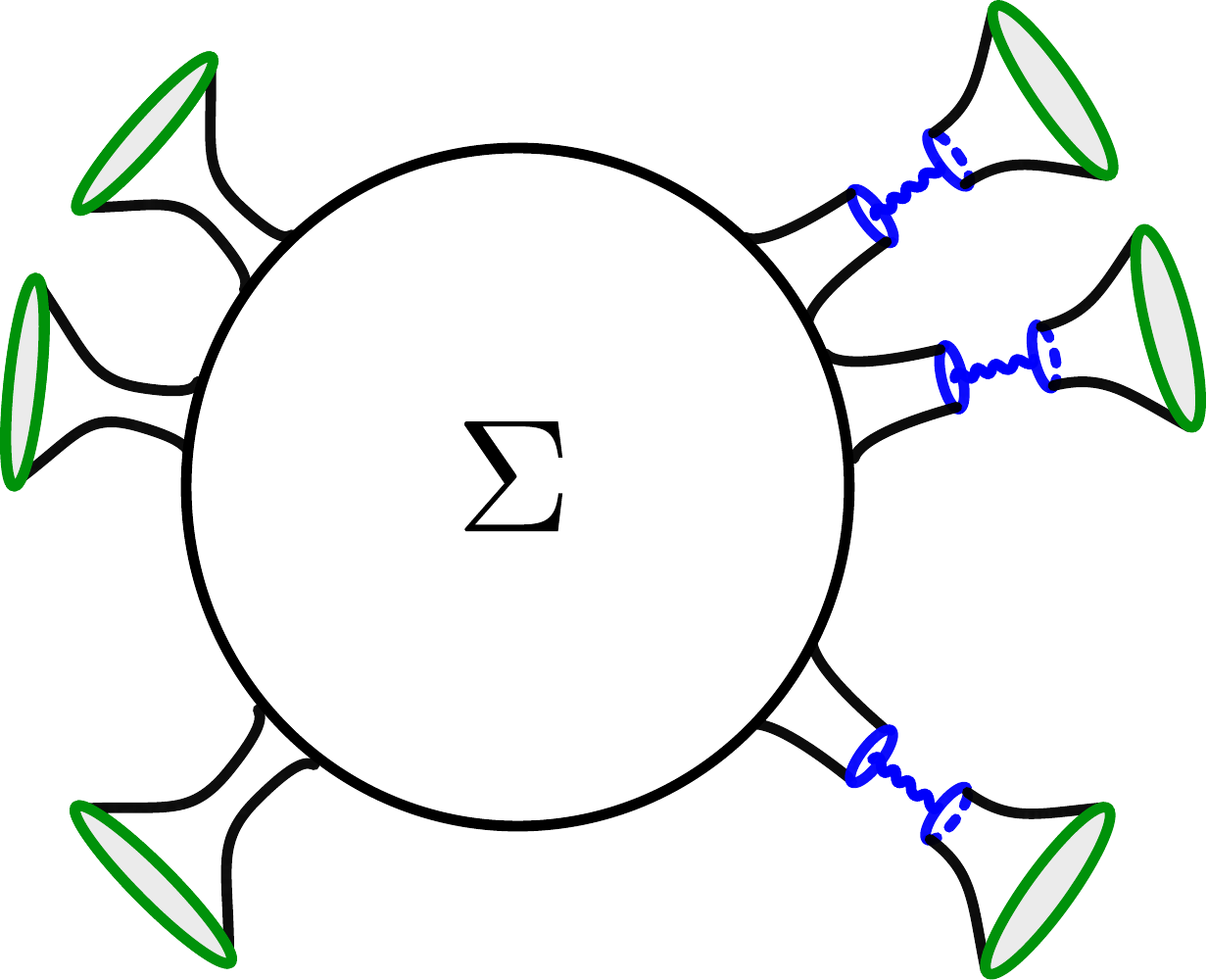}} at (0,0);
  \end{tikzpicture} \hspace{-0.3cm}  
  \quad\;\; = \quad \begin{tikzpicture}[baseline={([yshift=-.5ex]current bounding box.center)}, scale=0.90]
 \pgftext{\includegraphics[scale=0.38]{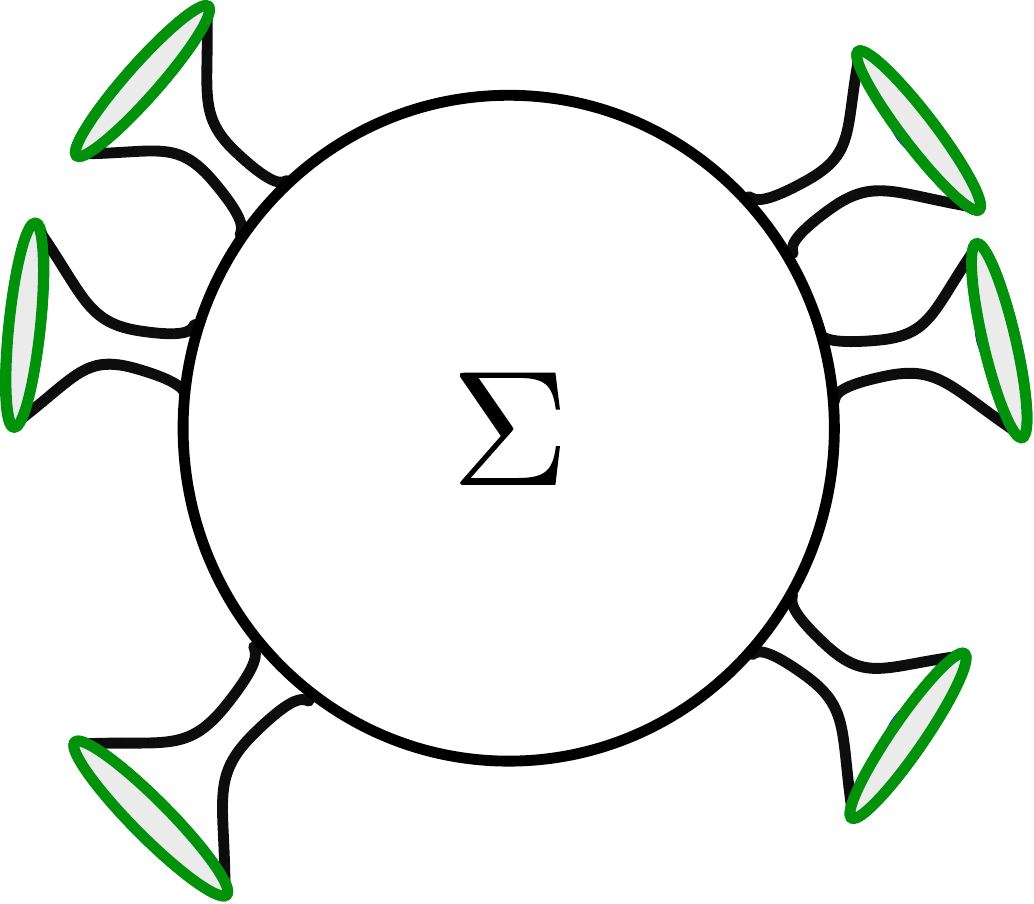}} at (0,0);
  \end{tikzpicture} \; \sum_{k=0}^n (-1)^k{n \choose k}=0\,.
\label{eq:connected-geom-grouping}
\ee
Here, $\Sigma$ could contain any number of handles and any numbers of branes that could either be connected or disconnected. Using the explicit form of the bilocal interaction \eqref{eq:non-local-deformation}, one finds that one can replace each connected brane homotopic to an asymptotic boundary by minus a wormhole 
\be 
\begin{tikzpicture}[baseline={([yshift=1.7ex]current bounding box.center)}, scale=1.0 ]
 \pgftext{\includegraphics[scale=0.14]{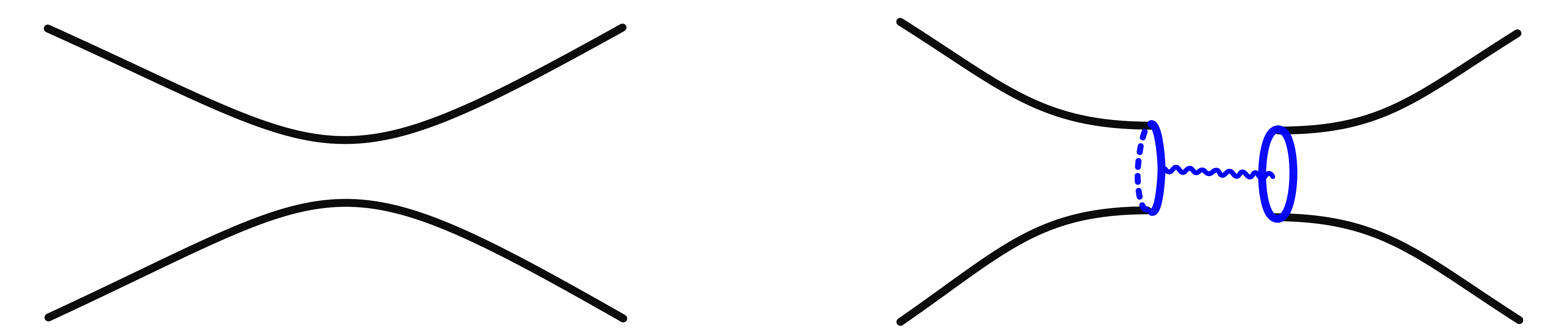}} at (0,0);
  \draw (-0.1, 0) node {=\,\,\,\,-};
  \end{tikzpicture}
\ee
From this, the rewriting of the sum as in \eqref{eq:connected-geom-grouping} follows and therefore, the connected contribution to the gravitational path integral vanishes, meaning that the theory factorizes
\be 
\label{eq:factorization-property}
Z(\beta_1, \,\dots,\,\beta_n) = Z(\beta_1) \,\dots\,Z(\beta_n) \,.
\ee

Applying this cancellation mechanism between wormhole geometries and bilocal interactions, one finds that the computation of the gravitational path integral with a single asymptotic boundary reduces to only the disk and the spacetime with a single brane (called half-wormhole in \cite{Saad:2021rcu,Saad:2021uzi})
\be 
\label{eq:disk+half-wormhole}
Z(\beta)=\quad\raisebox{-10mm}{\includegraphics[scale=0.35]{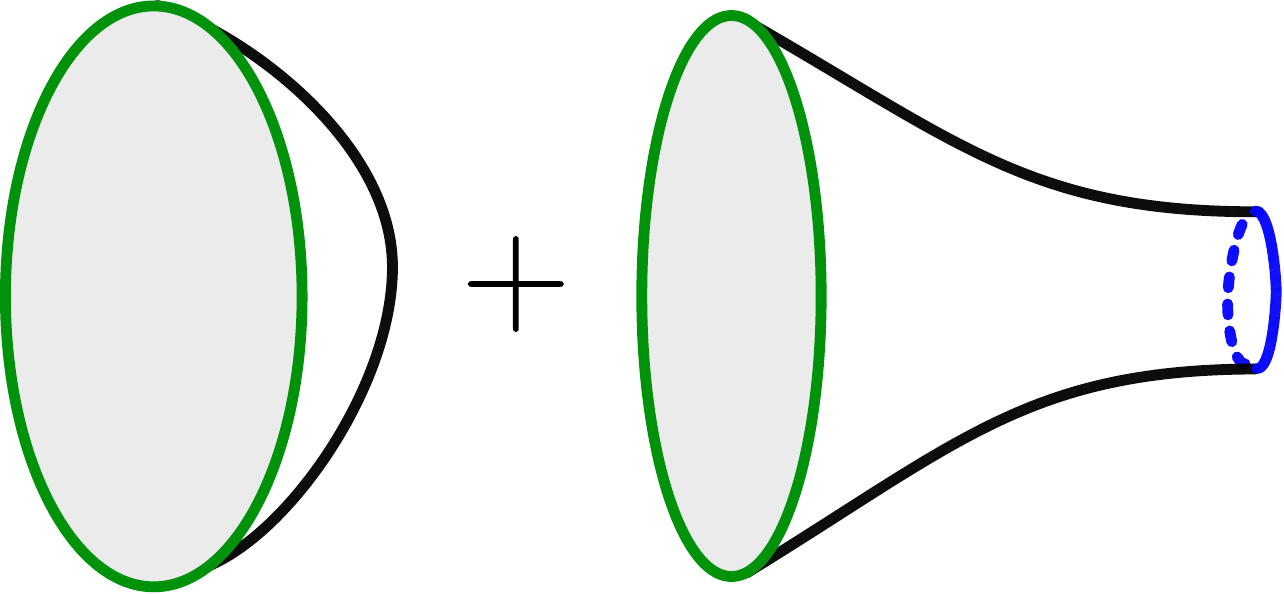}}+\quad \text{non-perturbative corrections.} 
\ee
The local dilaton potential $U_{\rm local}(\Phi)$ encoded in $Z_\text{brane}(b)$ \eqref{eq:local-deformation} enters only through this half-wormhole.

One might want to argue at this point that adding the bilocal term to the action is equivalent to an ad-hoc removal of all wormhole geometries.\footnote{We thank Juan Maldacena for bringing up this point.} We emphasize that this is not the case. The mechanism is a numerical cancellation between wormhole amplitudes, and Feynman diagrams for bilocal interactions. Before expanding out the interaction vertices in \eqref{eq:grav-path-integral-local-and-nonlocal} we just have a non-local interacting field theory and a sum over geometries. 
The fact that, in the end, the calculation of expectation values in this theory simplifies to only two integrals \eqref{eq:disk+half-wormhole} is a happy and surprising conclusion. We hope that this effective field theory which solves factorization provides an exploration ground to understand the origin of the degrees of freedom that were integrated out, in particular, those responsible for the bi-local interaction.

Perhaps more importantly, there is an important difference in formulas between our theory and the ad-hoc theory with no wormholes. In our case one finds non-perturbative corrections to \eqref{eq:disk+half-wormhole}. Indeed, as we now review, the deformation \eqref{eq:grav-path-integral-local-and-nonlocal} has an equivalent description in the matrix integral \cite{Saad:2019lba}. One finds that the spectrum of \eqref{eq:disk+half-wormhole} is \emph{discrete} because of the bilocal deformation, regardless of the specific form of $Z_\text{brane}(b)$. This non-perturbative effect can, as far as we can tell, not be derived in the ad-hoc theory. We need wormholes, albeit cancelling ones, and their associated matrix integral description for that.


\subsection{Matrix integral argument}\label{sect:2.2}
Observables in (undeformed) JT gravity can be computed using a matrix integral \cite{Saad:2019lba}
\be 
\mathcal Z_{\rm JT} = \int \d H\, e^{-L \Tr V_\text{JT}(H)} = \prod_{i=1}^L\int_\mathcal{\cont}\d\lambda_i\exp\bigg(-L\sum_{i=1}^L V_\text{JT}(\l_i)+\sum_{i\neq j}^L\log(\l_i-\l_j)\bigg)\,,
\label{eq:JT-matrix-integral}
\ee
where $\lambda_i$ are the random eigenvalues of $H$. JT gravity is a large $L$ double scaling limit of this \cite{Saad:2019lba}, but that does not play an important role in this paper. 

The dictionary between gravity and matrix integrals is that inserting boundaries in the gravitational path integral corresponds with computing expectation values of single-trace operators in the matrix integral - for instance
\be
Z(\beta_1\dots\beta_n) = \frac{1}{\mathcal{Z}_\text{JT}}\int \d H\,\Tr\left(e^{-\beta_1 H}\right) \dots \Tr \left(e^{-\beta_n H}\right) e^{-L \Tr V_\text{JT}(H)}=\average{\Tr( e^{-\beta_1 H}) \,\dots \, \Tr( e^{-\beta_n H}) }\,.
\ee
In particular, inserting a geodesic boundary in the gravity path integral is dual to inserting \cite{Blommaert:2021fob,Goel:2021wim} 
\be 
\mo_{\rm G}(b)=\frac{2}{b}\Tr \cos(b H^{1/2})-\int_0^\infty \d E\,\rho_{0, {\rm JT}}(E)\,\frac{2}{b}\cos( b E^{1/2})\quad \Leftrightarrow\quad \raisebox{-5.5mm}{\includegraphics[scale=0.12]{Geobdy.pdf}}\,\,\,b \label{eq:obdef}
\ee

So the expansion of the deformations in \eqref{eq:grav-path-integral-local-and-nonlocal} corresponds in the matrix integral with expanding out
\begin{align}
&e^{-I_{\rm local} - I_{\rm nonlocal}} \nonumber\\&\Leftrightarrow\quad \exp\bigg( \int_0^\infty \d b\, b\,\mo_{\rm G}(b)\,z_\text{brane}(b)+\frac{1}{2}\int_0^\infty \d b_1\,b_1\int_0^\infty \d b_2 \,b_2\,\mo_{\rm G}(b_1) \mo_{\rm G}(b_2)\,z_\text{brane}(b_1,b_2) \bigg)\,,\label{deform}
\end{align}
where the correspondence between the smearing functions $z_{\rm brane}(b)$ and $z_{\rm brane}(b_1, b_2)$ and the local and nonlocal dilaton potentials, written in terms of $\mo_\text{G}(b,\Phi)$ as in \eqref{eq:non-local-deformation} and \eqref{eq:local-deformation}, is to be determined.

Naively this looks trivial, however there is one key subtlety in the dictionary between gravity and random matrix theory. In the gravitational path integral there are no degenerate cylinders (i.e.~cylinders of zero surface that end on two geodesics of equal length). However, in the matrix integral, degenerate cylinders  contribute contact terms in  the correlators of the operator $\mo_{\rm G}(b)$.\footnote{The matrix integral assigns the first term in $\average{\mo_{\rm G}(b_1) \mo_{\rm G}(b_2)}_\text{conn}=\delta(b_1-b_2)/b_1+O(e^{-2\S})$ to a degenerate cylinder.} Consequently, they give a non-trivial contribution in the expansion of \eqref{deform} in the matrix integral \eqref{eq:JT-matrix-integral}. One should therefore view $z_\text{brane}(b_1,b_2)$ as a \emph{bare} brane propagator, while the propagator between branes seen through \eqref{eq:non-local-deformation} should be understood as the \emph{dressed} propagator. The dressed propagator \eqref{eq:non-local-deformation} is obtained from the bare propagator $z_{\brane}(b_1,b_2)$ by resumming the Dyson series of bare propagators connected via degenerate cylinders
\begin{equation}
\begin{tikzpicture}[baseline={([yshift=-.5ex]current bounding box.center)}, scale=0.6]
 \pgftext{\includegraphics[scale=0.50]{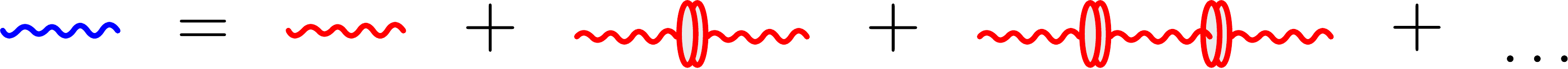}} at (0,0);
  \end{tikzpicture}\;
  \label{eq:geometric-relation-blue-to-red-prop}
\end{equation}
In formulas this becomes
\begin{align}
-\frac{1}{b_1}\delta(b_1-b_2)= z_{\brane}(b_1,b_2) + \int_0^{\infty} \d b_3 b_3\, z_{\brane}(b_1,b_3)z_{\brane}(b_3,b_2)+\dots 
\end{align}
whose solution one finds to be $z_{\brane}(b_1,b_2) = -q \delta(b_1-b_2)/b_1$ with $q\to \infty$. The degenerate cylinders similarly affect the relation between $z_{\rm brane}(b)$ (red) and $Z_{\rm brane}(b)$ (blue), one obtains the Dyson equation
\be 
\begin{tikzpicture}[baseline={([yshift=-.5ex]current bounding box.center)}, scale=0.6]
 \pgftext{\includegraphics[scale=0.50]{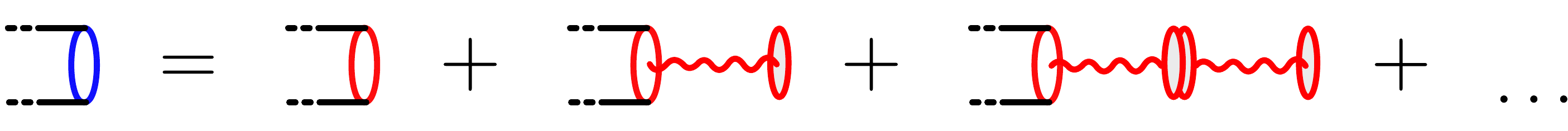}} at (0,0);
  \end{tikzpicture}\;
\ee
Plugging in the the solution for $z_\text{brane}(b_1,b_2)$ this equation becomes 
\be 
Z_{\brane}(b) = \lim_{q\to \infty} \frac{1}{1+q}\, z_{\brane}(b)\quad \Rightarrow\quad z_\text{brane}(b)=(q+1)\,Z_\text{brane}(b)\,.\label{2.20}
\ee 
Thus according to \eqref{deform} we should compute observables in the deformed matrix integral 
\begin{align}
    \mathcal Z^{\text{factorized}} &= \int \d H e^{-L \Tr V_\text{JT}(H) + (q+1) \int_0^\infty \d b\, b\,\mo_{\rm G}(b)\,Z_\text{brane}(b) - \frac{q}{2}\int_0^\infty \d b \, b\,\mo_{\rm G}(b) \mo_{\rm G}(b)\,} 
\label{eq:deformed-matrix-integral}
\end{align}
Looking at this one might think that the perturbative expansion in $q$ is horrible, but it is not, precisely because we can always resum the Dyson series \eqref{eq:geometric-relation-blue-to-red-prop}.\footnote{We thank Steve Shenker for emphasizing this point.} One then just recovers the gravitational genus expansion of the previous subsection, as in \eqref{eq:blue-connections}.

More importantly, instead of expanding out this matrix integral deformation perturbatively, we can enforce the limit $q\to\infty$ directly in the matrix integral action. Because we are being forced to take the $q \to \infty$ limit in order to match the geometric expansion in the gravitational path integral, the matrix integral in \eqref{eq:deformed-matrix-integral} \emph{localizes} at solutions to its saddle-point equations. Doing the integrals over $b$ in \eqref{eq:deformed-matrix-integral}, we find that the large $q$ saddle-point equations can be explicitly written as 
\be 
 \forall\, i=1\dots L: \quad   LV'[Z_\text{brane}](\l_{i})=\frac{1}{2\l_{i}}+2\sum_{j\neq i}^L\frac{1}{\l_{i}-\l_{j}}\,,\quad V[Z_\text{brane}]=V_\text{JT}+V_\text{local}[Z_\text{brane}]\,.
\label{eq:saddle-point-equation-matrix-integral}
\ee
The local deformation $V_\text{local}$ of the dilaton potential for generic $Z_\text{brane}$ is worked out in section \ref{sect:4.1}, where we need it explicitly. This set of $L$ equations has (typically) a unique dominant saddle for the $L$ eigenvalues $\l_i$. The matrix integral localizes to this saddle, so when we compute the partition function $Z(\beta)=\average{\Tr e^{-\beta H}}$ we find the discrete answer expected of a conventional QM system
\begin{equation}
    Z(\beta)=\sum_{i=1}^Le^{-\beta\l_i}\,,
\end{equation}
where $\l_i$ are the saddle point of \eqref{eq:saddle-point-equation-matrix-integral}, these depend on the brane one-point function through $V[Z_\text{brane}]$. It is important that when computing observables in matrix integrals, we normalize by $\mathcal{Z}_\text{factorized}$ from \eqref{eq:deformed-matrix-integral} before taking $q\to \infty$, such that the $q$-and $\l_i$ dependent one-loop determinants cancel in the expectation value of all observables.


\section{Alpha states in JT gravity}\label{sect:alpha}
In this section we explain how the discrete and factorizing models of dilaton gravity with deformation $I_\text{local}+I_\text{nonlocal}$ \eqref{eq:grav-path-integral-local-and-nonlocal} give rise to the alpha-states of pure JT gravity (the theory with no deformations). We give explicit expressions for these states in dilaton gravity variables, but use the non-perturbative matrix integral setup to show that they indeed form a basis of the baby universe Hilbert space. 

Before getting there it will be useful to quickly review the construction of the baby universe Hilbert space applied in AdS/CFT \cite{Marolf:2020xie}. The idea is that one can interpret multi-boundary gravitational path integrals as expectation values of boundary creating operators in a no-boundary state $\ket{\text{HH}}$. Explicitly in JT multi-boundary partition functions can be viewed as
\begin{equation}
    Z(\beta_1\dots \beta_n)=\bra{\text{HH}}\hat{Z}(\beta_1)\dots \hat{Z}(\beta_n)\ket{\text{HH}} \,\,\sim \sum_{\substack{\text{geometries with}\\\text{$n$ boundaries}}}
    \int\mathcal{D}g\,\mathcal{D}\Phi\,e^{-I_\text{JT}}
    \,,\label{gpi}
\end{equation}
with $\hat{Z}(\beta)$ an operator that creates a boundary with usual asymptotic AdS$_2$ boundary conditions \cite{Maldacena:2016hyu,Engelsoy:2016xyb,Jensen:2016pah}. Other boundary conditions \cite{Goel:2020yxl} have their own operators, they are linear combinations of $\hat{Z}(\beta)$. This is similar to how we compute correlators in QFT in the path integral by summing over Feynman diagrams (left), or by computing expectation values of $\hat{\phi}$ in a vacuum by expanding out in creation and annihilation operators (right) \cite{Anous:2020lka}. One key difference is that here all creation operators $\hat{Z}(\beta)$ are assumed to commute \cite{Marolf:2020xie}. 

The states
\begin{equation}
    \hat{Z}(\beta_1)\dots \hat{Z}(\beta_k)\ket{\text{HH}}\,,\label{span}
\end{equation}
span the Hilbert space of baby universes \cite{Marolf:2020xie} and all gravitational path integrals \eqref{gpi} can be viewed as inner products between two states in this space. Since the $\hat{Z}(\beta)$ commute, they can be simultaneously diagonalized by so-called alpha-states
\begin{equation}
    \hat Z(\beta) \ket{\alpha} = Z_\alpha (\beta) \ket{\alpha}\,.\label{15}
\end{equation}
Gravity path integrals \eqref{gpi} then indeed naturally acquire the structure of an ensemble average
\begin{equation}
     Z(\beta_1\dots \beta_n)=\bra{\text{HH}}\hat{Z}(\beta_1)\dots \hat{Z}(\beta_n)\ket{\text{HH}} = \sum_\alpha P_\alpha\, Z(\beta_1\dots \beta_n)_\alpha\,,\quad P_\alpha=|\bra{\alpha}\ket{\text{HH}}|^2\,,\label{16}
\end{equation}
and gravitational amplitudes within one alpha-state trivially factorize
\begin{equation}
    Z(\beta_1\dots \beta_n)_\alpha=\bra{\alpha}\hat{Z}(\beta_1)\dots \hat{Z}(\beta_n)\ket{\alpha}=Z_\alpha(\beta_1)\dots Z_\alpha(\beta_n)\,.
\end{equation}
Therefore, formally, the factorization puzzle can be resolved by stating that factorizing gravity theories should correspond to alpha-states \cite{Marolf:2020xie}. The difficulty has simply been shifted to concretely construct the alpha-states in a given model, and to understanding their gravitational interpretation. We will do so now.\footnote{Eigenbranes are another way to construct alpha-states in JT \cite{Blommaert:2019wfy,Blommaert:2020seb}, but their semiclassical interpretation is unclear, and null states are not obvious. The current approach is more intuitive and should generalize better to higher dimensions.}

A second important point in constructing a baby universe Hilbert space is to ensure that the inner product on it (provided by the gravitational path integral as in \eqref{gpi}) is non-degenerate. This means that if there are null states, they should be quotiented out. This will be the subject of section \ref{sect:null}.

\subsection{Constructing alpha-states}\label{sec:cons_alpha}

 Following the results of \cite{Blommaert:2021fob} as reviewed above, it is straightforward to see how we can construct explicit alpha-states. The only thing one has to do is to start with the state without any boundaries, the Hartle-Hawking state, and act on it such that we create the correlated geodesics boundaries discussed in section \ref{sect:review}. Specifically, we define an operator $\hat{Z}_\text{G}(b)$ in the baby universe Hilbert space, whose action is to create a geodesic boundary of the type \eqref{eq:OGb}
 
\be 
\hat{Z}_\text{G}(b)\quad \text{creates}\quad \raisebox{-5.5mm}{\includegraphics[scale=0.125]{Geobdy.pdf}}\,\,\,b
\ee
These operators allow us to construct a more convenient span for the baby universe Hilbert space than \eqref{span}
\begin{equation}
    \hat{Z}_\text{G}(b_1)\dots \hat{Z}_\text{G}(b_k)\ket{\text{HH}}\,.
\end{equation}
We can now construct normalized alpha-states in gravity as follows, 
\be 
\label{eq:alpha-state-definition}
\ket{\alpha_{\Hh}}= \mathcal{N}_\Hh e^{-\hat{I}_\text{deform}(\Hh)}\ket{\text{HH}}\,,
\ee
where the normalization $\mathcal{N}_{\Hh}$ is determined in section \ref{sect:norm} and $\hat{I}_\text{deform}(\Hh)$ is just the deformation in the gravity action $I_\text{local}(\Hh)+I_\text{nonlocal}$ \eqref{eq:local-deformation} translated to the baby universe operator formalism
\be 
\label{eq:boundary-creation-operator}
-\hat{I}_\text{deform}(\Hh) = -\frac{1}{2} \int \d b \, b\, \hat{Z}_{G}(b)\,\hat{Z}_{G}(b) + \int \d b \, b\,Z_{\brane}(b,\Hh)\, \hat{Z}_{G}(b)\quad \Leftrightarrow\quad -I_\text{local}(\Hh)-I_\text{nonlocal}\,.
\ee 
This is the explicit realization of the mapping between alpha-states and factorizing theories of dilaton gravity that we announced in \eqref{eq:O-alpha-to-alpha}. The operator $\hat{Z}_\text{G}(b)$ in the second term creates a single geodesic boundary of length $b$ which is smeared by the function $Z_{\brane}(b,\Hh)$. This function is chosen such that the corresponding factorizing theory has a spectrum given by the eigen-energies of a given Hamiltonian $\Hh$.\footnote{One concrete choice that reproduces the spectrum of a theory with Hamiltonian $\Hh$ is \cite{Blommaert:2021fob}
\begin{equation}
    Z_\text{brane}(b,\Hh)=\frac{2}{b}\Tr \cos(b \Hh^{1/2})-\int_0^\infty \d E\,\rho_{0, \rm{JT}}(E)\,\frac{2}{b}\cos( b E^{1/2})\,.\label{31}
\end{equation}
In section \ref{sect:null} we point out that this choice of $Z_{\brane}(b,\Hh)$ is not unique, large classes of other functions $Z_{\brane}(b,\Hh)$ result in the same spectrum.}  The operator $\hat{Z}_{G}(b)\hat{Z}_{G}(b)$ in the first term creates a correlated set (or an entangled set in the baby universe Hilbert space) of two geodesic boundary conditions each with length $b$. Schematically, we can thus represent $\ket{\alpha_{\Hh}}$ as
\be 
\ket{\alpha_{\Hh}} = \cN_{\Hh}\exp\bigg(\hspace{0.2cm}\begin{tikzpicture}[baseline={([yshift=-0.5ex]current bounding box.center)}, scale=1.0 ]
 \pgftext{\includegraphics[scale=0.35]{2Branebdy.pdf}} at (0,0);
  \end{tikzpicture}  \hspace{0.1cm} + \, \begin{tikzpicture}[baseline={([yshift=-0.5ex]current bounding box.center)}, scale=1.00 ]
 \pgftext{\includegraphics[scale=0.12]{onebrane.pdf}} at (0,0);
  \end{tikzpicture}\hspace{0.05cm} \bigg) \ket{{\rm HH}}\,,
\label{eq:geometric-representation-alpha states}
\ee 
where the blue line connecting the two branes is given by the integral in \eqref{eq:boundary-creation-operator}.

To prove that \eqref{eq:alpha-state-definition} are alpha-states it suffices to prove that
\begin{equation}
    \bra{\text{HH}}\hat{Z}(\beta_1)\dots \hat{Z}(\beta_n)\ket{\alpha_\Hh}=\Tr \left(e^{-\beta_1\Hh}\right)\dots \Tr \left(e^{-\beta_n\Hh}\right)\bra{\text{HH}}\ket{\alpha_\Hh}\,.\label{37}
\end{equation}
Indeed, because $\hat{Z}(\beta_1)\dots \hat{Z}(\beta_n)\ket{\text{HH}}$ for all values of $n$ and $\beta_1$,\dots,$\beta_n$, span the baby universe Hilbert space, proving \eqref{37} for all $n$ is equivalent to proving that $\ket{\alpha_\Hh}$ are eigenstates of $\hat{Z}(\beta)$, which is indeed the defining property of alpha-states,
\begin{equation}
    \hat{Z}(\beta)\ket{\alpha_\Hh}=\Tr( e^{-\beta\Hh})\ket{\alpha_\Hh}\,.
\end{equation}

As discussed in the introduction, inner products in the baby universe Hilbert space are obtained from the gravitational path integral with boundary conditions associated to the boundary creation operators evaluated inside the inner-product. For the inner product \eqref{37}, modulo the $\mathcal{N}_{\Hh}$ in the alpha-state, the left-hand side is given by the Euclidean path integral with asymptotic boundaries $\b_1,\dots,\b_n$, closing off smoothly or ending on any of the boundaries in the alpha-state \eqref{eq:geometric-representation-alpha states}. As we saw in section \ref{sect:review}, this path integral is that of the deformed theory \eqref{eq:grav-path-integral-local-and-nonlocal}. Thus, we find that 
\begin{align}
   \label{eq:alpha-state-HH-inner-product}
   \bra{\text{HH}}\hat{Z}(\b_1)\cdots \hat{Z}(\b_n) \ket{\alpha_{\Hh}} &= \frac{\mathcal N_{\Hh}}{\mathcal Z_{\rm JT}} \sum_{\substack{\text{geometries}\\ \text{with $n$ asymptotic} \\ \text{boundaries}}} \int  \mathcal{D}g\,\mathcal{D}\Phi\, e^{-I_\text{JT} - I_{\rm local} - I_{\rm nonlocal}}\,. 
   \end{align}
where we normalized $\ket{\text{HH}}$ such that $\bra{\text{HH}}\ket{\text{HH}}=1$ and
\be 
\mathcal Z_{\rm JT} = \sum_{\substack{\text{geometries}\\ \text{without asymptotic} \\ \text{boundaries}}} \int   \mathcal{D}g\,\mathcal{D}\Phi e^{-I_{JT} }\,. 
\ee
Using the fact that this gravitational path integral \eqref{eq:alpha-state-HH-inner-product} factorizes with the only remaining contributions being given by disks and half-wormholes as well as the fact that the matrix integral yields a theory  whose partition function is $Z(\beta) = \Tr(e^{-\beta\Hh})$, we obtain the desired answer
   \begin{align}
 \bra{\text{HH}}\hat{Z}(\b_1)\cdots \hat{Z}(\b_n) \ket{\alpha_{\Hh}} 
   &= \bra{\text{HH}} \ket{\alpha_{\Hh}}  \bigg[\bigg(\begin{tikzpicture}[baseline={([yshift=1.2ex]current bounding box.center)}, scale=0.45 ]
 \pgftext{\includegraphics[scale=0.48]{Halfwormhole_contribution.pdf}} at (0,0);
  \draw (-1.9, -2.2) node {$\beta_1$};
         \draw (2.3, -2) node {};
          \draw (3.7, 0) node {};
  \end{tikzpicture} \hspace{-0.42cm} \bigg) \times \dots \times \bigg(\begin{tikzpicture}[baseline={([yshift=1.2ex]current bounding box.center)}, scale=0.45 ]
 \pgftext{\includegraphics[scale=0.48]{Halfwormhole_contribution.pdf}} at (0,0);
  \draw (-1.9, -2.2) node {$\beta_n$};
         \draw (2.3, -2) node {};
          \draw (3.7, 0) node {};
  \end{tikzpicture} \hspace{-0.42cm}  \bigg) \bigg] \, \nn \\ & =   \bra{\text{HH}}\ket{\alpha_{\Hh}} \Tr(e^{-\beta_1 \Hh})\cdots \Tr(e^{-\beta_n \Hh})\,.
  \label{eq:factorization-proof}
\end{align}
Here the factor $\average{\text{HH}|\alpha_{\Hh}}$ (modulo a factor $\mathcal{N}_\Hh$ which is identical on the left and right in this equation) corresponds to the contribution of closed universes without asymptotic boundaries, which is \eqref{eq:alpha-state-HH-inner-product} with $n=0$. The factors of $1/\mathcal Z_{\rm JT}$ are identical left and right too.

This proves that \eqref{eq:alpha-state-definition} are indeed the alpha-states of JT gravity. Since all possible resulting spectra can be obtained by appropriately choosing $Z_{\rm brane}(b,\,\Hh)$ through the explicit construction in \eqref{31}, these alpha-states span the baby universe Hilbert space.\footnote{One can check that the alpha-states are also eigenstates of $\hat{Z}_\text{G}(b)$. This is a bit subtle due to the ambiguity related to the presence of degenerate cylinder. The unambiguous way to check this is to reintroduce the degenerate cylinders that feature in the matrix integral and carefully take $q\to\infty$.} 

To summarize, the operators $e^{-\hat{I}_\text{deform}(\Hh)}$ used to construct $\ket{\alpha_{\Hh}}$ can each be mapped to one of the  factorizing theories discussed in section \ref{sect:review}. In turn,  each such theory can be mapped to a matrix integral deformation that localizes the integral over all matrices to $\Hh$. Thus, we obtain the triality\footnote{The matrix integral will be defined further on as $\int \d H e^{-L \Tr V_{\text{JT}}(H) + q I(H,\Hh)}$.}
\begin{align}
 &\text{BU Hilbert space}\quad &&\hat I_{\rm deform}(\Hh) = \frac{1}{2} \int \d b \, b\, \hat{Z}_\text{G}(b)\,\hat{Z}_\text{G}(b) - \int \d b \, b\, \hat{Z}_\text{G}(b)\,Z_{\text{brane}}(b,\Hh)\nn \\ \Leftrightarrow \quad & \text{dilaton gravity}\quad &&I_\text{deform}(\Hh)= \frac{1}2 \int \d b\, b\, \mo_\text{G}(b,\Phi)\, \mo_\text{G}(b,\Phi) -\int \d b \,b\, \mo_\text{G}(b,\Phi)\, Z_{\brane}(b,\Hh) \nn \\ \Leftrightarrow \quad &\text{matrix integral} \quad &&q I(H,\Hh) = -\frac{q}2 \int \d b\, b\, \mo_\text{G}(b)\, \mo_\text{G}(b) +(q+1) \int \d b\, b\, \mo_\text{G}(b)\,Z_{\brane}(b,\Hh)
    \label{eq:alpha-grav-vs-matrix}
\end{align}
where the objects $\hat{Z}_\text{G}(b)$, $\mo_\text{G}(b,\Phi)$ and $\mo_\text{G}(b)$ create geodesic boundaries in the different languages
\begin{align}
& \text{BU Hilbert space}\quad  &&\hat{Z}_{G}(b) \,\text{creates geodesic boundary of length $b$}\nn \\ 
  \Leftrightarrow\quad &\text{dilaton gravity}\quad &&\mo_\text{G}(b,\Phi) = e^{-\S}\int_\Sigma \d^2 x\sqrt{g(x)}\,e^{-2\pi \Phi(x)}\,\cos(b\Phi(x))\nn\\ \Leftrightarrow\quad &\text{matrix integral}\quad &&\mo_\text{G}(b) =\frac{2}{b}\Tr \cos(b H^{1/2})-\int_0^\infty \d E\,\rho_{0,{\rm JT}}(E)\,\frac{2}{b}\cos( b E^{1/2}) \,.
  \label{mapoperators}
\end{align}


\subsection{Checking orthogonality}\label{sect:norm}

The alpha-states we have just constructed are eigenstates of $\hat{Z}(\b)$ and should thus form a complete set of states for the baby universe Hilbert space. This means the overlap between two alpha-states should be a delta function with unit coefficient. Geometrically this is subtle to confirm. However, orthogonality can easily be seen directly at the level of the matrix integral, where the inner product is\footnote{Taking $q$, $q'$ to not be independent but rather be the same parameter in fact yields similar results. } 
\be 
\bra{\a_\Hh}\ket{\a_\Hh'} =\cN_{\Hh}^* \cN_{\Hh'} \lim_{q,q'\to \infty} 
\int \d \L\, e^{-L\tr V_{\rm JT}(\L)} e^{q I(\L,\Hh) + q'I(\L,\Hh')} \,,
\ee
where the deformation of the action $I(\L,\Hh)$ is
\be 
q I(\L,\Hh) = \frac{q}{2} \sum_{i\neq j}^L \log(\l_i - \l_j) + \frac{q}{2} \sum_{i=1}^L P(\l_i,\l_i) - (q+1)\sum_{i,j=1}^L P(E_i,\l_j) + L \sum_{i=1}^L V_{\rm JT}(\l_i)\,,
\ee
and $P(E,\l)$ given in \eqref{eq:def-P-regulated} and $E_i$ are the eigenvalues of $\Hh$. Let us now take $q$ to infinity first and then take $q'$ to infinity. Taking $q$ large localizes the eigenvalue integral $\L$ to the eigenvalues of $\Hh$ (denoted by $\L_0$), so we get 
\begin{align}
\lim_{q,q'\to \infty} 
\int \d \L e^{-L\tr V_{\rm JT}(\L)} e^{q I(\L,\Hh) + q'I(\L,\Hh')} = \lim_{q \to \infty}(\mathcal Z_{\Hh}^\text{factorized}) \lim_{q'\to \infty}\left( 
\int \d \L e^{q'I(\L,\Hh')}\delta(\L - \L_0)\right)
\end{align}
where $Z_{\Hh}^\text{factorized}$ is defined in \eqref{eq:deformed-matrix-integral} for a single-trace deformation that results in a spectrum associated to the Hamiltonian $\Hh$. 
Now taking $q'$ large as well, we obtain 
\be 
\average{\a_\Hh|\a_{\Hh'}} 
= \cN_{\Hh}^* \cN_{\Hh'} \frac{\lim_{q \to \infty}(\mathcal Z_{\Hh}^\text{factorized}) \lim_{q' \to \infty}(\mathcal Z_{\Hh'}^\text{factorized})}{\mathcal Z_{\rm JT}^2}\frac{\delta(\Hh - \Hh')}{P_{\rm JT}(\Hh)},
\ee
where
\be 
P_{\rm JT}(\Hh) = \frac{1}{\mathcal{Z}_\text{JT}}e^{-L \Tr V(\Hh)}\,, \quad \mathcal Z_{\rm JT} = \int \d \Lambda e^{-L \Tr V_{\rm JT}(\Lambda)}\,.
\ee
is the (normalized) probability distribution of the original JT matrix integral. To get a unit coefficient of the delta function we fix the normalisation factor $\mathcal{N}_{\Hh}$ to be (up to an irrelevant phase)
\be 
\mathcal{N}_{\Hh} = \frac{\sqrt{P_{\rm JT}(\Hh)  } \mathcal Z_{\rm JT}}{\lim_{q \to \infty}(\mathcal Z_{\Hh}^\text{factorized})}\,,
\ee
which thus implies that 
\be 
\average{\a_\Hh|\a_{\Hh'}}  = \delta(\Hh - \Hh')\,.
\ee
This answer makes a lot of sense, since it means that finding a particular alpha-state in the Hartle-Hawking state is given by $P_{\rm JT}(\Hh)$ \cite{Marolf:2020xie}
\be 
|\average{\alpha_{\Hh}|\text{HH}}|^2  =|\mathcal{N}_{\Hh}|^2  \left(\frac{\lim_{q \to \infty}(\mathcal Z_{\Hh}^\text{factorized})}{\mathcal Z_{JT}}\right)^2 = P_{\rm JT}(\Hh) \,,
\ee
which is indeed how one should interpret the Hartle-Hawking state. It is the state in the baby universe Hilbert space that gives the ensemble, i.e. correlators evaluated in that state are given by the pure JT gravity path integral. Written differently, we have 
\be 
\label{eq:HHasH0}
\ket{\text{HH}} = \int \d \Hh \sqrt{P_{JT}(\Hh)}\ket{\alpha_{\Hh}}.
\ee


\subsection{The Coleman-Giddings-Strominger mechanism revisited}

In this context, we want to revisit the mechanisms discussed by Coleman, Giddings and Strominger to account for the effects of wormholes. Our results makes precise how ensemble averaging can either be used to “integrate out” wormholes, through the mechanism discussed in our previous paper \cite{Blommaert:2021fob}, which we called ``fighting ensemble with ensemble”. Alternatively, we can “integrate in” wormholes, through the mechanism similar to that discussed by  Coleman, Giddings and Strominger \cite{coleman1988black, giddings1988loss, Marolf:2020xie, Marolf:2020rpm}, though in our example the distribution of the required ensemble can be precisely determined and shown to be non-Gaussian.

\subsection*{Integrating out wormholes: fighting ensemble with ensemble}

Any factorizing theory with the universal bilocal can alternatively be written as a bulk ensemble average \cite{Blommaert:2021fob} over a Hubbard-Stratanovich field $Q(b)$\footnote{Here the $Q$ integral is along the imaginary axis. This is similar to what was obtained in \cite{Mukhametzhanov:2021hdi} in the context of SYK.}:
\begin{align}
\label{eq:HS-deformation}
   \int \mathcal{D}g\, \mathcal{D}\Phi\, e^{-I_{\rm JT}-I_{\rm local} - I_{\rm nonlocal} } &=  \int \mathcal{D}g\, \mathcal{D}\Phi\,  \mathcal{D}Q(b)\, e^{-I_{\rm JT}+\int_0^\infty \d b\, b\,Q(b)\,\mo_{\rm G}(b, \Phi)}\,e^{ \frac{1}{2}\int_0^{\infty} \d b\,(Q(b)-Z_\text{brane}(b))^2}\nn\\&=\average{\int \mathcal{D}g\, \mathcal{D}\Phi\, e^{-I_{\rm JT}+ \frac{1}{2}\int \d^2 x\sqrt{g}\,U_{\rm local}(\Phi,Q(b))}}_{\text{couplings}}\,,
\end{align}
where the local dilaton potential is now given $U_{\rm local}(\Phi,Q(b)) = e^{-2\pi \Phi}\int db\,b Q(b) \cos(b \Phi(x))$ and we have to ensemble average over the coupling  $Q(b)$ with a Gaussian weight centered around $Z_\text{brane}(b)$.

Thus, all the factorizing theories described in this paper can alternatively be viewed as Gaussian ensembles of different local bulk theories of dilaton gravity. The integral over $Q(b)$ then has the role of integrating-out wormholes: while each theory in the ensemble has wormholes in its geometric expansion, the resulting theory has only disks and half-wormhole remain in its geometric expansion. Once again, all wormhole contributions cancel. This implies that each alpha-states  can also be viewed, not only to correspond to a single gravitational theory that is non-local and factorizable, but also as an ensemble average over bulk theories:
\be 
\average{\int \mathcal{D}g \mathcal{D}\Phi e^{-I_{\rm JT}+ \frac{1}{2}\int \d^2 x\sqrt{g}\,U_{\rm local}(\Phi,Q(b))}}_{\text{couplings}}\, \Leftrightarrow \qquad \ket{\alpha_\Hh} = \average{e^{ \int \d b \, b\,Q(b)\, \hat{Z}_{\text{G}}(b)}}_{\text{couplings}} \ket{\text{HH}}\,.
\label{eq:ensemble-average-over-couplings}
\ee

\subsection*{Integrating in wormholes: an ensemble average over alpha-states}

Alternatively, we can instead start with the alpha-states, which are described by the factorizing theory with all wormhole contributions cancelling, and try to integrate-in the wormholes. This is again given by an ensemble average that is no longer Gaussian, but instead is given by the JT gravity matrix integral itself. Explicitly, starting with the partition function $Z_\text{JT}(\beta_1,\dots, \beta_n) = \bra{\text{HH}}\hat{Z}(\b_1)\cdots \hat{Z}(\b_n)\ket{\text{HH}}$ and introducing the resolution of the identity $\mathbb{1} = \int \d \Hh \ket{\alpha_{\Hh}}\bra{\alpha_{\Hh}}$ we can write,
\begin{align}
&Z_{\rm JT}(\b_1,\dots,\b_n) = \bra{\text{HH}}\hat{Z}(\b_1)\cdots \hat{Z}(\b_n)\ket{\text{HH}} \nonumber\\
&\,\,\,\,= \int \d \Hh\, P_{\rm JT}(\Hh)\, Z(\b_1,\Hh)\cdots Z(\b_n,\Hh), \nn \\
&\,\,\,\,= \int \underbrace{\d \Hh P_{\rm JT}(\Hh)}_{\substack{\text{Average over}\\\text{bulk couplings}}} \,\,\overbrace{\bigg(\hspace{0.0cm}\begin{tikzpicture}[baseline={([yshift=1.2ex]current bounding box.center)}, scale=0.45 ]
 \pgftext{\includegraphics[scale=0.48]{Halfwormhole_contribution.pdf}} at (0,0);
  \draw (-1.9, -2.2) node {$\beta_1$};
    \draw (5.5, 0) node {$Z_{\rm brane}(b, \Hh)$};
         \draw (2.3, -2) node {};
          \draw (3.7, 0) node {};
  \end{tikzpicture} \hspace{-0.1cm} \bigg) \times \dots \times \bigg(\hspace{0.0cm}\begin{tikzpicture}[baseline={([yshift=1.2ex]current bounding box.center)}, scale=0.45 ]
 \pgftext{\includegraphics[scale=0.48]{Halfwormhole_contribution.pdf}} at (0,0);
  \draw (-1.9, -2.2) node {$\beta_n$};
    \draw (5.5, 0) node {$Z_{\rm brane}(b, \Hh)$};
         \draw (2.3, -2) node {};
          \draw (3.7, 0) node {};
  \end{tikzpicture} \hspace{-0.1cm} \bigg)}^{\text{Factorizing theories with couplings fixed by }\Hh} \,.\label{eq:average-over-alpha states}
\end{align}
Each term in the above expansion is a factorized theory in which all wormholes can be cancelled, these theories depend via the brane coupling $Z_{\rm brane}(b, \Hh)$ on $\Hh$. The integral over $\Hh$ can  be viewed as an ensemble over couplings in the factorized theories that is needed in order to ``integrate in'' the wormholes present in the geometric expansion of $Z_{\rm JT}(\b_1,\dots,\b_n)$. This is the mechanism discussed by Coleman, Giddings and Strominger \cite{coleman1988black, giddings1988loss} made precise in the context in JT gravity. 

 The results obtained above are not very surprising, because we start with a factorizing theory from the beginning, so averaging over $\Hh$ again should give the usual JT answers back. From this, one can already infer \eqref{eq:average-over-alpha states}. Nevertheless it is amusing to see that using the matrix integral (together with the degenerate cylinders that cause the appearance of $q$) we can derive these results in a very concrete way. 

Let us emphasize the difference between the ensemble averages that integrate out  \eqref{eq:ensemble-average-over-couplings} and integrate in the wormholes \eqref{eq:average-over-alpha states}: while the former is a simple universal Gaussian path integral (whose origin is the universal form of the bilocal interaction), the latter is a complicated non-Gaussian matrix integral where we have to consider $e^{\S}$ couplings. Integrating-in the wormholes is much more complicated.\footnote{On a technical level, this is due to the fact that the contribution of geometries of a certain genus in JT gravity depend on the complicated quotient by the mapping class group that has to be performed in the gravitational path integral. In contrast, when integrating out the wormholes the form of the mapping class group for a manifold with wormholes of a given topology is unimportant, as discussed in \cite{Blommaert:2021fob}.  }

\subsection{The baby universe Hilbert space of each factorizing model is one dimensional}\label{sect:one}

We should contrast the description found above for the baby universe Hilbert space of JT gravity, $\mathcal{H}_{\text{BU}}^{\text{JT}}$, with that of the baby universe Hilbert space,  $\mathcal{H}_{\text{BU}}^{\rm factorizing}$, of the factorizing theories found in \cite{Blommaert:2021fob}. In the latter, for a factorizing theory with some local deformation $I_{\rm local}$ and the universal bi-local deformation $I_{\rm nonlocal}$, inner products with the Hartle-Hawking state $\ket{\fHH}$ are computed as
\begin{align}
  \frac{  \bra{\fHH} \hat Z(\beta_1) \dots \hat Z(\beta_n) \ket{\fHH}}{ \bra{\fHH}\ket{\fHH}} &= \frac{1}{\mathcal Z^{\rm factorizing}} \sum_{\substack{\text{geometries} \\ \text{with $n$ asymptotic} \\ \text{boundaries}}}\int \mathcal{D}g\,\mathcal{D}\Phi\,  e^{-I_{\rm JT}-I_{\rm local}-I_{\rm nonlocal}} \nn \\ 
    & = \Tr(e^{-\beta_1\Hh})\dots\Tr(e^{-\beta_n\Hh})\,,
    \label{eq:fact-BU-Hilbert-space}
\end{align}
where we suppressed the fact that the definition of $\ket{\fHH}$ depends on $\Hh$. Consequently, in contrast to the case of undeformed JT gravity, \eqref{eq:fact-BU-Hilbert-space} shows that $\ket{\fHH}$ is an alpha-state in $\mathcal{H}_{\text{BU}}^{\rm factorizing}$.

Are there other alpha-states or does $\ket{\fHH}$ span all of $\mathcal{H}_{\text{BU}}^{\rm factorizing}$? To find the dimension of $\mathcal{H}_{\text{BU}}^{\rm factorizing}$, we can compute the inner-product between two arbitrary states in this Hilbert space\footnote{We thank D.~Stanford and Z.~Yang for useful comments in this direction.}  
  \be 
    \<\Psi_1| \Psi_2\> =  \alpha^\dagger_1\cdot \mathcal Z\cdot \alpha_2 \,,\qquad \text{ where } \qquad |\Psi_a\> = \sum_{i=0}^{n_\text{bdies}} \alpha^i_a\,\hat Z(\beta)^i|\fHH\>\,,
    \ee
where $n_\text{bdies}$ is the maximum number of boundaries involved in the state $|\Psi_a\>$ (which can be taken $\infty$) and where, for simplicity, we restrict to creating asymptotic boundaries with the same $\beta$. \eqref{eq:fact-BU-Hilbert-space} implies that the inner-product matrix (whose dimension is $n_\text{bdies}\times n_\text{bdies}$) is
\begin{equation}
    \mathcal{Z}_{i j}=\Tr(e^{-\beta\Hh})^{i+j}
\end{equation}
The rank of $\mathcal Z$ gives the dimension of $\mathcal H_{\text{BU}}^{\rm factorizing}$, and the above matrix has rank one, because it can be written as the square of a vector. Following a similar construction for inner products of states created by inserting asymptotic boundaries with different $\beta_i$, one finds that $\mathcal H_{\text{BU}}^{\rm factorizing}$ is one-dimensional. So $|\fHH\>$ is the unique state in $\mathcal H_{\text{BU}}^{\rm factorizing}$, which is radically different from the infinite dimensional $\mathcal H_{\text{BU}}^{\rm JT}$. The factorizing theory with BU Hilbert space $\mathcal H_{\text{BU}}^{\rm factorizing}$ is associated with one specific alpha-state $\ket{\alpha_\Hh}$ in $\mathcal H_{\text{BU}}^{\text{JT}}$, therefore $\mathcal H_{\text{BU}}^{\rm factorizing}$ is the subspace of $\mathcal H_{\text{BU}}^{\text{JT}}$ spanned by $\ket{\alpha_\Hh}=\ket{\fHH}$.

To emphasize, the non-trivial point result is that we find that the dimension of the BU Hilbert space in the factorizing models of \cite{Blommaert:2021fob} is one dimensional, a property that had been predicted for UV complete theories in \cite{Marolf:2020xie,McNamara:2020uza}.


\section{Null states in JT gravity}\label{sect:null}

In the previous section we constructed explicit alpha-states and showed that they form a complete basis for the baby universe Hilbert space, because they cover the whole ensemble of random Hamiltonians $\Hh$. When constructing any Hilbert space one needs to quotient by potential null states. However, since we used the matrix integral to show that the alpha-states \eqref{eq:alpha-state-definition} form a complete basis, there are simply no null states to account for in the matrix integral. In a sense, they have already been eliminated since each different matrix in the ensemble has a different spectrum. This leads to the conclusion that the matrix integral only knows about the Hilbert space \emph{after} the quotienting has been performed, also called GNS Hilbert space.

Nevertheless, in situations where we do not have the luxury of considering the matrix integral or something equivalent as for instance would be the case in higher dimensions, the question remains whether there is a geometric or perhaps semiclassical understanding of the null states. In general this is a hard question to answer and hints at the mechanism emphasized in \cite{Saad:2021rcu}, where it was argued in the context of the $G\Sigma$ action of SYK that there could be multiple bulk descriptions which ultimately give the same physics. In our model we can make this mechanism very precise by studying two alpha-states that within the matrix model give the same boundary spectrum, but when going to a geometric picture or in a naive semiclassical approximation, there is a clear difference between the two. In fact, we will show that there is an infinite number of geometric theories that give rise to the same boundary spectrum. Subtracting (or taking the appropriate linear combinations of) alpha-states associated to theories that have the same spectrum yields null states that are not part of the baby universe Hilbert space.
\subsection{Discreteness implies redundancies}\label{sect:4.1}

The key to understanding why two superficially different alpha-states give the same boundary spectrum is the discreteness of the spectrum. Namely, this implies that as long as we localize on the chosen spectrum at large $q$ we can change the deforming operator in the matrix integral to whatever way we wish. A bit more concretely, when we alter $Z_\text{brane}$ in such a way that the localization equations \eqref{eq:saddle-point-equation-matrix-integral} remain the same and we do not spoil the stability of the saddle (i.e. the Hessian around the saddle has the correct sign for all its eigenvalues), the effect in the matrix integral is absolutely nothing. No observable will change whatsoever, we call such changes in $Z_\text{brane}$ \emph{null deformations}. The name of the game is then to see whether we can translate this change back to an ordinary geometric understanding. 

To do so, let us first map out a subset of all the null states and translate those back to deformations of the Euclidean path integral. To simplify the discussion we consider as our initial theory JT gravity with just the bilocal deformation turned on, so $I_{\rm local}=0$ in \eqref{eq:local-deformation}. Generalizations to cases with non-zero $Z_{\rm brane}$ as the initial theory are straightforward. We then consider general deformations around this theory, and seek for deformations that do not affect the spectrum. 

General deformations around the theory with $I_{\rm local}=0$ (which we call canonical JT gravity) correspond with the following insertion in the JT matrix integral 
\begin{align}
\exp\bigg((q+1) \int_0^\infty \d b\, b\,\mo_{\rm G}(b)\,\delta Z_\text{brane}(b)-\frac{q}{2}\int_0^\infty \d b\, b\,\mo_{\rm G}(b) \mo_{\rm G}(b)\bigg)\,,\label{bb}
\end{align}
with $q\to\infty$ and $ \delta Z_\text{brane}(b)$ parameterized as\footnote{The dilaton potential and the genus zero spectral density are linearly related when the latter vanishes at $E=0$ \cite{Turiaci:2020fjj, Witten:2020wvy}. In those cases $\delta\rho_0(E)$ is the change in the genus zero spectral density if one would not turn on the bilocal. In other cases, the would-be genus zero spectral density $\rho_{0,\text{JT}}(E)$ would receive subleading corrections in the deformation of the action. None of this is relevant in our factorizing models, where localization determines the discrete spectrum.}
\begin{equation}
    \delta Z_\text{brane}(b)=\int_0^\infty \d E\,\delta\rho_0(E)\,\frac{2}{b}\cos(b E^{1/2})\,.\label{42}
\end{equation}
Here $\delta Z_{\rm brane}$ is actually of the most general form, since $b\,Z_{\rm brane}(b)$ needs to be an even function of $b$. For this particular deformation to not change anything about the spectrum $\delta Z_{\rm brane}$ needs to satisfy some conditions. These conditions can be derived by considering the large $q$ saddle-point equations for the discrete set of eigenvalues $\l_i$\footnote{Here $P( \l,E)$ is a regularized version of $2\log\abs{\l-E}$ with the regulator only relevant when $\delta\rho_0(E)$ has delta spikes \cite{Blommaert:2021fob}
\begin{align} 
   P(\l, E) = 
 \log\Big(\left(E_1^{1/2} - E_2^{1/2}\right)^2 + \varepsilon^2\Big) + \log\Big(\left(E_1^{1/2} + E_2^{1/2}\right)^2 + \varepsilon^2\Big)\,,
   \label{eq:def-P-regulated}
\end{align}}
\begin{equation}
    LV_\text{JT}'(\l_i)+\int_{0}^{\infty} \d E\,\delta\rho_0(E)\,\partial_{\l}P(\l,E)\rvert_{\l=\l_i}=\frac{1}{2\l_i}+2\sum_{j\neq i}\frac{1}{\l_i-\l_j}\,.\label{44}
\end{equation}
One can view the second term as introducing a change in the electrostatic potential for the eigenvalues
\begin{equation}
    L( V_\text{JT}'(\l_i)+ \delta V'(\l_i))=\frac{1}{2\l_i}+2\sum_{j\neq i}\frac{1}{\l_i-\l_j}\,.\label{eom}
\end{equation}

For smooth functions $\delta\rho_0(E)$ you recover the standard relation \cite{Saad:2019lba,brezin1993planar,Blommaert:2021fob} between changes in the matrix integral potential and changes in the disk spectral density
\begin{equation}
    L\delta V'(\l)=2\fint_0^\infty \d E\,\delta\rho_0(E)\,\frac{1}{\l-E}\,.\label{46}
\end{equation}
where $\fint$ is the principal value integral. 

Let us define the eigenvalues $\l_{\text{JT}\,i}$ of a matrix $\HJT$ to be the solutions of the electrostatics problem for canonical JT gravity
\begin{equation}
    LV_\text{JT}'(\l_{\text{JT}\,i})=\frac{1}{2\l_{\text{JT}\,i}}+2\sum_{j\neq i}\frac{1}{\l_{\text{JT}\,i}-\l_{\text{JT}\,j}}\,.\label{vjt}
\end{equation}
Suppose now that we consider deforming this theory by adding fine-tuned changes in the potential that depend specifically on the $\l_{\text{JT}\,i}$ in such a way that the deformation satisfies
\begin{equation}\label{tosatisfyfornull}
    L \delta V'(\l_{\text{JT}\,i})=0\,,
\end{equation}
whilst not spoiling the signs of the eigenvalues of the Hessian, then the electrostatics problem \eqref{eom} has exactly the same solutions $\l_{\text{JT}\,i}$. This means that the large $q$ matrix integrals with and without these \emph{null deformation} are equivalent, as they localize onto the same eigenvalues. In particular all observables calculated in the matrix model at large $q$ will agree, for instance in both theories
\begin{equation}
    Z(\beta_1\dots\beta_n)=\Tr(e^{-\beta_1\HJT})\dots \Tr(e^{-\beta_n\HJT})\text{ with or without null deformation}\,.\label{49}
\end{equation}

This means that in the large $q$ matrix integral with the bilocal interaction, there is a vast redundancy in description when specifying the theory by its potential $V(\l)$
\begin{align}
    V_\text{JT}(\l)\quad \text{ indistinguishable from }\quad V_\text{JT}(\l)+\delta V_\text{null}(\l)\,\label{410}
\end{align}
The fundamental reason for this redundancy is that in the large $q$ matrix integral we are only imposing a \emph{discrete} set of equations of motion \eqref{eom}. For instance, a non-trivial solution can be obtained by solving \eqref{46}
\begin{equation}
    \fint_0^\infty \d E\,\delta\rho_{0\,\text{null}}(E)\,\frac{1}{\l-E}= G(\l)\,,\quad G(\l_{\text{JT}\,i})=0\,.\label{411}
\end{equation}
If we find a function $G(\l)$ with zeros on the spectrum $\l_{\text{JT}\,i}$ we can invert this relation using standard methods for singular integral equations \cite{estrada2012singular} provided $G$ decays sufficiently fast to zero at large $\l$, from that we can determine a null deformation $Z_\text{brane\,null}$ using \eqref{42}, which in turn determines a $I_\text{local\,null}$ in the dilaton gravity action via \eqref{eq:local-deformation}. 

We show in section \ref{sect:4.3} that there is one deformation $\delta\rho_{0\,\text{null}}(E)$ corresponding to the trivial solution $G(\l)=0$. All other null deformations will be immediate consequences of discreteness, the fact that we can have many nontrivial functions $G(\l)$ that vanish on a \emph{discrete} set of points $\l_{\text{JT}\,i}$.
\subsection{Null states}
\label{subsect:null states}

In section \ref{sect:alpha}, we have described how $\mathcal H_{\rm BU}^{\rm JT}$ can be constructed by acting on $\ket{\rm HH}$ with boundary creation operators and then taking linear combinations of the resulting states, 
\be 
\hat Z_G(b_1) \dots \hat Z_G(b_n) \ket{\rm HH}\,.
\ee
As pointed out by Marolf and Maxfield \cite{Marolf:2020xie}, in this construction of $\mathcal H_{\rm BU}^{\rm JT}$, we need to be careful to eliminate the linear combinations that have zero norm, i.e.~the possible null states in this construction.

The deformations of the matrix integral that we have described above, that leave the localization locus invariant, have a natural interpretation in terms of null states. As mentioned earlier, to appreciate this relation with null states it is essential to distinguish the large $q$ matrix integral perspective, where degenerate cylinders are included, from the spacetime path integral, where degenerate cylinders do not exist. In the large $q$ matrix integral, we are interested in the deformation
\begin{equation}\label{nulldeformationsMM}
    q I(H,\HJT,\mathsf{s})=-\frac{q}{2}\int_0^\infty \d b\, b\,\mo_{\rm G}(b)\,\mo_{\rm G}(b)+(q+1)\int_0^\infty \d b\, b\,\mo_{\rm G}(b)\,Z_{\text{brane}\,\text{null}}(b,\HJT,\mathbf{s})\,,
\end{equation}
with
\be 
Z_{\text{brane}\,\text{null}}(b,\HJT,\mathbf{s}) = \sum_{a} s_a\, Z^{(a)}_{\text{brane}\,\text{null}}(b,\HJT)
\ee
with $a$ labelling all solutions $\delta\rho^{(a)}_{0\,\text{null}}$ of \eqref{411} that gives the null-state deformation $Z^{(a)}_{\text{brane}\,\text{null}}(b,\HJT)$, and $s_a$ some complex numbers whose range is constrained so as to ensure stability of the large $q$ saddle point. In the baby universe Hilbert space language these correspond via \eqref{eq:alpha-grav-vs-matrix} with the states
\be 
\ket{\a_{\HJT},\mathbf{s}} = \mathcal{N}_{\HJT\,\mathsf{s}} e^{-\hat I_{\rm deform}(\HJT,\mathsf{s})}\ket{\text{HH}}\,,\label{4.14}
\ee
with
\begin{equation}
    \hat I_{\rm deform}(\HJT,\mathsf{s}) = \frac{1}{2} \int_0^\infty \d b \, b\, \hat{Z}_\text{G}(b)\,\hat{Z}_\text{G}(b) -\int_0^\infty \d b \, b\, \hat{Z}_\text{G}(b)\, Z_{\text{brane}\,\text{null}}(b,\HJT,\mathbf{s})\label{4.15}
\end{equation}

The null states $\ket{\psi}$ are hence given by differences of alpha-states with different values of parameters $s_a$ 
\begin{equation}
    \ket{\a_\HJT,\mathbf{s}}-\ket{\a_\HJT,\mathbf{s}'}\,\text{ is a  null-state if $\mathbf{s}\neq \mathbf{s}'$}\,,
    \label{eq:simplified-null-state}
\end{equation}
or more generally if there are $k$ parameters $s_a$ and with $\G$ denoting the stable deformations
\begin{equation}
    \int_{\G} \d^k\mathbf{s}\,f(\mathbf{s})\ket{\a_\HJT,\mathbf{s}}\,\text{ is a null-state if }\int_{\G} \d^k\mathbf{s}\,f(\mathbf{s})=0\,.
    \label{eq:general-null-state}
\end{equation}
This is simply because 
\be 
 \int_{\G} \d^k\mathbf{s}\,f(\mathbf{s}) \bra{{\rm HH}} \hat Z(\beta_1) \dots  \hat Z(\beta_n) \ket{\a_\HJT,\mathbf{s}} =\Tr \left(e^{-\beta_1 \HJT}\right) \dots \Tr \left(e^{-\beta_n \HJT}\right)\int_{\G} \d^k\mathbf{s}\,f(\mathbf{s}) = 0 \,,
\ee
for all $n$. Once again, using the fact that $\hat Z(\beta_1),\, \dots,\,  \hat Z(\beta_n) \ket{{\rm HH}}$ span the Hilbert space, it then follows that \eqref{eq:simplified-null-state} and \eqref{eq:general-null-state} are null.

We remark here that even though from the matrix model perspective (as $q \to \infty$) the $\alpha$-states $\ket{\a_{\HJT},\mathbf{s}}$ do not depend on $s_a$, they will actually lead to \emph{inequivalent} gravitational descriptions of the bulk. Said differently, since in gravity we do not consider the degenerate cylinders, the parameter $q$ does not appear and the operators with or without null deformations turned on are different, as in section \ref{sect:4.3}.

These null states described here are quite similar to the examples constructed by Marolf and Maxfield \cite{Marolf:2020xie} in pure $2$d Einstein-Hilbert gravity. In that model one can construct null deformations analogous to our \eqref{4.15} by inserting in the gravity path integral for an alpha-state the operator
\begin{equation}
   \exp\bigg(\i\sum_{n=1}^\infty s_n\, 2\pi n \hat{Z}\bigg)\,.
\end{equation}
These are null deformations because
\begin{equation}
    \hat{Z}\ket{m}=m\ket{m}\,.
\end{equation}
whose analog we will describe shortly. In both our and their models we can have null deformations because there is a fundamental discreteness in the spectrum of the theory. Nevertheless, in our case we will see a rich set of null states that can depend on the exact details (not only discreteness) of the alpha-states whose null deformations we study.


\subsection{Examples of null states and their spacetime interpretation}\label{sect:4.3}
One benefit that we have, as compared to the simple model of \cite{Marolf:2020xie}, is that we can map null deformations to changes in the dilaton gravity potential. Indeed, using \eqref{eq:alpha-grav-vs-matrix} we see that there are inequivalent dilaton gravity actions associated with our class of matrix integrals \eqref{nulldeformationsMM}
\begin{equation}
    I_\text{deform}(\HJT,\mathsf{s})= \frac{1}2 \int \d b\, b\, \mo_\text{G}(b,\Phi)\, \mo_\text{G}(b,\Phi) -\int \d b \,b\, \mo_\text{G}(b,\Phi)\, Z_{\brane}(b,\HJT,\mathsf{s})\,,
\end{equation}
which can be written explicitly in dilaton and metric variables as in \eqref{eq:non-local-deformation} and \eqref{eq:local-deformation} using
\begin{align}
\mo_\text{G}(b,\Phi) =  e^{-\S}\int \d^2 x \sqrt{g}\, e^{-2\pi \Phi(x)} \cos\left(b\Phi(x)\right)\,.
\end{align}
The change in the dilaton action always involves the bilocal term to ensure factorization and is universal, but changes to the local dilaton potential itself can be null. This means that gravitational path integrals with two potentials related by a null deformation are inequivalent perturbatively, but once the matrix model non-perturbative effects are included they are equivalent. 

For instance, if one wants to calculate the semiclassical entropy of the black hole in two null-related dilaton gravities, one would obtain identical answer to leading order, but they would start deviating at subleading orders in $e^{\S}$. We expand more on this in section \ref{subsec:bulk-interp-of-null states}.


\subsection*{A first example: intuitive null operators and null states}

A\textit{ null operator} can be constructed by using the discreteness of the spectrum and inserting,\footnote{We thanks Douglas Stanford and Henry Maxfield for bringing this up.}
\begin{equation}
e^{2\pi \i k \hat N(E_1, E_2)}=  
    \exp\bigg( 2\pi\i k \int_{E_1}^{E_2}\d E\,\hat{\rho}(E)\bigg)\,\qquad \text{ for } k \in \mathbb{Z}\,,
    \label{eq:exp-of-number-op}
\end{equation}
When acting on any alpha-state we expect this operator gives one because the spectrum is discretized and consequently the eigenvalues of $ \hat N(E_1, E_2)$ are given by $N(E_1, E_2) \in \mathbb Z^+$. 
Consequently, 
\be 
\label{eq:null-states-from-null-op}
\left(e^{2\pi \i k \hat N(E_1, E_2)} - \mathbb 1\right) \ket{\alpha_\Hh} \sim 0\,,
\ee
is null for all $\Hh$ for which $E_1$ and $E_2$ do not belong to the spectrum.\footnote{We should be careful when $E_1$ and $E_2$ are part of the spectrum of $\Hh$. In such corner cases, the operator in \eqref{eq:null-states-from-null-op} might not be null depending on the definition of the integral in \eqref{eq:exp-of-number-op}. Thus, perhaps it would be better to call this operator \textit{almost null } since there is a set of measure zero set of states for which its action might not result in a null state. }

We can now see where this operator comes from in our formalism and whether the null states that it generates can indeed be though of as those described in section \ref{sect:4.1} and \ref{subsect:null states}. First, to define $\hat{\rho}$ we take an inverse Laplace transform of $\hat{Z}(\b)$, 
\be 
\hat{\rho}(E) = \rho_{0,{\rm JT}}(E) + \int_0^{\infty} b \d b \frac{\cos(b \sqrt{E})}{2\pi \sqrt{E}} \hat{Z}_G(b)
\ee
where we remind the reader that $\rho_{0,{\rm JT}}(E)$ is the leading density of states in JT gravity given by the disk contribution.  If we now insert this in \eqref{eq:exp-of-number-op} we can read off what $\delta Z_{\rm brane}(b)$ needs to be by brining the integral over $E$ inside. The constant coming from the smooth disk density of states will give a constant that is the same for all our alpha-states and can be factored out. The result for $\delta Z_{\rm brane}(b)$ is
\be 
\delta Z_{\rm brane}(b) = 2\i \,k\,  \frac{\sin(b \sqrt{E_1}) - \sin(b \sqrt{E_2})}{b}.
\ee
Inverting the relation \eqref{42} we get a $\delta \rho_0(E)$ given by
\be \label{deltarhoExpN}
\delta \rho_0(E) = \frac{4\i\,k\,\sqrt{E}}{\pi} \left[ \frac{\sqrt{E_2}}{E_2 - E} - \frac{\sqrt{E_1}}{E_1 - E} \right].
\ee
To check whether this is indeed a null deformation as we have defined it in \eqref{tosatisfyfornull}, we simply need to calculate \eqref{46} and show that it gives a constant shift in the potential. This is indeed the case as long as $\lambda \neq E_i$, which is always the case since we want to integrate the density in \eqref{eq:exp-of-number-op} always in between the delta peaks not on top of them and so we always have $\delta V'(\l_{\text{JT}\, i}) = 0$ as required. When $k \in \mathbb Z$, the state $e^{2\pi \i k \hat N(E_1, E_2)}\ket{\alpha_\Hh} $ is unit normalized and thus the general framework described in section section \ref{sect:4.1} and \ref{subsect:null states} indeed describes the null states in \eqref{eq:null-states-from-null-op}.  More generally, we can construct null operators by considering functions of $\hat N(E_1, E_2)$ that vanish on the integers - since such functions can be constructed by taking linear combinations of the operators in  \eqref{eq:exp-of-number-op} the null states described by such operators are simply linear combinations of those found in \eqref{eq:null-states-from-null-op}.   However, not all null states can be obtained from null operators and we give two examples of such cases below.\footnote{This is the case even when taking $k \,\slashed{\in}\, \mathbb{Z}$. In such a case, \eqref{eq:null-states-from-null-op} needs to be modified by multiplying $e^{2\pi \i k \hat N(E_1, E_2)}$ by a normalization factor that is dependent on the specific alpha-state upon which we act. }

\subsection*{A second trivial solution}
To get some feeling for the null deformations, we first consider the case $G(\l)=0$ which has solutions \cite{estrada2012singular}
\be 
\delta \rho_{0\,\text{null}}(E) = s\,\frac{1}{E^{1/2}} .
\ee
Plugging this in \eqref{411}, this gives zero, so that this particular $\delta \rho$ induces a constant shift of the potential
\be
\delta V(\lambda) = \text{ constant }\quad \Rightarrow \quad  \delta V'(\l) = 0 \,,
\ee 
so there is no change to the saddle point equations and the spectrum on which we localize is not affected. The brane one-point function associated to this deformation of the matrix model potential is
\be 
Z_{\rm brane\, null}(b) = s\,\int_0^{\infty} \d E\,\frac{2}{b} \frac{1}{E^{1/2}} \cos(b E^{1/2}) = s\, \frac{\delta(b)}{b}.
\ee
In the dilaton gravity action this deformation gives rise to a change of the dilaton potential of the form
\be 
\label{eq:change-in-the-dilaton-example}
\delta U(\Phi) = s\, e^{-\S}\int_{\Sigma} \d^2 x \sqrt{g} \, e^{-2\pi \Phi}\,. 
\ee
Geometrically, this corresponds to the insertion of a gas of cusps. The point is that we have shown here that non-perturbatively, when we include the bilocal interaction, this gas of cusps no effects whatsoever on the resulting discrete spectrum.

Following the description in section \ref{subsect:null states}, from the deformation \eqref{eq:change-in-the-dilaton-example} we find the following null states for example
\be
 \left(\mathcal{N}_{\HJT\,0},\,e^{-\frac{1}{2} \int \d b \, b \,\hat{Z}_{G}(b)\,\hat{Z}_{G}(b) } - \mathcal{N}_{\HJT,\,s}\,e^{-\frac{1}{2} \int \d b \, b \,\hat{Z}_{G}(b)\,\hat{Z}_{G}(b) +s\,\hat{Z}_\text{G}(0)} \right)\ket{\text{HH}}\sim 0\,.
 \label{eq:null states-very-explicit}
\ee
While the operator acting on $\ket{\text{HH}}$ is not null, the resulting state is. 
Such states again have to be eliminated in order to construct $\cH^{\text{JT}}_{\text{BU}}$, which practically means we should include only one of the states \eqref{4.14} because  all resulting theories have the same spectrum.

\subsection*{Non-trivial null states: the determinant solution}
Another example that comes to mind is by taking $G_{\rm JT}(\l)$ in \eqref{411} to be the characteristic polynomial of $\HJT$
\begin{equation}
    \fint_{0}^{+\infty} \d E\,\delta\rho_{0\,\text{null}}(E)\,\frac{1}{\l-E}=\det(\l-\HJT)\,e^{-\frac{L}{2}V_\text{JT}(\l)}\,.\label{428}
\end{equation}
To get something useful out of this we can use the following fact, proven in section \ref{sec:gravity_without_chaos} below. To leading order in large $L$, the characteristic polynomial of $\HJT$ equals the ensemble average of the characteristic polynomial of $H$ in the original undeformed JT gravity ensemble
\begin{equation}
    \det(\l-\HJT)=\average{\det(\l-H)}_\text{JT}\,.
\end{equation}
The potential term in \eqref{428} makes for a finite double scaling limit, upon which the determinant reduces to the Baker-Akhiezer function $\psi_\text{JT}(\l)$ of JT gravity ($x=0$ in \eqref{532} below). The leading order behavior of the Baker-Akhiezer functions can be obtained using a WKB or disks and cylinders approximation \cite{Maldacena:2004sn,Saad:2019lba} and one finds up to a proportionality factor (that can be absorbed in the parameter $s$ multiplying the null deformation)
\begin{equation}
    \psi_\text{JT}(\l)\sim \frac{1}{\l^{1/4}}\cos\bigg(\pi \int_0^\l 
    \d E\,\rho_{0,\text{JT}}(E)-\frac{\pi}{4}\bigg)\,,\label{432}
\end{equation}
and so
\be 
 \fint_{0}^{+\infty} \d E\,\delta\rho_{0\,\text{null}}(E)\,\frac{1}{\l-E} = \frac{1}{\l^{1/4}}\cos\bigg(\pi \int_0^\l \d E\,\rho_{0,\text{JT}}(E)-\frac{\pi}{4}\bigg)\,.
\label{eq:delta-rho-to-BA}
\ee

Using standard techniques for singular integral equations one can solve for $\delta \rho_{0,\rm null}$ and deduce from that the explicit null deformation of the dilaton potential. The most practical way to do so is to write the Baker-Akhiezer function in the Fourier domain, using a generalization of the Kontsevich \cite{KontsevichModel} integral representation of the Airy function built to reproduce the WKB approximation above. Inverting \eqref{eq:delta-rho-to-BA}, one finds that at large $\int_0^E \d M\,\rho_{0,{\rm JT}}(M)$ the approximate null deformation is given by 
\begin{equation}
     \delta \rho_{0\,\text{null}}(E)=s\, \frac{1}{E^{1/4}}\sin\bigg(\pi \int_0^E \d M\,\rho_{0,{\rm JT}}(M)-\frac{\pi}{4}\bigg)\,.
\end{equation}
This can be generalized by inserting polynomial functions of the determinant as left-hand side in \eqref{428}. One ultimately finds the following null deformations of the dilaton potential (to leading order)
\begin{equation}
     \delta U(\Phi)=e^{-\S}\,\Phi \sum_{a=1}^\infty  s_a\, \sin\bigg(a\pi \int_0^{\Phi^2} \d M\,\rho_{0,{\rm JT}}(M)-\frac{\pi}{4}\bigg)\,e^{-2\pi\Phi}\,.\label{nullCH}
\end{equation}
Turning on these deformations would semiclassically appear to have non-trivial effects (see section \ref{subsec:bulk-interp-of-null states}), but non-perturbatively they do not change the theory at all (as long as we have the bilocal interaction).


\section{Black holes without quantum chaos}
\label{sec:gravity_without_chaos}

We saw in section \ref{sect:null} that there is a spectrum $\textsf{H}_{\rm JT}$ that is intrinsic to JT gravity: it is uniquely determined by the matrix integral potential $V_\text{JT}(\l)$ since it is the spectrum of the theory of dilaton gravity whose action is
\be 
\label{eq:bulk-canonical-theory}
I = I_{\rm JT} + I_{\rm non-local}\,.
\ee
As mentioned in the introduction, due to the lack of a correction to the local dilaton potential, the geometric expansion in this model, seen in \eqref{eq:disk+nothing}, only contains the disk contribution as the half-wormhole contribution vanishes due to $Z_{\rm brane} = 0$. Notably this system does not require fine-tuning $\sim e^{\Ss}$ parameters in the dilaton-gravity action, unlike the models studied in \cite{Blommaert:2021fob}. 

We now study the properties of this spectrum $\HJT$, and through this analysis, present a dual QM description to the bulk theory \eqref{eq:bulk-canonical-theory}.

Let us summarize the main steps that result in this QM description.
\begin{enumerate}
    \item The spectrum $\l_\text{JT}$ of $\HJT$ are the solutions to the electrostatic problem \eqref{vjt}. We will prove that (to leading order in large $L$) the solutions are the zeros of the orthogonal polynomial $P_{\text{JT}\,L}(\l)$
    \begin{equation}
    LV_\text{JT}'(\l_{\text{JT}\,i})=\frac{1}{2\l_{\text{JT}\,i}}+2\sum_{j\neq i}^L\frac{1}{\l_{\text{JT}\,i}-\l_{\text{JT}\,j}}\quad \Leftrightarrow \quad P_{\text{JT}\,L}(\l_{\text{JT}\,i})=0\,.\label{51}
    \end{equation}
    \item These zeros are approximately evenly spaced with coarse grained spectral density $\rho_{0,\text{JT}}(\l)$, this is called \emph{clock behavior}, and should be contrasted with the fluctuations in level spacing in quantum chaotic systems (which follows the Wigner surmise, or has random matrix statistics \cite{Mehta_1994,Haake:1315494})
    \begin{equation}
        \l_{\text{JT}\,i+1}-\l_{\text{JT}\,i}=\frac{1}{\rho_{0,\text{JT}}(\l)}\,.
    \end{equation}
    
    \item The orthogonal polynomials with respect to the JT potential $P_{\text{JT}\,n}(\l)$ satisfy a recursion relation, which can be written as a matrix equation featuring an $\infty$ dimensional matrix $Q$
    \begin{equation}
        \l\,P_{\text{JT}}(\l)=Q\cdot P_{\text{JT}}(\l)\,.\label{rec}
    \end{equation}
    Imposing the quantization condition \eqref{51} $P_{\text{JT}\,L}(\l)=0$ reduces this to an eigenvalue equation for the reduction $Q_L$ of $Q$ to its first $L$ rows and columns, thus the eigenvalues of $Q_L$ are the zeros of $P_{\text{JT}\,L}(\l)$, which in turn (as we saw above) are the spectrum of $\HJT$. So $Q_L$ and $\HJT$ have identical eigenvalues, and can be identified to leading order in large $L$
    \begin{equation}
        P_{\text{JT}\,L}(\l)=\det(\l-Q_L)=\det(\l-\HJT)\quad \Leftrightarrow\quad Q_L=\HJT\,.
    \end{equation}
    \item Upon double scaling \cite{Saad:2019lba}, orthogonal polynomials become Baker-Akhiezer functions and the matrix $Q$ becomes the operator $\hat{Q}$ in the Lax formalism \cite{Eynard:2015aea,Maldacena:2004sn,Okuyama:2019xbv,DiFrancesco:1993cyw,Dijkgraaf:1991qh}. The recursion relation \eqref{rec} becomes a Schrodinger equation, and crucially \eqref{51} becomes a Dirichlet boundary condition which \emph{discretizes} the spectrum
    \begin{equation}
        \l\, \psi_\text{JT}(x,\l)=\hat{Q}\,\psi_\text{JT}(x,\l)\,,\quad \hat{Q}=\hat{H}_\text{JT}=-\frac{1}{e^{2\Ss}}\partial_x^2+u_\text{JT}(x)\,,\quad \psi_\text{JT}(0,\l)=0\,.
    \end{equation}
    The partition function $Z(\beta)=\Tr\big(e^{-\beta\hat{Q}}\big)$ involves no projection operators, therefore this QM is a dual description of gravity - unlike when this $\hat{Q}$ appears in the Lax description of matrix integrals. 
\end{enumerate}
In the remainder of this section we provide more details and interpretation for these steps, and we drop the subscripts JT for notational comfort.


\subsection{Orthogonal polynomials}

Let us quickly review some facts about orthogonal polynomials on the real line, but see \cite{DiFrancesco:1993cyw, Eynard:2015aea} for more details. Consider polynomials $P_n(\l)$ orthogonal with respect to the measure $e^{-LV(\l)}\d \l$\footnote{We take the measure to be normalized such that $P_0(\l)=1$.}
\be 
\int_{-\infty}^{+\infty} \d \l\;  e^{-LV(\l)} P_n(\l)P_m(\l) = \delta_{n m}\,,\quad P_n(\l)=\frac{1}{\sqrt{h_n}}\l^n+\text{lower degree}\,.\label{56}
\ee
We are interested in large $L$ large and in even polynomials $V(\l)$. We can expand $\l P_n(\l)$ in $P_m(\l)$ with $m\leq n+1$, because of the maximal degree of each polynomial
\begin{equation}
    \l\,P_n(\l)=\sum_{m=0}^{n+1}Q_{n m}\,P_m(\l)\,.\label{57}
\end{equation}
Because the inner product of $\l\,P_n(\l)$ and $P_m(\l)$ is symmetric under exchanging $n$ and $m$, $Q$ is symmetric and we have only three terms in the expansion. Moreover the diagonal is zero, as the integral of $\l\,P_n(\l)^2$ vanishes for even potentials, leaving only $Q_{n n+1}=Q_{n+1 n}$ as nonzero matrix elements. Comparing the terms of highest degree gives finally $Q_{n n+1}=\sqrt{h_{n+1}}/\sqrt{h_n}$, such that the recursion relation is\label{recursionRel}
\be
\l P_{n}(\l) = a_{n+1}P_{n+1}(\l) + a_{n}P_{n-1}(\l)\,,\quad a_n=\frac{\sqrt{h_n}}{\sqrt{h_{n-1}}}\,.\label{58}
\ee
Very explicitly the so-called Jacobi matrix $Q$ reads
\be 
Q = \begin{pmatrix}
0 & a_1 & 0   & 0& \dots\\
a_1 & 0 & a_2 & 0     & \dots\\
0   & a_2 & 0 & a_3   & \dots \\
0   & 0 & a_3 & 0   & \dots \\
\vdots & \vdots & \vdots & \vdots & 
\end{pmatrix}\,.
\ee

The polynomials can thus be found recursively from knowledge of $a_n$, which in term can be found recursively from the potential $V(\l)$. Indeed, we obtain the following equation, sometimes referred to as the discrete string equation \cite{Eynard:2015aea}
\begin{align}
0 &= \int \d\l\, \frac{\d}{\d \l} \left( P_n(\l)P_{n-1}(\l) e^{-LV(\l)}\right) \nn\\
&= \int \d\l \left( P_n'(\l)P_{n-1}(\l) e^{-LV(\l)} + P_n(\l)P_{n-1}'(\l) e^{-LV(\l)} - LV'(\l) P_n(\l)P_{n-1}(\l)e^{-LV(\l)}\right)\nn\\
&= \frac{n}{a_n} - L V'(Q)_{n n-1}\,.
\end{align}
In the first and second term we used \eqref{56} and the definition of $a_n$ in \eqref{58} to expand the derivative of an orthogonal polynomial in terms of orthogonal polynomials of lower degree as $P'_n(\l)=n/a_n P_{n-1}(\l)+\text{lower degree}$, then most terms except one cancel using orthogonality. For the third term we can Taylor expand $LV'(\l)$ in $\l$ and then use that powers of $Q$ act as powers of $\l$ on the polynomials \eqref{57}. Since $Q$ depends only on $a_n$, the above is an equation that can be used to solve for the $a_n$. For instance when $V(x) = x^4/4 - t\, x^2/2$ this becomes 
\be \label{anQuartic}
\frac{n}{L a_n} =-t\, a_n+a_n(a_{n+1}^2 + a_n^2 + a_{n-1}^2)\,,
\ee
which can be solved recursively for $a_n$. This is then enough to find the orthogonal polynomials recursively from knowledge of the potential $V(\l)$.


\subsection*{Electrostatics and zeros of orthogonal polynomials}

To prove \eqref{51}, we need to take a different route. The proof follows essentially by deriving a second order differential equation for orthogonal polynomials, which was achieved by \cite{ismail2000electrostatics,chen1997ladder}. Though elementary, the derivation would distract a bit too much from our point, so we simply state the result
\be \label{diffeqpn}
P_n''(\l) - \bigg( L V'(\l) + \frac{A_n'(\l)}{A_n(\l)} \bigg) P_n'(\l) + S_n(\l) P_n(\l) = 0\,.
\ee
The form of $S_n(\l)$ will prove to be irrelevant, and $A_n(\l)$ is explicitly \cite{ismail2000electrostatics,chen1997ladder}
\be 
A_n(\l)= \int_{-\infty}^{+\infty} \d \mu \frac{V'(\l)-V'(\mu)}{\l-\mu}P_n(\mu)^2\, e^{-L V(\mu)}\,.
\ee
This differential equation can be used to obtain an equation for the zeros of the orthogonal polynomials \cite{ismail2000electrostatics}. Suppose that $\l_i$ are the zeros of $P_L(\l)$, then the derivatives at those zeros become
\be 
\sqrt{h_L}\,P_L(\l) = \prod_{i=1}^L(\l-\l_i)\,,\quad \sqrt{h_L}\, P_L'(\l_i) = \prod_{j\neq i}^L (\l_j-\l_i)\,,\quad \sqrt{h_L}\,P_L''(\l_i) = 2\sum_{k=1}^L \prod_{i\neq j\neq k}^L (\l_j - \l_i)
\ee
Evaluating \eqref{diffeqpn} at these zeros (for which the last term vanishes) results in the equation
\be 
2\sum_{i\neq j} \frac{1}{\l_i - \l_j} = L V'(\l_j) + \frac{A_L'(\l_j)}{A_L(\l_j)} \label{515}
\ee
To leading order in large $L$ we can ignore the second term, which is order one, and the same is true for the $1/2\l_i$ term in \eqref{51}. Therefore to leading order in large $L$ we indeed find that the solutions to our electrostatics problem (the eigenvalues to which the matrix integral localizes when we only turn on the bilocal deformation) are the zeros or the orthogonal polynomial $P_L(\l)$ of the original matrix ensemble
\begin{equation}
    2\sum_{i\neq j} \frac{1}{\l_i - \l_j} = L V'(\l_j)\quad\Leftrightarrow\quad P_L(\l_i)=0\,.
\end{equation}

This means in particular that the zeros of this polynomial match the eigenvalues of $\HJT$ (everything from hereon is implicitly to leading order in large $L$)
\begin{equation}
    \sqrt{h_L}\,P_L(\l)=\det(\l-\HJT)\,.\label{517}
\end{equation}


\subsection*{Jacobi matrix as the Hamiltonian}
Now we want to prove that the spectrum of $\HJT$ matches that of $Q_L$. For this we reconsider the recursion relation \eqref{57} but with the constraint that $P_L(\l)=0$. Now we see that the recursion relation closes on the polynomials $P_0(\l)\dots P_{L-1}(\l)$, in particular equation \eqref{58} fixes $P_{L-1}(\l)$ in terms of just $P_{L-2}(\l)$, resulting in the matrix equation
\begin{equation}
    \l\,P(\l)=Q_L\cdot P(\l)\,,\quad P_L(\l)=0\,,
\end{equation}
where $P(\l)$ is a column with entries $P_0(\l)\dots P_{L-1}(\l)$. This is now just a standard eigenvalue problem, searching for the eigenvalues of $Q_L$. The $L$ solutions are by construction the $L$ zeros of $P_L(\l)$, so that one can expand this polynomial as
\begin{equation}
    \sqrt{h_L}\,P_L(\l)=\det(\l-Q_L)\,.
\end{equation}
Up to unitary transformations\footnote{We care only about eigenvalues of $\HJT$ throughout this paper, for some discussion on the gravitational interpretation of the eigenvectors (or unitary transformations) see \cite{Blommaert:2020seb,Blommaert:2021etf,Saad:2019pqd}.} combined with \eqref{517} this means that the restriction of the Jacobi matrix to the first $L$ rows and columns is actually the Hamiltonian of the QM that we are localizing to in the matrix integral, so the Jacobi matrix has an important physical meaning
\begin{equation}
    \HJT=Q_L\,.
\end{equation}
Using the same argument one proves that 
\begin{equation}
    \sqrt{h_n}\,P_n(\l)=\det(\l-Q_n)\,.
\end{equation}

We also note that the orthogonal polynomials can can alternatively be computed as the expectation values of determinants in $n$ dimensional random matrix theory, using the same potential $V(\l)$
\begin{equation}
    \sqrt{h_n}\,P_n(\l)=\average{\det(\l-H)}_{n}\,,\label{orthopol}
\end{equation}
which means that the eigenvalues of $\HJT$ are the typical zeros of the determinant of $H$ in random matrix theory, which is not the same as the typical eigenvalues of $H$. Indeed, as we now discuss, the eigenvalue statistics of $\HJT$ is quite \emph{atypical} for the random matrix ensemble.


\subsection{No random matrix statistics}
\label{subsec:no-random-stats}

Now that we have a realization of $\HJT$ we want to know the properties of its spectrum. The level density follows to leading order in large $L$ the averaged level density $\rho_{0,{\rm JT}}(\l)$ in the original ensemble, we choose to present the proof of this only in the double scaling limit in section \ref{sect:ds}. Here instead we focus on the correlation between nearby energy levels, which has more surprising properties. The discussion of this section carries over unmodified to gravity in the double scaling limit, the only thing that changes really upon double scaling is the averaged level density.

\begin{figure}
    \centering
    \includegraphics[scale=0.6]{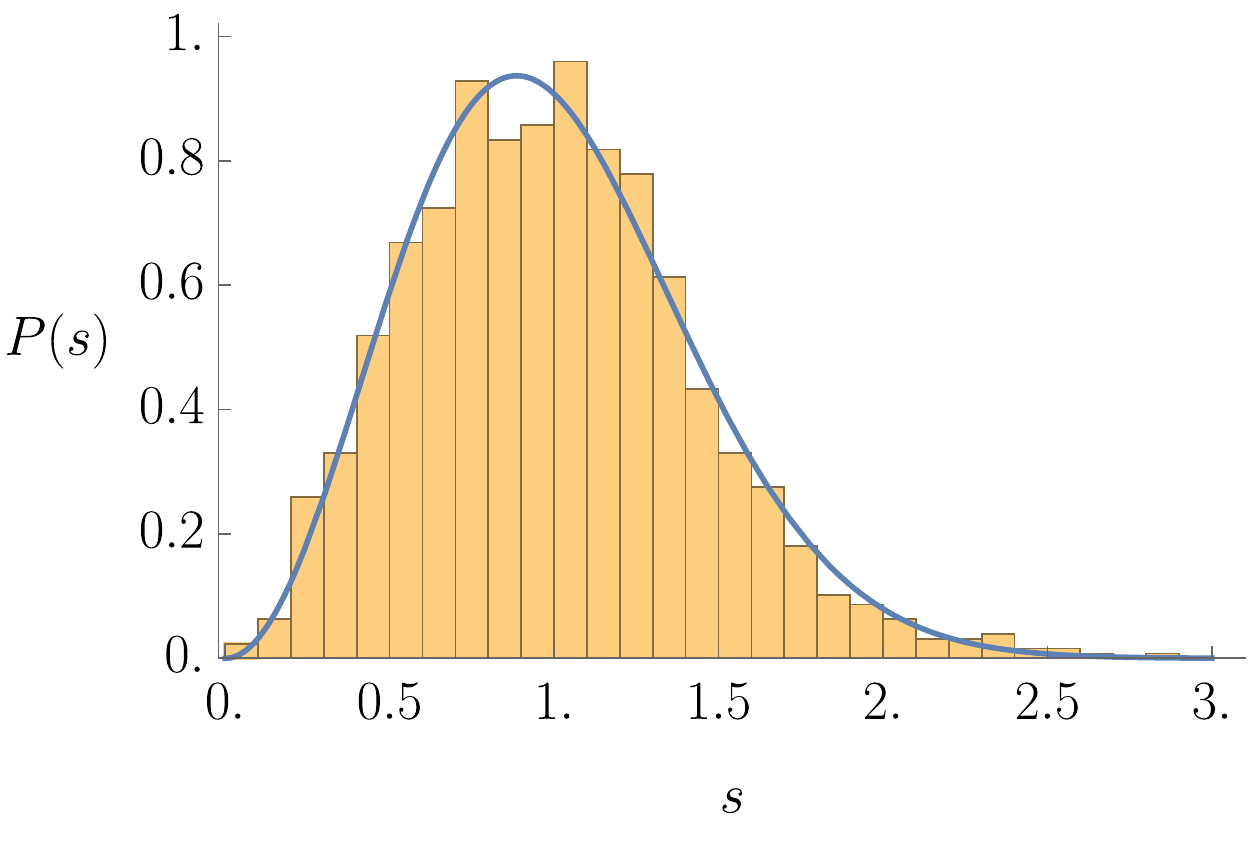}
    \caption{The spacing distribution $P(s)$ of adjecent energy levels obtained from a Gaussian Hermitian random matrix at $L = 10^4$. The histogram is a function of the difference between adjecent eigenvalues, denoted by $s$ and we focussed on $1272$ eigenvalues $E$ within the bin $[-1/10,1/10]$. The solid blue curve is the infinite $L$ result given by $\frac{32s^2}{\pi^2}e^{-4s^2/\pi}$. }
    \label{fig:surmise}
\end{figure}

In GUE random matrix theory there is quadratic repulsion between neighboring energy levels, which is most clearly visible as a quadratic zero in the so-called Wigner surmise \cite{Mehta_1994,Haake:1315494} as shown in Figure \ref{fig:surmise}. 

This plot indicates that eigenvalues repel each other, it is unlikely to find eigenvalues very close together in quantum chaotic systems. This should be contrasted with Poisson behavior, where eigenvalues have a certain density but are otherwise uncorrelated, and can lump together.

Quadratic level repulsion is actually the bell cow of GUE random matrix statistics, in the literature it is often used as \emph{the} defining property of quantum chaotic systems \cite{Haake:1315494,Mehta_1994}. Black holes on the other hand are certainly classically chaotic systems, which are characterized by Lyapunov growth - exponential sensitivity to changes in initial conditions, and observations in black hole backgrounds have this property, because early perturbations get exponentially blueshifted as they fall in black holes \cite{Sekino:2008he,Shenker:2013pqa,Shenker:2013yza,Roberts:2014isa,Roberts:2014ifa,Jackson:2014nla,Shenker:2014cwa,Maldacena:2015waa}. 

\begin{figure}
    \centering
    \includegraphics[scale=0.65]{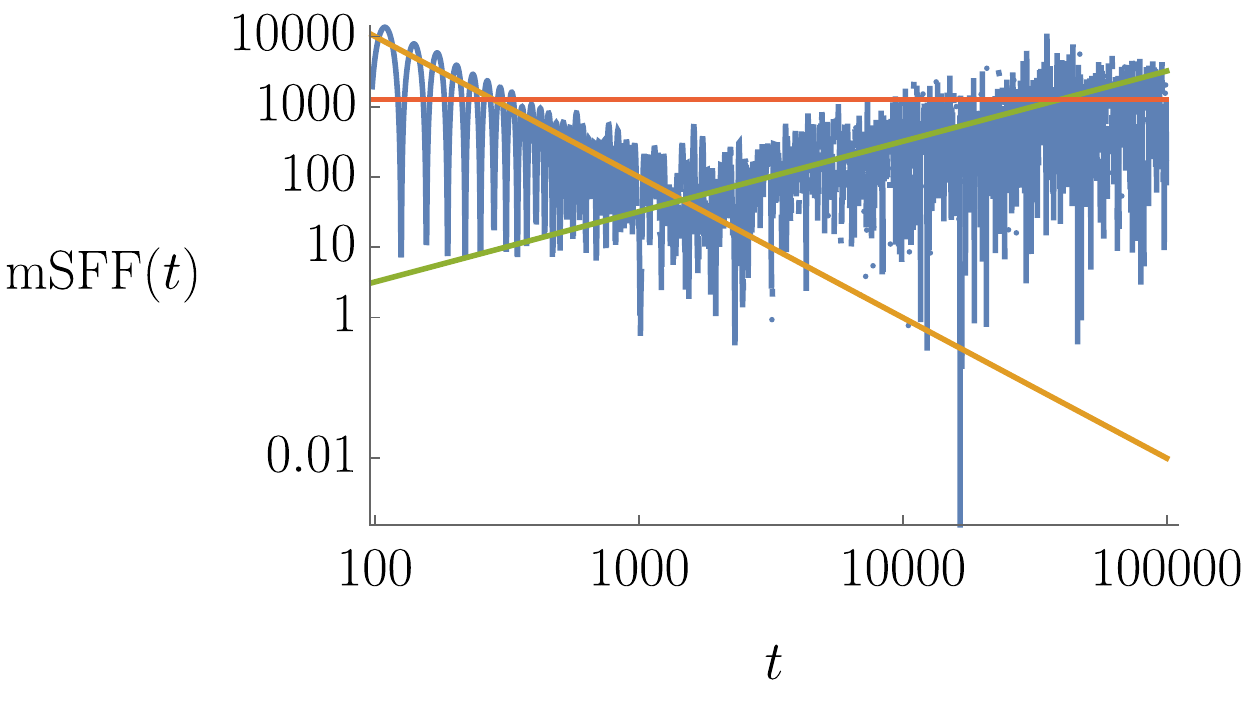}
    \includegraphics[scale=0.65]{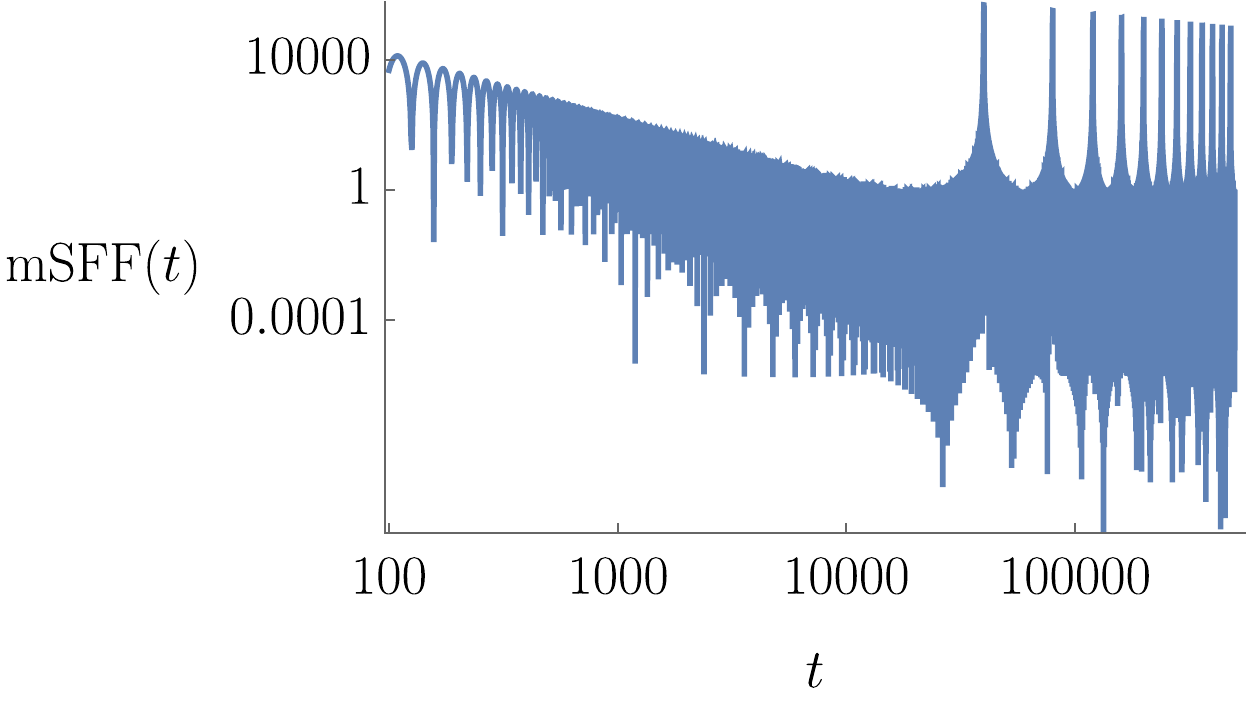}
    \caption{\textbf{Left}: Microcanonical spectral form factor for GUE as a function of time for a bin of size $1/10$ centered around zero energy. \textbf{Right}: Same microcanonical spectral form factor but with a spectrum obtained from the electrostatics problem \eqref{51} with a Gaussian potential $V = 1/2 x^2$. We clearly see a huge difference between the plot on the right, not only are there the Poincare recurrences, but there is no ramp region as well. In both plots we used $L = 10^4$ and the bin contained $1272$ eigenvalues. We verified that the recurrence time is indeed $2\pi \rho(0) = 4\times 10^4$ even if appearing non-uniform in the figure above due to the log-log scale.}
    \label{fig:SFF_Gaussianstuff}
\end{figure}

It then seems logical to assume that systems which exhibit classical chaos, such as black holes, also exhibit quantum chaos. This has led to the idea that black holes should have GUE random matrix level statistics \cite{Cotler:2016fpe}. One way to visualize random matrix level statistics, is to plot the spectral form factor in some microcanonical energy window $\delta$ around $\l$
\begin{equation}
    \text{mSFF}(t)=\sum_{i,j\in\delta}e^{\i t(\l_i-\l_j)}\,.
\end{equation}
We can compare this quantity for a typical draw of random matrix theory with the spectrum obtained by solving our electrostatics problem \eqref{51}, see Figure \ref{fig:SFF_Gaussianstuff}

At early times these are identical, but at exponentially late times they are very different, systems with random matrix statistics have a linear ramp followed by a plateau starting at $t=2\pi \rho (\l)$, whereas our system has almost perfect Poincare recurrences at times $t=2\pi n \rho (\l)$ and no visible ramp.

The reason for this is that the spectrum of our theory $\HJT$ is approximately equally spaced, on scales where $\rho(\l)$ is approximately constant, and this periodicity is responsible for the Poincare recurrences. This is already obvious when we consider the electrostatic problem \eqref{51}. This model has an attractive force coming from the potential and a repulsive force coming from the bilocal deformation. The potential is a large distance force, saying how much $\l_i$ should be packed in some smaller interval. Within those smaller intervals electrostatic repulsion wins, and the eigenvalues are \emph{maximally} separated. So locally the eigenvalues are equally spaced, and globally there is an envelope determined by the potential. Let us now show this in more detail by studying the distribution of zeros of $P_L(\l)$.

\begin{figure}
    \centering
    \includegraphics[scale=0.5]{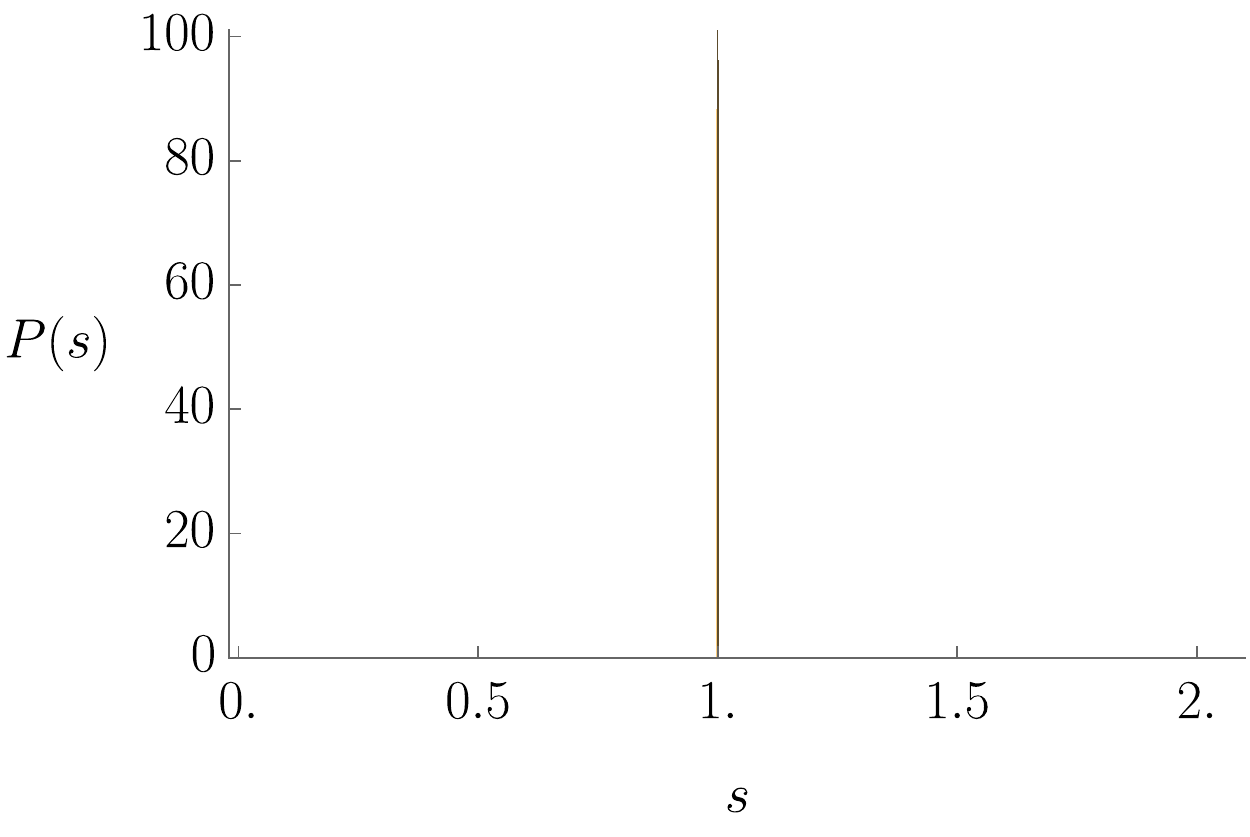}\qquad
     \includegraphics[scale=0.5]{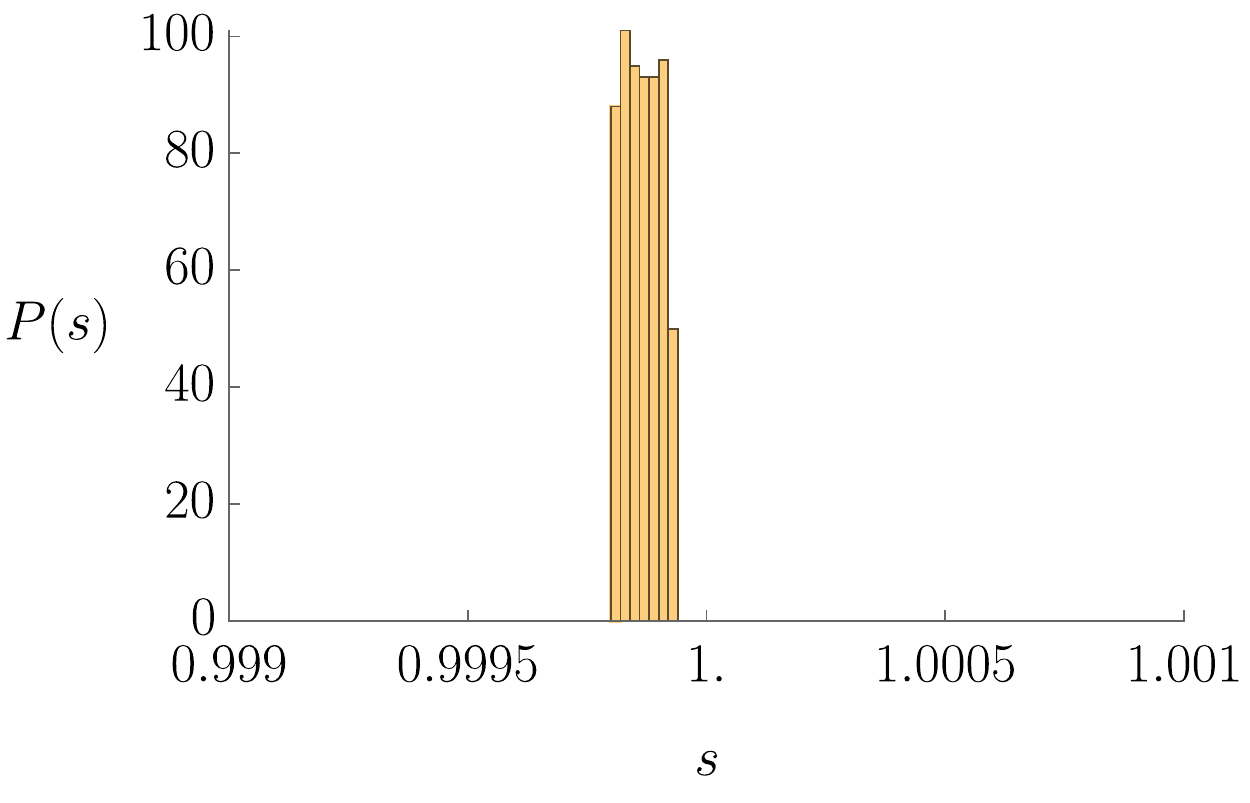}
    \caption{Spacing distribution typical for \emph{clock behaviour}. The right is a zoom-in of the left plot that emphasizes the small spread of the distribution. Here we used the electrostatics problem for a quartic potential $V = -1/2x^2 + x^4$, again we used the binsize $1/10$ centered around zero and took $L=10^4$. }
    \label{fig:Clock}
\end{figure}

Consider the orthogonal polynomials $P_L(\l)$ associated with some potential $V(\l)$. The zeros of $P_L(\l)$ are always in some finite region (or set of regions) along the real axis\footnote{At least for the potentials that have certain growth properties at large $|\l|$ \cite{Eynard:2015aea}} and in particular when $L$ becomes larger, the zeros of $P_L(\l)$ become more and more tightly packed. Let us consider some $\l_0$ and consider a set of zeros $\l_0^{(j)}$ around this value,
\be\label{setofzeros}
\dots < \l_0^{(-1)} < \l_0 = \l_0^{(0)} < \l_0^{(1)} < \dots 
\ee
To study the spacing of these zeros, we follow Lubinsky \cite{lubinsky2007new} (see also \cite{BarryProgressReport}) and consider the Christoffel-Darboux kernel associated to the orthogonal polynomials
\be 
K_L(\l_1,\l_2) = a_L \frac{P_{L}(\l_1)P_{L-1}(\l_2) - P_{L-1}(\l_1)P_L(\l_2)}{\l_1-\l_2}.\label{527}
\ee
This object has two interesting properties. First, if $P(\l_1)=0$ then $K_L(\l_1,\l_2)=0$ if $P(\l_2)=0$ and $\l_1\neq \l_2$. Second, if $\l_1-\l_2\sim 1/\rho(\l)\sim 1/L\to 0$ the Christoffel-Darboux kernel becomes the sine kernel
\be\label{sinekernel}
\lim_{L\to \infty} \frac{1}{\rho(\l_0)}K_L\left(\l_0 + \frac{a}{ \rho(\l_0)},\l_0 + \frac{b}{\rho(\l_0)}\right) = \frac{\sin \pi(a-b)}{\pi (a-b)}
\ee
The relation \eqref{sinekernel} holds for any potential and is the manifestation of random matrix theory universality. The sine kernel vanished for $a-b\in \mathbb{Z}_0$ and so combining this with the first property, we see that when we have a zero at $\l_0^{(j)}$, then the at large $L$ the next zero $\l_0^{(j+1)}$ lies as
\be 
\l_0^{(j+1)} - \l_0^{(j)}=\frac{1}{\rho(\l_0)}\label{clock}
\ee
Thus the spectrum is locally equally spaced and in the orthogonal polynomial literature it is called \emph{clock behaviour}. This rigidity is also clearly visible when we plot the average level separation, the analogue to the Wigner surmise, see Figure \ref{fig:surmise}, where indeed we see this is highly peeked around $1/\rho(\l_0)$ as shown in Figure \ref{fig:Clock}

This leads us to a rather surprising result when we consider the double scaling limit to gravity. By construction we are studying a theory of black holes, after all we are investigating pure JT gravity with a bilocal interaction turned on. The spectrum $\HJT$ described the exact quantum spectrum of these black holes, and we found that this spectrum has clock statistics, \emph{not} random matrix statistics. So there are theories of black holes without random matrix statistics, which seems to contradict at least superficially the claim of \cite{Cotler:2016fpe}.

Some comments are in order.

\begin{enumerate}
    \item Clock behaviour of the zeros of orthogonal polynomials follows from the sine-kernel and is thus not only a very universal property, but also a non-perturbative (in $L$) one. To see this, notice that the energies $a$ and $b$ are proportional to $\rho(\l)\sim L$ so we have an oscillating exponential of $L$. Furthermore, universality also tells us that when we go to the gravitational description, i.e. when we double scale, clock behaviour will persist, the sine kernel is always there \cite{Mehta_1994,Haake:1315494,Altland:2020ccq}.
    \item The ramp has an interpretation in gravity as due to a wormhole stretching between two asymptotic boundaries in ordinary JT gravity (without local or bilocal deformations) \cite{Saad:2019lba,Saad:2019pqd,Saad:2018bqo}. This computation is a good approximation for a typical member of the random matrix ensemble. Clearly it is \emph{not} a good approximation to our canonical version of JT gravity with only the bilocal turned on, since we see no ramp even under time averaging. That is fine, because this is an atypical draw from the point of random matrix theory (even though, in some sense, it is the most likely draw). But this atypical draw is very natural from the bulk point of view.
    \item When adding matter in the probe limit, our theory still gives the Lyapunov growth of the OTOC, simply because the coarse grained spectrum is the same as in ordinary JT gravity. The scrambling time scale is too short to distinguish our discretized spectrum from the continuous one.
    \item The argument that black holes are quantum chaotic seems more robust in higher dimensions, simply because there are more degrees of freedom, so this could be a pathology of our lower dimensional setup. It is fathomable that the clock behavior is not robust under such deformations.
\end{enumerate}


\subsection{Quantum mechanical dual of canonical JT gravity}\label{sect:ds}

We now consider the double scaling limit where one takes $L$ to infinity and scales towards the spectral edge of the spectrum \cite{Saad:2019lba}. The precise double scaling procedure is not very important, here we will limit ourselves to sketching how different quantities map to one another and state the corresponding double scaled structures, which we then check gives the desired answers.


\subsection*{The auxiliary quantum mechanics reloaded}

In the double scaling limit orthogonal polynomials become the Baker-Akhiezer functions of the relevant double scaled matrix model \cite{Maldacena:2004rf}
\begin{equation}
    P_n(\l)\to \psi(x,\l)\,,\label{map}
\end{equation}
where $L-n$ becomes proportional to a rescaled version of the (now continuous) coordinate $x$, and $\l$ has been scaled towards the edge, furthermore we left out prefactors for comfort. For large $e^{\Ss}$, which replaces $L$ upon double scaling, we have the following approximation for the Baker-Akhiezer functions (up to normalization that should be fixed later) for $\l>u(x)$ \cite{Maldacena:2004sn,Johnson:2022wsr}
\begin{equation}
    \psi(x,\l)=\frac{1}{(\l-u(x))^{1/4}}\cos\bigg(e^{\Ss}\int_{u(x)}^\l dv\,\mathcal{F}'(v)(\l-v)^{1/2}+\frac{\pi}{4}\bigg)\,,\quad x = \frac{\sqrt{u}}{2\pi}I_1(2\pi \sqrt{u})= \mathcal{F}(u)\,,\label{532}
\end{equation}
where the last expression is the leading approximation to the JT string equation \cite{Okuyama:2019xbv}. For $\l>u(x)$ this function is exponentially decaying. Perhaps the easiest way to understand the mapping \eqref{map} is that Baker-Akhiezer functions are by definition expectation values of determinants in double scaled matrix integrals, just like the orthogonal polynomials \eqref{orthopol}. This definition leads straight to \eqref{532} \cite{Maldacena:2004sn,Moore:1991ir}.

Baker-Akhiezer functions satisfy two differential equations (this is the Lax formalism), one of which is \cite{Maldacena:2004sn,Okuyama:2019xbv}
\begin{equation}
        \l\, \psi(x,\l)=\hat{Q}\,\psi(x,\l)\,,\quad \hat{Q}=-\frac{1}{e^{2\Ss}}\partial_x^2+u(x)\,.\label{qm}
\end{equation}
One recognizes \eqref{532} indeed as the WKB approximation for the solutions to this differential equation. To leading order $u(x)$ satisfies $x=\mathcal{F}(u)$, in general is is a more complicated function, see section \ref{subsect:higherorder}. The parameter $x$ is the coupling constant $t_0$ of the KdV hierarchy, which usually is not thought about as physically relevant, all dilaton gravity models sit at $t_0=0$.

\begin{figure}
    \centering
    \includegraphics[scale=0.2]{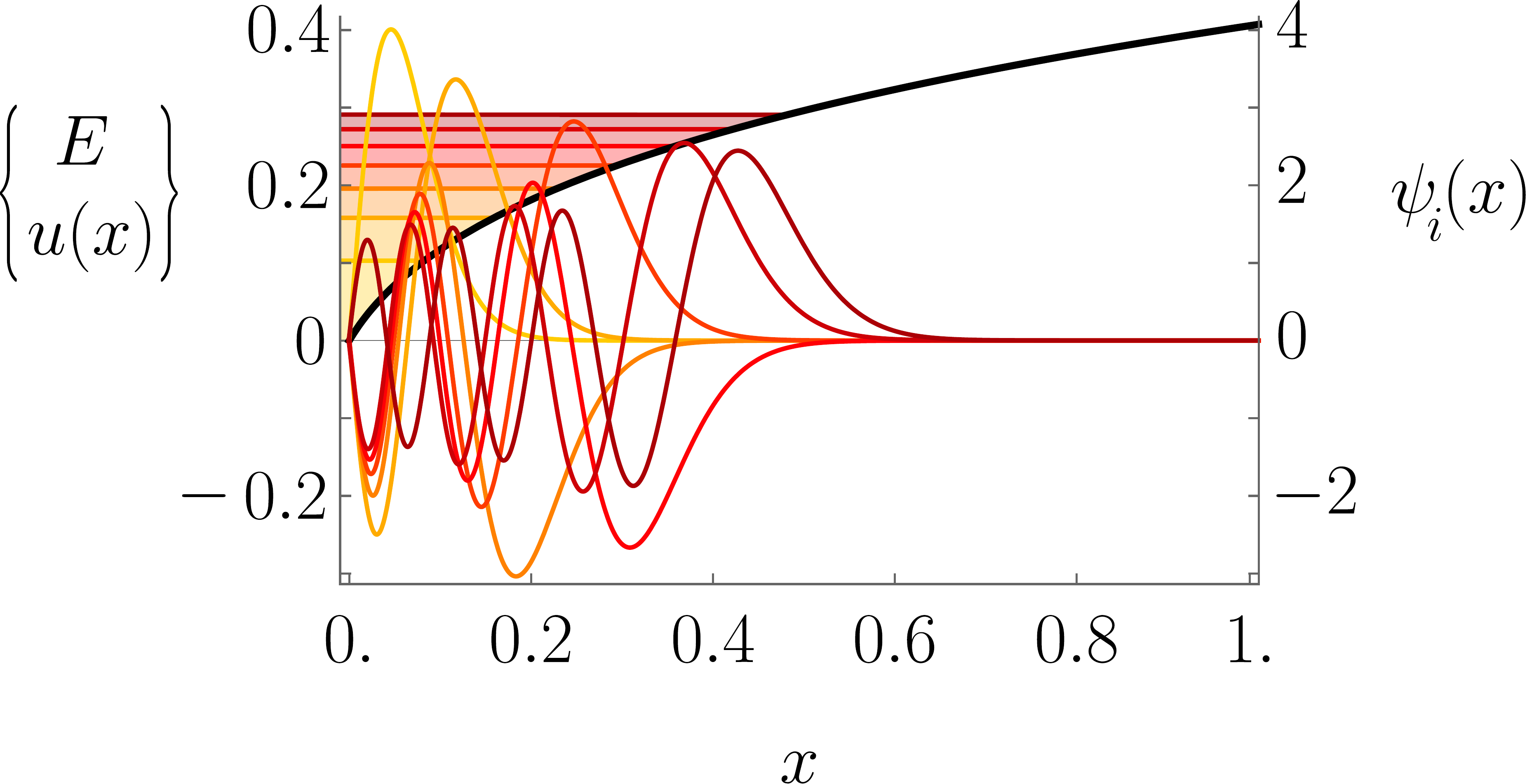}\qquad
    \includegraphics[scale=0.2]{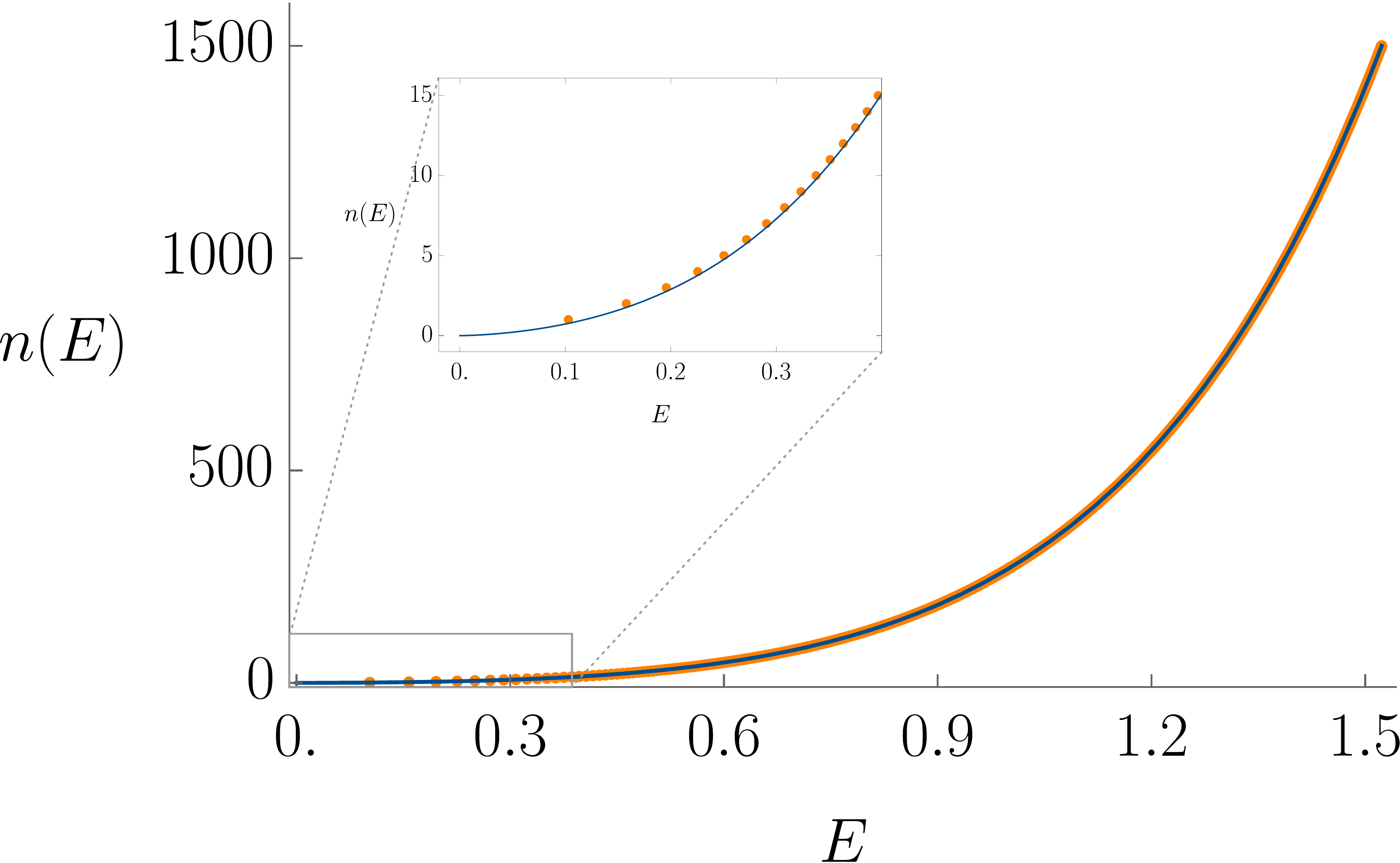}
    \caption{\textbf{Left}: First seven normalized eigenstates $\psi_i(x)$ and energies of $\hat{Q}$ for $\S = 5$. We also plotted the potential $u(x)$ in black. The more red the coloring the higher the energy. The left vertical axis is for $E_i$ and $u(x)$, whereas the right one is for $\psi_i$. \textbf{Right}: Integrated spectral density $n(E) = \int^E_0 \rho(E')\d E'$ for the first 1500 eigenvalues of $\HJT$ (orange dots) and the JT gravity result (blue).}
    \label{fig:SpecHJT}
\end{figure}

The point is that the differential operator $\hat{Q}$ can be obtained directly by double scaling the Jacobi matrix $Q$, in the usual way that difference equations such as \eqref{58} become differential equations when $n$ becomes continuous. The constraint $P_L(\l)=0$, which according to \eqref{51} is equivalent to finding the eigenvalues of $\HJT$, gets replaced with a Dirichlet boundary condition upon double scaling
\begin{equation}
    \psi(0,\l)=0\,.\label{dirichlet}
\end{equation}
The spectrum of $\hat{Q}$ with these boundary conditions will then, by construction, reproduce the spectrum of $\HJT$. This means that we can interpret this QM system as a \emph{dual} description of JT gravity with only the bilocal deformation, which we refer to as canonical JT gravity. See figure \ref{fig:SpecHJT} for the spectrum, wavefunctions and the potential $u(x)$. \footnote{The spectrum being given by the zero's of the Baker-Akhiezer functions also appeared in a recent work by Clifford Johnson \cite{Johnson:2022wsr} (based on his earlier work \cite{Johnson:2019eik, Johnson:2020exp, Johnson:2021zuo}). There it was noted that this spectrum is naturally embedded in the matrix integral by studying the probability distribution of the first, second, etc. eigenvalues of the matrix in the double scaling limit. The peaks of these probability distributions then coincide (approximately) with the zeros of the Baker-Akhiezer function. It was then argued that on the disk this discrete spectrum should be present, which at large energy is just the continuous Schwarzian density of states. Higher genus corrections would then fill in the gaps and produce a continuous spectrum. This is not what we are doing here. We have the spectrum given by the zeros of the Baker-Akhiezer function to be the non-perturbative density of states in the sense that when we compute any correlator of $\rho(E)$ in the matrix integral we get, first of all, a factorizing answer, and second the density is a sum of delta functions on this spectrum. To get these two features it is essential to deform the matrix integral (and gravity theory) with the bilocal deformation as shown in \cite{Blommaert:2021fob}.}
The partition function of canonical JT can be computed literally as
\begin{equation}
    Z(\beta)=\Tr\big(e^{-\beta\hat{Q}}\big)\,,
\end{equation}
the spectral form factor is $\Tr\big(e^{-\beta_1\hat{Q}}\big)\Tr\big(e^{-\beta_2\hat{Q}}\big)$ etcetera. One can check that wavefunctions for different solutions $\l_i$ and $\l_j$ are orthogonal and can be normalized to one on $x>0$ using formula (3.37) in \cite{Okuyama:2020ncd}, which is a continuous version of the Christoffel-Darboux formula \eqref{527}
\begin{equation}
    \int_0^\infty \d x\,\psi(x,\l_i)\psi(x,\l_j)=\frac{\psi(0,\l_i)\partial_x \psi(0,\l_i)-\psi(0,\l_j)\partial_x \psi(0,\l_i)}{\l_i-\l_j}=0\text{ when }\l_i\neq \l_j\,,
\end{equation}
where one uses that $\psi(0,\l_i)=\psi(0,\l_j)=0$ according to the Dirichlet boundary condition \eqref{dirichlet}.

There are several ways to confirm that the clock behavior \eqref{clock} is respected in this double scaled theory. One is to use that the Christoffel-Darboux kernel reduces to the sine kernel for nearby energies. Another is to use that $\psi(0,\l)$ according to \eqref{532} becomes \eqref{432}, the spacing of zeros follows then from the quasi-periodicity of the cos. The simplest though, it to simply use Bohr-Sommerfeld quantization for the Schrodinger equation \eqref{qm} with Dirichlet boundary conditions
\be 
\pi (n+1/2) = e^{\S} \int_0^{\mathcal{F}(E)} \d x\, \sqrt{E - u(x)}=e^{\Ss}\int_0^E \d u\,\mathcal{F}'(u)\,\sqrt{E - u}\,,
\ee
where the $x=0$ turning point comes from the hard Dirichlet wall. The coarse grained density of states is $\d n/ \d E$, which results in
\be 
\rho(E) =\frac{\d n}{ \d E}= \frac{e^{\S}}{2\pi} \int_0^E \d u \frac{\mathcal{F}'(u)}{\sqrt{E-u}} = e^{\S}\frac{\sinh{2\pi \sqrt{E}}}{4\pi^2},\label{538}
\ee
which reproduces indeed the coarse grained spectral density of our canonical JT gravity, which should be indeed that of the original ensemble.

We stress that the true spectrum of our QM is \emph{discrete}, this sets it aside from, say, Schwarzian quantum mechanics \cite{Stanford:2017thb}, but also from the auxiliary quantum mechanical description of matrix integrals in the Lax formalism, where the same Hamiltonian $\hat{Q}$ features, but without Dirichlet boundary conditions. To drive home the point that discreteness is conserved upon double scaling, consider for instance the fact that double scaling the Hermite polynomials results in the Airy function, which does not have a continuum of zeros (and is the simplest Baker-Akhiezer function). 


\subsection*{Path integral formulation}\label{subsect:higherorder}

It is straightforward, at least to leading order in $e^{\S}$ to write down an action for our quantum mechanical problem, which is describing a particle moving in a potential $u(x)$. The partition function is
\be 
Z(\beta) = \int \mathcal{D}x\,\exp\bigg( -\int_0^\b \d \t \bigg( \frac{e^{2\S}}{4}\,\dot{x}^2 + u(x) \bigg) \bigg)\,,\label{pathint}
\ee
where the sum is over periodic paths $x(\tau + \beta) = x(\tau)$ and we have the constraint $x(\tau)>0$ for all $\tau$. This is the way to enforce Dirichlet boundary conditions in path integrals, the wavefunction must vanish at the boundary, so the particles can never travel beyond $x=0$. One can think of the potential $u(x)$ as being infinite for $x<0$. 

To leading order in $e^{\S}$ we can solve this path integral by saddle-point. As there is no $e^{\S}$ multiplying the potential, the leading order equation of motion are simply $\ddot{x}=0$, so the unique periodic solutions are constants
\be 
x(\tau) = x_0\,,\quad 0<x_0<\infty
\ee
Notice that $x_0$ is a parameter of the on-shell solution and we need to integrate over it in order to sum over all saddles. To leading order we can ignore the effects of the constraint $x(\tau)>0$ on the Gaussian fluctuations around the saddle, and we can evaluate $u(x)$ on shell, so we just need to compute standard Gaussian integrals\footnote{The $\sqrt{\b}$ comes from choosing an orthonormal basis for $x$ on the circle, the path integral measure is then $d x_0 \sqrt{\beta}$.}
\be 
Z(\beta) = \sqrt{\b}\int_0^{\infty} \d x_0\, e^{-\b u(x_0)} \int_{-\infty}^{+\infty} \prod_{n>0} \d a_n \d b_n \exp\bigg(-\frac{e^{2\S} \pi^2 n^2}{\b^2}(a_n^2+b_n^2)\bigg)\label{541}
\ee
After zeta regularization this results indeed in the Laplace transform of \eqref{538}
\be 
Z(\beta) = \sqrt{\b}\int_0^{\infty} \d x_0\, e^{-\b u(x_0)}\frac{e^{\S}}{\b \sqrt{4\pi}}= \frac{e^{\S}}{4\sqrt{\pi \beta}} \int_0^{\infty} \d u\, \mathcal{F}'(u)\, e^{-\b u}=\frac{e^{\S}}{4 \pi^{1/2} \b^{3/2}} e^{\frac{\pi^2}{\b}}\,.
\ee
An alternative way to obtain this is by writing \eqref{pathint} as a phase space path integral (introducing a field $p(\tau)$), and rescaling $\tau$ by $e^{\S}$ such that $e^{\S}$ stands everywhere where one usually has $1/\hbar$ in QM. Then taking $\hbar\to 0$ localizes this phase space path integral to a classical phase space integral over constants $p_0$ and $x_0$ as explained in \cite{Mertens:2018fds}. With the phase space measure $d p_0\, d x_0/ 2\pi \hbar$ one recovers the above answer
\begin{equation}
    Z(\beta)=\frac{e^{\S}}{2\pi}\int_0^\infty \d x_0\int_{-\infty}^{+\infty}\d p_0\,e^{-\beta(p_0^2+u(x_0))}\,.
\end{equation}

So where is the discreteness in this calculation? Even for extremely large $e^{\S}$, where we can certainly trust this saddle-point, our quantum mechanics still has a discrete spectrum, which we remind the reader is a direct consequence of the Dirichlet boundary conditions \eqref{dirichlet}. This sits in the fact that we have ignored the effects of the constraint $x(\tau)>0$ on the fluctuations, instead we have essentially solved the path integral with $x_0>0$ constraints. This path integral will never result in a discrete spectrum, not even if we include the expansion of $u(x)$ around the saddle-point to arbitrary order in the fluctuations.

So discreteness must trickle in through the constraints that $x(\tau)>0$ puts on the integration range in the mode expansion, the allowed integration range of $x_0, a_n$ and $b_n$ in \eqref{541} becomes a complicated mess, which somehow should organize into a discrete spectrum. Solving path integrals in quantum field theory with positivity constraints on the fields is a notoriously difficult problem, this is essentially the reason why string theory in Rindler space is hard \cite{Witten:2018xfj}. It would be interesting to see how discreteness arises directly in the path integral, without resorting to solving the Schrodinger equation with Dirichlet boundary conditions in canonical quantization.

Regardless of that subtlety we should note another source for corrections in the $e^{\S}$ expansion, which comes from the fact that actually the $u(x)$ that features in the Schrodinger equation \eqref{qm} is only the solution to $x=\mathcal{F}(u)$ to leading order in $e^{\S}$. Including corrections, $u(x)$ is the solution to a differential equation
\be 
\sum_k t_k R_k[u,e^{\S}] = x\,,\label{conconcon}
\ee
with $R_k$ the Gelfand-Dickii differential operators and the $t_k$ are known constants for JT \cite{Okuyama:2019xbv,eynard2021natural}. These operators contain derivatives of $u$ wrt to $x$ and so to construct an action that works to higher orders in $e^{\S}$ as well, those need to be converted into $\tau$ derivatives by using $\partial_x = (\dot{x})^{-1}\partial_\tau$. So the Gelfand-Dickii differential operators should be viewed as functions of (derivatives of) both $u$ and $x$ here $R_k[u,x,e^{\S}]$. We can still write up a quantum mechanics by introducing a Lagrange multiplier $\l(\tau)$ that forces the constraint \eqref{conconcon}\footnote{There is also a nontrivial measure on the path integral over $u(\tau)$ which compensates the Jacobian from the delta.}
\be 
I[u,x,\l] = \int_0^{\b} \d \tau \bigg(  \frac{e^{2\S}}{4}\,\dot{x}^2 + u - i \l \bigg( \sum_k t_k R_k[u,x,e^{\S}] - x \bigg)\bigg)
\ee

Before turning to the conclusions let us say a few words about the difference between our system, and the more familiar appearance of the Schrodinger equation \eqref{qm} in the matrix integral literature. Usually \cite{Eynard:2015aea,Maldacena:2004sn,Okuyama:2019xbv,DiFrancesco:1993cyw,Dijkgraaf:1991qh} one views this quantum mechanics as auxiliary, and one does not have the Dirichlet boundary conditions $\psi(0,\l)=0$, nor the interpretation that $\hat{Q}$ is literally the Hamiltonian of the gravity theory. Instead, the computation of the partition function involves a projector on positive $x$ at one point in time
\begin{equation}
    Z(\beta)=\Tr\big(\Pi\,e^{-\beta\hat{Q}}\big)\,,\quad \Pi=\int_0^\infty \d x \ket{x}\bra{x}\,.
\end{equation}
In the path integral formulation this boils down to summing over all periodic paths with the constraint $x(\tau_0)>0$ for one chosen time $\tau_0$. This does not result in a discrete spectrum in $Z(\beta)$, moreover we are not computing the partition function of the QM (which has a flat spectrum), because of the projector. Therefore there is no sense in which the matrix integral and the auxiliary QM are holographic duals; in our case, products of partition functions map to partition functions, so we \emph{do} have a shot at describing a genuine holographic duality. We comment more on this below.

\subsection*{Aside: non-perturbative effects can imply perturbative corrections in $e^{-\S}$ expansion}

The geometric expansion of canonical JT gravity only contains the disk. The corrections to the density of states of this geometry are non-perturbative effects in the original matrix integral, namely, as discussed in section \ref{subsec:no-random-stats}, they are given by the electrostatic problem whose spacing between eigenenergies are given by the vanishing values of the sine-kernel (the leading non-perturbative contribution in the spectral density two-point function for $\lambda_1\neq \lambda_2$). However, the fact that only non-perturbative effects are present (aside from the disk contribution) does not imply that in the partition function $Z(\beta) = \Tr e^{-\beta \HJT}$ there are no perturbative corrections in an $e^{-\S}$ expansion.\footnote{We thank Don Marolf for raising this point.}

In matrix integrals non-perturbative effects can, at least naively, manifest themselves in observables at perturbative order in $e^{-\S}$. For instance, even in the undeformed matrix integral, in the spectral density correlator non-perturbative effects give rise to a term which is only suppressed by $e^{-\S}$. Indeed $\average{\rho(\lambda_1) \rho(\lambda_2)} \supset \rho_{0,{\rm JT}}(\lambda_1) \delta(\lambda_1 - \lambda_2)\sim e^{\S}$ which has no geometric origin in the topological expansion (since the leading disconnected geometric contribution scales as $e^{2\S}$ and the subleading one as well as the leading connected contribution scales as $O(1)$). Canonical JT gravity takes this to the extreme.  Due to the appearance of the parameter $q$ from the solution of the Schwinger-Dyson equations reviewed in section \ref{sect:review}, the topological recursion relations are no longer valid once the double-trace deformation (associated to the bilocal deformation in gravity) are turned on. This is because the starting assumption needed to derive these relations is that the spectrum at the saddle-point can be approximated by a continuum;  with the deformation turned on, this is a bad approximation since we have $q \gg e^{\S}$. Instead, the spacing in the spectrum that we find above is given by the values of the energies at which the sine-kernel from the original JT gravity matrix integral vanishes, from which we find that $\lambda_0^{(j+1)}-\lambda_0^{(j)} \sim e^{-\S}$ in the double scaled limit. Such small differences nevertheless contribute at perturbative order in $ e^{-\S}$ to $Z(\beta)$, which thus has a whole perturbative series in $e^{-\S}$ that is non-vanishing. Thus the perturbative corrections in $e^{-\S}$ are akin to the $\sim e^{\S} \delta(E-E')$ contribution in the spectral density correlator - even though they both corrections naively appear perturbative, they all in fact have a non-perturbative (non-geometric) origin.

\subsection{Bulk interpretation of null states}
\label{subsec:bulk-interp-of-null states}

In the previous subsections we focused on the interpretation of the canonical JT theory by looking at the statistics of the eigenvalues $\HJT$. Here we want to study its semiclassics, borrowing also some ideas from null states in section \ref{sect:null}. 

In the canonical JT theory we deform the JT path integral just by a bilocal spacetime interaction. This causes the genus expansion to collapse to just the disk and from the matrix integral we know the spectrum is discrete, given by $\HJT$. However, using the null states, we can equally well describe this theory with a path integral that has both the bilocal and local deformation turned on. In particular we take the local deformation to be the one discussed in \cite{Blommaert:2021fob}, but now with $\Hh = \HJT$,
\begin{equation}
    Z_\text{brane}(b,\HJT)\quad\Leftrightarrow\quad\delta \rho_0(E,\HJT)= \Tr( \delta(E-\HJT))-\frac{e^{\Ss}}{4\pi^2}\sinh(2\pi E^{1/2})\,.\label{415}
\end{equation}
The contribution of this one-point function to the equations of motion is
\begin{align}
    \int_{0}^{\infty} \d E\,\delta\rho_0(E,\HJT)\,\partial_{\l}P(\l,E)\rvert_{\l=\l_i}\nn&=-2\fint_{0}^{\infty} \d E\,\frac{e^{\Ss}}{4\pi^2}\sinh(2\pi E^{1/2})\,\frac{1}{\l_i-E}+\frac{1}{2\l_i}+2\sum_{j\neq i}\frac{1}{\l_i-\l_j}\\&=-L V'_\text{JT}(\l_i)+\frac{1}{2\l_i}+2\sum_{j\neq i}\frac{1}{\l_i-\l_j}=0\text{ when }\l_i=\l_{\text{JT}\,i}\,,\label{416}
\end{align}
where in the second line we used the relation between the JT disk spectral density and the JT potential \cite{Saad:2019lba,Blommaert:2021fob}. Notice now that the second line in \eqref{416} is precisely of the form \eqref{vjt}, such that this contribution vanishes when $\l_i=\l_{\text{JT}\,i}$. In fact, we can multiply $Z_{\brane}(b,\HJT)$ with a parameter $s_0$ and still obtain the same equations of motion and solutions $\l_i = \l_{\text{JT}\,i}$, provided $0\leq s_0 \leq 1$ to ensure stability of the saddle. So turning on \eqref{415} is another example of the null deformations discussed in \ref{sect:null}.

\begin{figure}[t!]
    \centering
    \includegraphics[scale=0.35]{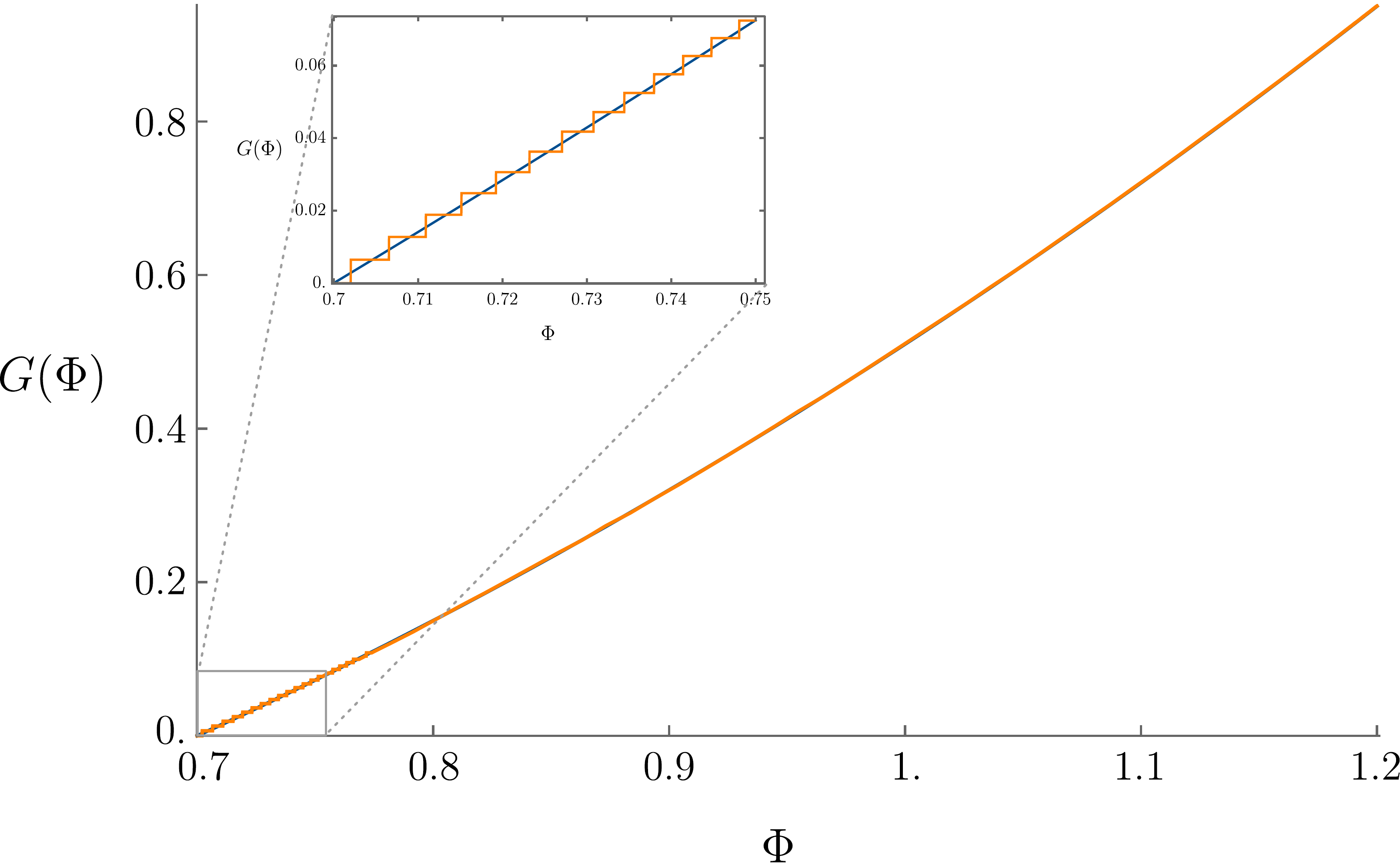}
    \caption{Metric function $G(\Phi)$ as a function of the radial coordinate (equal to the dilaton $\Phi$) for the potential $U(\Phi,s_0)$ with $s_0 = 1$ (orange) alongside with the undeformed metric potential $\Phi^2 - \Phi_h^2$. We used $\Phi_h = 0.7$ and took the first $1500$ eigenvalues of $\HJT$ at $S_0 = 5$. The inset shows a clear difference between the two cases, but on 'average' (average over a small $\Phi$ window) the blue and orange curve lie on top of each other.}
    \label{fig:metricnull}
\end{figure}

We now want to understand a bit better what the semiclassical interpretation of this null deformation is. In particular, we want to compare the case where we only have the bilocal deformation turned on and the cases where $s_0 \neq 0$. Working semiclassically means that we take only the order one and order $e^{-\S}$ corrections into account. Thus we neglect the bilocal, which although important for higher genus corrections, is not important semiclassically. 

The semiclassical physics of general dilaton gravity is most easily understood by going to a gauge where the dilaton is the radial direction and the metric takes the form \cite{Witten:2020ert}
\be 
\label{eq:bulk-metric-ansatz}
\d s^2 = G(\Phi) \d \t^2 + \frac{\d \Phi^2}{G(\Phi)}
\ee
with
\be 
G(\Phi) = \int_{\Phi_h}^{\Phi} \d\Phi' \, U(\Phi)
\ee
and $\Phi_h$ determined by demanding smoothness of the solution. For the case at hand the dilaton potential $U(\Phi)$ is given by 
\begin{align}
U(\Phi,s_0) &= 2\Phi + 4\pi s_0 e^{-\S}\int_0^{\infty} b\d b \, e^{-2\pi \Phi} \cos(b \Phi) Z_{\rm brane}(b,\HJT)\nn\\
&= 2\Phi + (4\pi)^2 s_0 \Phi e^{-2\pi \Phi}\left( e^{-\S}\Tr \delta(\Phi^2 - \HJT) -\frac{ \sinh(2\pi \Phi)}{4\pi^2} \right).
\end{align}
In the second line we used the explicit form of $Z_{\rm brane}$ \eqref{415} and performed the integrals over $b$ (and $E$). From this expression we clearly see that to leading order the dilaton potential is just the ordinary one, but there are $e^{-\S}$ corrections to it. It is amusing to see as well that amongst these null deformations, the one with $s_0 = 1$ has a dilaton potential which does not have the $2\Phi$ anymore 
\be 
U(\Phi,s_0 = 1) = 2\Phi e^{-4\pi \Phi} + 8\pi^2 e^{-2\pi \Phi} e^{-\S} \sum_{i=0}^{\infty} \delta(\Phi - \l_{{\rm JT}\,i}^{1/2}).
\ee
Inserting the leading order approximation \eqref{538} for the spectrum of $\HJT$, the $2\Phi$ reappears. So this is in some sense a discretization of the JT dilaton potential.

It is also interesting to note that since we can calculate the spectrum $\HJT$ numerically, we can simply plot the metric function $G(\Phi)$ with the deformation turned on. See figure \ref{fig:metricnull} for an example at $s_0 = 1$. The most notable feature is that we observe a raggedness of the metric in the deep IR, but the metric becomes smooth the closer to the boundary we get.

Analytically we have the following slightly draconian formula for $G(\Phi,s_0)$
\begin{align}
G(\Phi,s_0) &= (1-s_0)(\Phi^2 - \Phi_h^2) - \frac{s_0}{8\pi^2}\left( e^{-4\pi \Phi}(1+4\pi \Phi) - e^{-4\pi \Phi_h}(1+4\pi \Phi_h)\right)\nn\\
&\qquad + 8\pi^2 s_0 e^{-\S} \sum_{\Phi^2>\l_{{\rm JT} i}>0} e^{-2\pi \l_{{\rm JT} i}^{1/2} } - 8\pi^2 s_0 e^{-\S} \sum_{\Phi_h^2>\l_{{\rm JT} i}>0} e^{-2\pi \l_{{\rm JT} i}^{1/2} }\label{disc}
\end{align}
Consider two regimes: large $\Phi_h$ (large temperature) and the other is small $\Phi_h$ (small temperature). In the first case we expect no difference with the usual JT results, because at large temperature we probe the high energy part of the spectrum of $\HJT$, where the eigenvalues are dense and we can approximately coarse grain.

At small $\Phi_h$ though, the last term on the second line now does not approximate well the last term on the first line, as we sum over only few (sparse) states. Thus, for small $\Phi_h$ we see a difference in ADM energy (defined as the subleading piece at large $\Phi$) of the solution with or without the null deformation. 

Another important point that results from this analysis is that null deformations can change the details of the on-shell solution. For $s_0 = 0$ we just have the normal JT black hole metric, but as we see in \eqref{disc}, for $s_0 \neq 0$ there are corrections to it. In other words, semiclassical physics is \emph{not} invariant under null deformations. Schematically for the $s_0 = 1$ and $s_0 = 0$ theories, we can picture this as
\begin{equation}\underbrace{ \overbrace{\begin{tikzpicture}[baseline={([yshift=0cm]current bounding box.center)}, scale=0.7 ]
 \pgftext{\includegraphics[scale=0.45]{deformeddisk.pdf}} at (0,0);
  \end{tikzpicture}}^{\substack{\text{Leading }\\ \text{saddle with}\\ \text{null deformation}}} \hspace{0.3cm}+\hspace{0.3cm}\text{non-perturbative corr.~in }\S }_{ \begin{tikzpicture}[baseline={([yshift=0cm]current bounding box.center)}, scale=0.7 ]
 \pgftext{\includegraphics[scale=0.30]{Halfwormhole_contribution.pdf}} at (0,0);
  \end{tikzpicture}\hspace{0.2cm}+\hspace{0.15cm}\text{non-perturbative effects }}\quad
=
  \quad  \overbrace{\begin{tikzpicture}[baseline={([yshift=0cm]current bounding box.center)}, scale=0.7 ]
 \pgftext{\includegraphics[scale=0.45]{disk.pdf}} at (0,0);
  \end{tikzpicture}}^{\substack{\text{Leading }\\ \text{disk saddle}}} \hspace{0.2cm}+\hspace{0.15cm}\text{ non-perturbative effects }\,,
\end{equation}
where the theory deformed by a null deformation is represented on the left, while canonical JT gravity is represented on the right. 
The full answer on both sides is exactly identical as per the matrix model calculations, but as an expansion, we can shuffle things between the semiclassical geometries (the drawn disks) and non-perturbative corrections in $\S$. This leads to naively different semiclassical physics or simply different black holes geometries that are equivalent non-perturbatively. This is an explicit example in gravity of the mechanics that \emph{multiple bulk descriptions can coexist} \cite{Saad:2021rcu}. This is a direct consequence of null states and one of the major points we wanted to make in this work.


\section{Conclusion and outlook}\label{sect:disc}
 In this paper, we have given an explicit geometric description of all states in the baby universe Hilbert space of JT gravity. We have found that alpha-states can be obtained from factorizing theories of dilaton gravity which contain non-local terms. Null states on the other hand can be obtained by studying the redundancies in the UV completion of such theories and led us to the idea that different semiclassical descriptions of the bulk can be equivalent non-perturbatively. To explore the properties of such theories, we have focused on the simplest factorizing theory, that we called canonical JT gravity, a theory which only has the disk contribution in its geometric expansion. We have found that this theory can be described by a conventional quantum mechanical Hamiltonian, whose spectrum is regular and does not exhibit random matrix statistics. This points towards a holographic duality between a theory of AdS$_2$ quantum gravity, whose action is 
    \be 
\label{eq:canonical-JT}
I_{\text{canonical}} = I_\text{JT} + I_\text{nonlocal}\,,
\ee
and a non-gravitational conventional discrete quantum mechanics, whose action is given by
\be
\label{eq:conclusion-bdy-action}
I_\text{boundary} = \int_0^\b \d \t \bigg( \frac{e^{2\S}}{4}\,\dot{x}^2 + u(x) \bigg)\,,
\ee 
where the potential $u(x)$ is determined from the string equations in the original JT matrix integral.
    
Understanding such dualities has a long history in the literature since they are motivated by finding examples of AdS/CFT that are in some sense most applicable in our own universe: all extremal and near-extremal black hole solutions have an AdS$_2$ near-horizon region that appears in their decoupling limit. Nevertheless, there has also been an equally long list of confusions regarding these dualities, some of which are useful to recall. For instance, in \cite{Sen:2008vm} it was pointed out that a CFT$_1$ boundary dual should only capture the degeneracy of ground states that the black hole has at extremality. However, this assumed that a thermodynamics mass gap existed between the extremal states, which would have an exact degeneracy scaling as $e^{\S}$, and lightest near-extremal states in all such black holes. However, computations that showed that JT gravity serves as an effective theory for the near-horizon region of such black holes \cite{Almheiri:2014cka, Almheiri:2016fws, Maldacena:2016upp, Iliesiu:2020qvm}, show that, when accounting for the backreaction in the AdS$_2$ region, no gap is present at the predicted energy scale and there is no large degeneracy among extremal states (at least for non-supersymmetric extremal and near-extremal black holes \cite{Iliesiu:2020qvm, Heydeman:2020hhw, Iliesiu:2021are, Boruch:2022tno}). Thus, since this gap is absent, the boundary description should capture the contribution of both the extremal and near-extremal states, the latter of which explicitly break the $SL(2, \mathbb R)$ isometry in the near-horizon region \cite{Maldacena:2016upp}. One might consequently hope that instead of some CFT$_1$ that has an $SL(2, \mathbb R)$ conformal symmetry, the Schwarzian theory, which is the effective theory that explicitly captures the breaking of this near-horizon $SL(2, \mathbb R)$ isometry, provides the appropriate dual description for such black holes. However, the Schwarzian does not represent the full answer: firstly, it has a continuous spectrum which is not expected of black hole micro-states in a UV complete theory, and secondly, it does not capture any non-perturbative corrections from other geometric contributions to the gravitational path integral. The boundary quantum mechanics theory discussed in this paper addresses  both these issues, albeit in a bottom-up manner: by accounting for both the non-local interaction resulting from integrating out UV degrees for freedom and for the non-perturbative corrections that result from a sum over all topologies, we have found a theory with a discrete spectrum. At the same time, our theory \eqref{eq:conclusion-bdy-action} reproduces the results obtained from JT gravity at leading order in $e^{-\S}$. 

While our paper completes the program of concretely  characterizing the baby universe Hilbert space of JT gravity and points towards a new holographic duality, there are however numerous open questions, some of which we hope to address in future work.  
 
\subsection*{Completing the holographic dictionary}
 
In our bottom-up construction, we have integrated out all degrees of freedom with the exception of the 2d metric and dilaton in some UV complete theory of gravity. In the model that we called canonical JT gravity, we have chosen for concreteness the local dilaton potential to vanish. However, each near-extremal black hole whose origin is in a higher dimensional UV complete theory will come with its own local dilaton corrections -- for instance, there is no reason to expect that the local dilaton corrections for near-extremal black holes in an asymptotically $\mathbb R^{3,1} \times T^6$ spacetime is the same as those for black holes in $AdS_5 \times S^5$. These corrections will be determined by performing the dimensional reduction to the near-horizon AdS$_2$ and integrating out all other degrees of freedom in the theory (such as all matter fields, gauge fields or higher derivative interactions).\footnote{Integrating-out gauge fields decomposes the matrix integral description into multiple sectors, each associated to an irreducible representations of the bulk gauge group \cite{Iliesiu:2019lfc, Kapec:2019ecr}. Consequently, we expect the analysis in this paper to straightforwardly extend to the study of fixed irrep sectors of such black holes.  } Thus, we expect that each higher dimensional UV complete theory will have its own spectrum of near-extremal black hole micro-states determined by the corrections, in principle determined by the corrections to the local dilaton potential in the dimensionally reduced theory. Nevertheless, canonical JT gravity serves as a self-consistent example that  illustrates how discreteness and factorization arise together in the gravitational path integral, in a manner that we hope is generalizable to the more complicated examples arising in the dimensional reductions of stringy UV completions.

Additionally, since many interesting observables can be obtained from the insertion of (gravitationally dressed) matter fields, it would be fruitful to take one step-back, and not integrate-out some of the matter fields which we know are present in a theory that has its UV completion in string theory. In particular, it would be interesting to understand whether the presence of such matter fields, when no longer integrated out, can change the universal non-local interaction which we had to include in order to get a factorizing answer.

Furthermore, while the mapping between the spectra of the bulk \eqref{eq:canonical-JT} and boundary \eqref{eq:conclusion-bdy-action} theories is clear, the mapping of operators is not. In contrast to higher dimensional examples of AdS/CFT, in quantum mechanics there is no locality to guide us in finding a mapping between bulk and boundary operators. Nevertheless, one might still hope that the canonical operators of \eqref{eq:conclusion-bdy-action}, such as $\hat x$ and its conjugate $\hat p$, have a nice geometric meaning in the bulk.
 
Conversely, one might hope to determine the boundary meaning of the commonly discussed diffeomorphism invariant bulk operators. For instance, what boundary operator is dual to the insertion of bulk probe geodesics? To leading order in $e^{-\S}$ this should reproduce the expectation value of 
Schwarzian bilocals \cite{Yang:2018gdb, Saad:2019pqd,Mertens:2017mtv,Mertens:2018fds,Blommaert:2018iqz,Iliesiu:2019xuh,Blommaert:2018oro,Kitaev:2018wpr} which also capture the effect of inserting such geodesics. Nevertheless, identifying such operators in the theory \eqref{eq:conclusion-bdy-action} seems difficult even at leading order in $e^{\S}$. This is partly because  it is difficult to find even an approximate relation between $x(\tau)$ and the Schwarzian mode $f(\tau)$ \cite{Maldacena:2016upp,Engelsoy:2016xyb,Jensen:2016pah}. The difficulty in this identification can be observed by comparing the leading order computation of the thermal partition function using the path integral in section \ref{sect:ds} to that in the Schwarzian theory. While in the Schwarzian theory there is a unique saddle and the factor of $1/\beta^{3/2}$ comes from computing the one-loop correction around this saddle, the theory in \eqref{eq:conclusion-bdy-action} has an (approximate) moduli space of saddles that we have to integrate over in order to reproduce the factor of  $1/\beta^{3/2}$. 

A related step towards completing the holographic dictionary is in identifying the meaning of the wavefunctions $\psi_E(x)$. Such wavefunctions arise in the quantum mechanics by considering the double-scaling limit of the orthogonal polynomials associated to the JT matrix integral. In the bulk theory, one can then wonder whether $\psi_E(x)$ becomes a Hartle-Hawking wave-function that can be computed using a gravitational path integral. If so, the boundary conditions needed to specify such a path integral, would consequently reveal the bulk meaning of $\hat x$.

 \subsection*{Gauge invariant observables}

It would be interesting to explore whether the explicit presence of null states, and their corresponding null deformations, can explicitly affect the experience of observers in the bulk or boundary theory. To explore this question it is useful to draw an analogy between the null deformations in gravity and gauge transformations in gauge theory.
In quantum field theory, whenever we find a redundancy in description (gauge freedom) we are prompted to identify the subset of gauge-invariant (physical) observables: for instance, in electromagnetism we learn that the field strength $F$ is physical, but not the gauge field $A$.

In this analogy, the gauge field configurations corresponds to the space of theories of quantum gravity. Gauge equivalent configurations correspond to gravitational theories related by a null deformation. 
In both cases, redundancies imply that not all observables that we thought were physical, actually are. For instance, the spacetime action in the gravitational theory is not gauge invariant since it is affected by null deformations of the dilaton potential. However, the full path integral over such actions is gauge invariant since it is insensitive to the action of null deformations.   More generally, we want to identify observables that actually \emph{are} invariant under the null deformations described in this paper.

\begin{figure}[t!]
    \centering
    \includegraphics[scale=0.62]{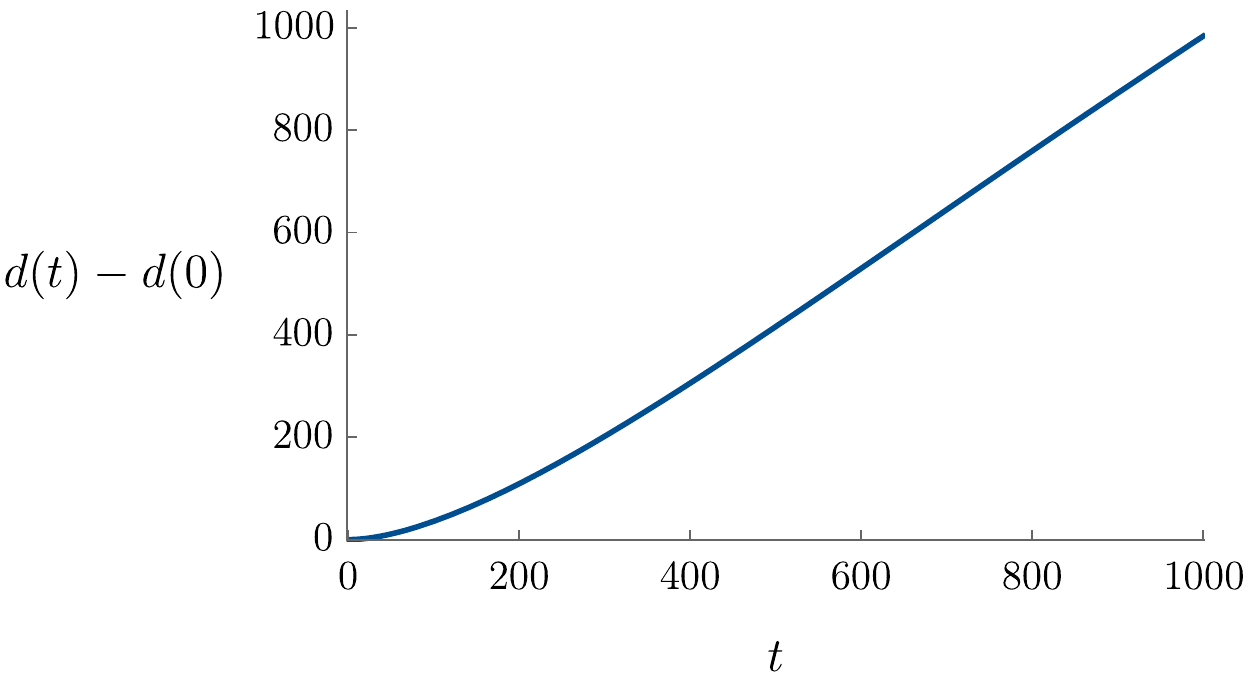}\qquad
     \includegraphics[scale=0.62]{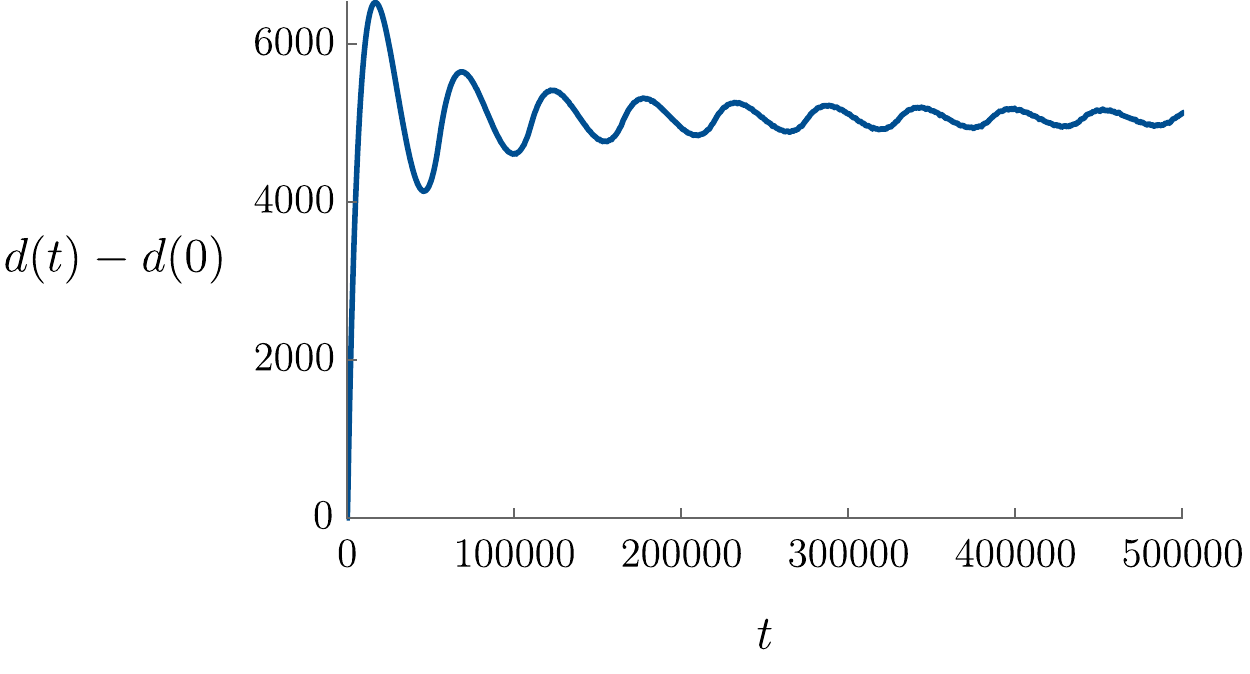}
    \caption{Length across the Einstein-Rosen bridge (or the volume of the black hole interior), including non-perturbative corrections, obtained by using \eqref{eq:spectral-decomp-geodesic-distance} for $u_{1}-u_{2} = \beta/2+i\,t$. \textbf{Left:} At early time our model exhibits a linear growth of the length, as expected in all gravitational theories. \textbf{Right:} At later times $\sim e^{\S}$, approximate recurrences appear due to the clock behavior of the energy spacing, for small energy differences. For higher values of $\beta$ the recurrences become more and more pronounced, and periods of linear growth and linear decay are more clearly visible. This should be contrasted with the results found in \cite{Iliesiu:2021ari} for theories with a chaotic spectrum for which the noise of the plateau was much smaller and much more erratic. }
    \label{fig:lengthER}
\end{figure}

In section \ref{subsec:bulk-interp-of-null states}, we have explicitly seen that the semiclassical description of the bulk is not ``gauge invariant'', in that there exist null deformations in the theory that change the leading classical solution but do not affect the ultimate exact spectrum of the theory. So only observables that can be defined non-perturbatively, in that they receive contributions from all geometries in the gravitational path integral, could be physical. To emphasize this point, consider the semiclassical definition of the distance between two boundary points, determined by their proper times $u_1$ and $u_2$. In the semiclassical description, this distance sensitively depends on the solution for $G(\Phi)$ in \eqref{eq:bulk-metric-ansatz}. Since $G(\Phi)$ depends on the exact form of the dilaton potential and thus is sensitive to null deformations, the geodesic distance also seems to naively depend on the choice of gauge for the dilaton gravity action. Assuming that $u_1$ and $u_2$ do not scale in any way with $e^{\S}$, an explicit calculation shows that the classical geodesic distance can change by an $e^{-\S}$ amount under null deformations of the dilaton potential.

Nevertheless, we can instead explicitly insert the geodesic distance between the two boundary points as an operator in the gravitational path integral
\be 
\label{eq:distance-in-grav-path-integral}
\average{d(u_1, u_2)}_\beta = \frac{1}{\mathcal Z^{\rm factorized}}\sum_{\substack{\text{geometries with}\\\text{one boundary}}}
    \int\mathcal{D}g\,\mathcal{D}\Phi\,d(u_1, u_2)\,e^{-I_{JT} - I_{\rm local} - I_{\rm nonlocal}}
    \ee
and once again expand the local and non-local dilaton potentials as brane insertions in the sum over geometries. Then, after some manipulations \cite{Blommaert:2021fob, Iliesiu:2021ari}, one finds that the distance $\average{d(u_1, u_2)}_\beta$ becomes
\be 
\label{eq:spectral-decomp-geodesic-distance}
\average{d(u_1, u_2)}_\beta \sim \text{constant } + {e^{-\S}}\sum_{i\neq j}^\infty \frac{e^{-(u_1-u_2)E_i - (\beta - u_1 - u_2)E_j} }{(E_i-E_j)\left(\cosh(2\pi \sqrt{E_i}) - \cosh(2\pi \sqrt{E_j})\right)}\,.
\ee
The answer \eqref{eq:spectral-decomp-geodesic-distance} only depends on the energies $E_i$ of the Hamiltonian $\Hh$ and, consequently, is insensitive to null deformations. For convenience we plot this answer (as a function of Lorentzian time $u_{1}-u_2 = \beta/2 + i t$) in figure \ref{fig:lengthER} to emphasize that this observable is explicitly computable in our model.  We thus see that while in the original theory of canonical JT gravity the saddle-point value of $d(u_1, u_2)$ is different than that in the null-deformed theory, the distances between $u_1$ and $u_2$ in both theories are in fact the same when accounting for all the corrections to the gravitational path integral  (due to the subleading non-local corrections and due to the explicit sum over all geometries).  

While invariance under null deformations can be checked on a case-by-case basis, we do not know of a general way to prove that all diffeomorphism invariant observables (in the sense of general relativity) are also gauge invariant (in the sense of the null deformations discussed above). This is because each observable requires a careful non-perturbative definition at the level of the gravitational path integral. This can be nontrivial, even in the definition of the distance \eqref{eq:distance-in-grav-path-integral} a choice which disallows self-intersecting geodesics in the measurement of distance has been made. While numerous boundary observables have non-perturbative definitions (for example, partition functions and correlators of probe matter fields) little is known about the non-perturbative definition of observables related to the experience of an in-falling observer (progress in this direction has been made in \cite{Jafferis:2020ora, Gao:2021tzr}), or even observables for static bulk observers (progress in this direction has been made in \cite{Blommaert:2019hjr,Blommaert:2020yeo,Blommaert:2020seb}). Without understanding whether such observables are gauge invariant in addition to diffeomorphism invariant, questions related to the
faith of an observer crossing the horizon are impossible to address. This problem is more pronounced for typical states which can for instance be obtained by  evolving a thermofield double state for a time scaling with $e^{\S}$, because of this large time non-perturbative corrections typically become dominant in the entire $e^{\S}$ expansion and thus quantities that are not gauge invariant could change by a wild amount under a null deformation. 

In summary, only observables that can be defined non-perturbatively in the matrix integral are gauge invariant, we need a non-perturbative definition to predict the outcome of an experiment.\footnote{Because, obviously, every experiment is by definition computing the expectation value of a gauge-invariant observable. By definition and by construction it is \emph{impossible} to do any measurement that detects a null deformation.} Semiclassics might be deceptive, especially inside black holes \cite{Almheiri:2021jwq}.

\section*{Acknowledgements}
We thank Don Marolf, Henry Maxfield, Steve Shenker, Douglas Stanford and Zhenbin Yang for valuable discussions. AB was supported by the ERC-COG Grant NP-QFT No. 864583 and thanks the SITP at Stanford for hospitality during part of this project. LVI was supported by the Simons Collaboration on Ultra-Quantum Matter, a Simons Foundation Grant with No. 651440. JK is supported by the Simons Foundation. This work was partly done at the  Aspen Center for Physics, which is supported by National Science Foundation grant PHY-1607611.  We furthermore thank the people of Stanford's SITP, Weizmann and Aspen Center for Physics for the opportunity to present pieces of this work at a preliminary stage, and for useful comments.

\bibliographystyle{ourbst}
\bibliography{Refs}

\end{document}